Accepted for publication in the ApJ: Dec 30, 2016

# GALAXY AND MASS ASSEMBLY (GAMA): EXPLORING THE *WISE* COSMIC WEB IN G12

T.H. Jarrett[1], M.E. Cluver[2], C. Magoulas[1], M. Bilicki[1,3,7], M. Alpaslan[4], J. Bland-Hawthorn[5], S. Brough[6], M.J.I. Brown[8], S. Croom[5], S. Driver[9,10], B. W. Holwerda[3], A. M. Hopkins[6], J. Loveday[11], P. Norberg[12], J.A. Peacock[13], C.C. Popescu[14,18], E.M. Sadler[5], E.N. Taylor[15], R.J. Tuffs[16], L. Wang[17]

*Accepted for publication in the ApJ: Dec 30, 2016*

## ABSTRACT

We present an analysis of the mid-infrared *WISE* sources seen within the equatorial GAMA G12 field, located in the North Galactic Cap. Our motivation is to study and characterize the behavior of *WISE* source populations in anticipation of the deep multi-wavelength surveys that will define the next decade, with the principal science goal of mapping the 3D large scale structures and determining the global physical attributes of the host galaxies. In combination with cosmological redshifts, we identify galaxies from their *WISE* W1 (3.4 $\mu$m) resolved emission, and by performing a star-galaxy separation using apparent magnitude, colors and statistical modeling of star-counts. The resultant galaxy catalog has $\simeq$590,000 sources in 60 deg$^2$, reaching a W1 5-$\sigma$ depth of 31 $\mu$Jy. At the faint end, where redshifts are not available, we employ a luminosity function analysis to show that approximately 27% of all WISE extragalactic sources to a limit of 17.5 mag (31 $\mu$Jy) are at high redshift, $z > 1$. The spatial distribution is investigated using two-point correlation functions and a 3D source density characterization at 5 Mpc and 20 Mpc scales. For angular distributions, we find brighter and more massive sources are strongly clustered relative to fainter and lower mass source; likewise, based on *WISE* colors, spheroidal galaxies have the strongest clustering, while late-type disk galaxies have the lowest clustering amplitudes. In three dimensions, we find a number of distinct groupings, often bridged by filaments and super-structures. Using special visualization tools, we map these structures, exploring how clustering may play a role with stellar mass and galaxy type.

## 1. INTRODUCTION

[1] Department of Astronomy, University of Cape Town, Private Bag X3, Rondebosch, 7701, South Africa
[2] Department of Physics and Astronomy, University of the Western Cape, Robert Sobukwe Road, Bellville, 7535, Republic of South Africa
[3] Leiden Observatory, University of Leiden, Netherlands
[4] NASA Ames Research Center, N232, Moffett Field, Mountain View, CA 94035, United States
[5] Sydney Institute for Astronomy (SIfA), School of Physics, University of Sydney, NSW 2006, Australia
[6] Australian Astronomical Observatory, PO Box 915, North Ryde, NSW 1670, Australia
[7] Janusz Gil Institute of Astronomy, University of Zielona Góra, ul. Szafrana 2, 65-516, Zielona Góra, Poland
[8] School of Physics and Astronomy, Monash University, Clayton, Victoria 3800, Australia
[9] International Centre for Radio Astronomy Research (ICRAR), University of Western Australia, 35 Stirling Highway, Crawley, WA 6009, Australia
[10] SUPA, School of Physics and Astronomy, University of St Andrews, North Haugh, St Andrews, Fife, KY16 9SS, UK
[11] Astronomy Centre, Department of Physics and Astronomy, University of Sussex, Brighton, BN1 9QH, U.K.
[12] Institute for Computational Cosmology, Department of Physics, Durham University, South Road, Durham DH1 3 LE, U.K.
[13] Institute for Astronomy, University of Edinburgh, Royal Observatory, Edinburgh EH9 3HJ, UK
[14] Jeremiah Horrocks Institute, University of Central Lancashire, PR1 2HE, Preston, UK
[15] School of Physics, the University of Melbourne, Parkville, VIC 3010, Australia
[16] Max Planck Institut fuer Kernphysik, Saupfercheckweg 1, 69117 Heidelberg, Germany
[17] SRON Netherlands Institute for Space Research, Landleven 12, 9747 AD, Groningen, The Netherlands
[18] The Astronomical Institute of the Romanian Academy, Str. Cutitul de Argint 5, Bucharest, Romania

The emergence of the elegant Universe, often portrayed as a great *cosmic web*, can be hierarchically cast with organization built from smaller particles colliding and coalescing into larger fragments, forming groups, clusters, filaments, walls and superclusters of galaxies (see the review in van de Weygaert & Bond 2005). The ultimate fate of the Universe, of the cosmic web, depends on the cosmological properties of the Universe, dominated by the elusive dark components of gravity and energy. Research efforts are focused on both the present day Universe (or the Local Universe, to indicate its physical and time proximity to us), and various incarnations of the early Universe from high redshift constructions of large scale structure (LSS), to the cosmic microwave background (CMB), linking the past to the present. Astronomers use galaxies as the primary observational marker or signpost by which to map out structure and study the dynamic Universe. Notwithstanding, galaxies are heterogeneous and time-evolving, observed to have a wide variety of shape, size, morphology and environmental influence; it is therefore central to any effort in precision cosmology to understand the diverse populations and key physical processes governing star formation, supernovae feedback, and black hole growth, for example.

In the last three decades, mapping and characterizing LSS, and its galaxy constituents has swiftly advanced chiefly due to wide-area redshift surveys, of notable reference the 2dF Galaxy Redshift Survey (Colless et al. 2001), Sloan Digital Sky Survey (SDSS, Eisenstein et al. 2011), 2MASS Redshift Survey (Huchra et al. 2012) and the 6dF Galaxy Survey (Jones et al. 2004, 2009).

These are relatively shallow surveys (sacrificing depth for breadth) that focus on the Local Universe, in contrast to the many pencil-beam (narrow, $<1$ deg$^2$) studies that extend large aperture-telescope spectroscopy to the early Universe. Bridging the gap between narrow and broad redshift surveys are the so-called "deep-wide" efforts, which attempt to push the sensitivity limits of moderately-sized telescopes using fast and efficient multi-object spectrographs.

One such effort is the Galaxy and Mass Assembly (GAMA; Driver et al. 2009, 2011) survey, which used the 2-degree field multi-object fibre-feed (2dF) to the AAOmega spectrograph on the Anglo-Australian Telescope (AAT) to efficiently target several large equatorial fields, building upon the SDSS measurements by extending $\sim$2 mags deeper with high completeness, devised to fully sample galaxy groups and clusters. Three primary fields: G09, G12, and G15, cover a total of 180 deg$^2$ and $\sim$200,000 galaxies (Hopkins et al. 2013), reaching an overall median redshift of $z \sim 0.22$, but with a significant high redshift (luminous) component. The survey was designed to survey enough area and redshift space, hence volume, to be useful for galaxy evolution, LSS and cosmological studies. In addition to crucial cosmological redshifts, GAMA has also collected and homogenized a vast multi-wavelength ancillary data from X-ray/ultraviolet to far-infrared/radio wavelengths, constructing a comprehensive database to study the individual and bulk components of LSS (Liske et al. 2015; Driver et al. 2016).

A number of detailed studies[19] have been published or are currently underway. One of which, Cluver et al. (2014), henceforth, referred to as Paper I), specifically studied GAMA redshifts combined with the ancillary mid-infrared photometry from the Wide-field Infrared Survey Explorer (*WISE*; Wright et al. 2010), focusing on the stellar mass and star formation properties of galaxies. *WISE* is particularly suited to this end; the 3.4 $\mu$m (W1) and 4.6 $\mu$m (W2) bands trace, with minimal extinction, the continuum emission from low-mass, evolved stars, similar to the near-infrared bands at low redshift, while also having longer wavelength bands that are sensitive to the interstellar medium and star formation activity (Jarrett et al. 2011): the 12 $\mu$m (W3) band is dominated by the stochastically-heated 11.3 $\mu$m PAH (polycyclic aromatic hydrocarbon) and 12.8 $\mu$m [NeII] emission features, while the 22 $\mu$m (W4) light arises from a combination of warm, small grain, and cold, small grain in equilibrium, dust continuum that is reprocessed radiation from star formation and AGN activity (see for example Popescu et al. 2011). Combining the GAMA stellar masses (Taylor et al. 2011) and H$\alpha$ star formation rates (SFR) with the *WISE* luminosities, Paper I derived a new set of scaling relations for the dust-obscured SFRs and the host stellar mass-to-light (M/L) ratios.

Paper I demonstrated the diverse applications of combining mid-IR *WISE* photometry with redshift measurements. It did not, however, focus on the distribution of *WISE* sources within the GAMA G12 redshift range; instead it is this current work that extends the GAMA-*WISE* analysis to consider the 3D distribution and the nature of the *WISE* source population, including those beyond the Local Universe. This dual approach is motivated by the fact that *WISE* is a whole sky survey. The next generation large-area surveys, including the SKA-pathfinder radio HI (e.g. WALLABY, Koribalski 2012) and continuum surveys (MIGHTEE, Jarvis 2012, EMU, Norris et al. 2011), LSST (Ivezic et al. 2008), VIKING (Edge et al. 2013) and KiDS (de Jong et al. 2013), will require ancillary multi-wavelength data to make sense of their new source populations, and all-sky surveys such as *WISE* are particularly useful to this end. Consequently, it is vital that we understand the *WISE* source population and its suitability of probing clustering on small and large scales, from local galaxies to those in the Early Universe that drive the key science goals of deep radio surveys. Recent studies (e.g., Jarrett et al. 2011; Assef et al. 2013; Yan et al. 2013), attempted to characterize *WISE* sources using multi-wavelength information (using for example SDSS). Due to its volume that extends beyond the Local Universe, to z $\sim$0.5, and high completeness, $>$95% to r=19.8, of GAMA, we can study evolutionary changes in the host galaxies. Moreover, *WISE* is sensitive to galaxies well beyond these limits, as we show in this current study, to the epoch of active galaxy formation at z $\sim$1 to 2.

In this study, we focus on the source count distributions, galaxy populations, angular correlations and the 3D LSS of *WISE*-detected sources cross-matched with GAMA redshifts in the 60 deg$^2$ region of G12. G12 was chosen because it is one of the most redshift-complete fields of GAMA, and is located near the North Galactic Cap, which complements a study currently underway of the South Galactic Cap (see below). In the case of source counts and *WISE* photometric properties, this study is similar to that of Yan et al. (2013) who characterized the *WISE*-SDSS combination, except that in our case the GAMA redshifts extend to much greater depths and we attempt to map the LSS. This study considers the nature of sources that are well beyond the detection limits of either redshift survey, probing to depths beyond $z \sim 1$.

Our central goal is to map the LSS and the clustering characteristics in terms of the spatial attributes, flux (counts) and the fundamental galaxy properties. Recent studies that use GAMA to study clustering (e.g., Farrow et al. 2015; McNaught et al. 2014), and galaxy groups (e.g., Alpaslan et al. 2015) are more comprehensive to the specific topic, for example, employing the two-point correlation function, cluster finding methods, environmental influence on luminosity function evolution; whereas the study presented in this work has a broader perspective on the cosmic web contained within the G12 cone. At the opposite end of the sky, we are also looking at the South Galactic Pole, using similar methods to understand the nature and distribution of sources, but without GAMA information; the results will be presented in an upcoming publication (Magoulas et al. *in preparation*). Finally, at the largest angular scales, and using GAMA to produce detailed training sets, we are combining *WISE* and SuperCOSMOS to produce a 3D photo-z view of $3\pi$ sky (Bilicki et al. 2014, 2016).

This paper is organized as follows. The *WISE* and GAMA datasets are introduced in §2. Source properties, such as photometry, number counts, redshift distributions and spatial projections are presented in §3, where we focus on resolved sources in *WISE* – which require careful measurements – and star-galaxy separation since a large fraction of field-sources are in fact Galactic in nature. Constructing a *WISE*-GAMA galaxy catalog, §4 then presents the properties of the galaxies, including SFR and stellar masses, clustering and overdensities at $\sim$few Mpc and larger scales, angular and radial correlations, and finally 3D maps of the region, followed by a summary.

[19] http://www.gama-survey.org/pubs/

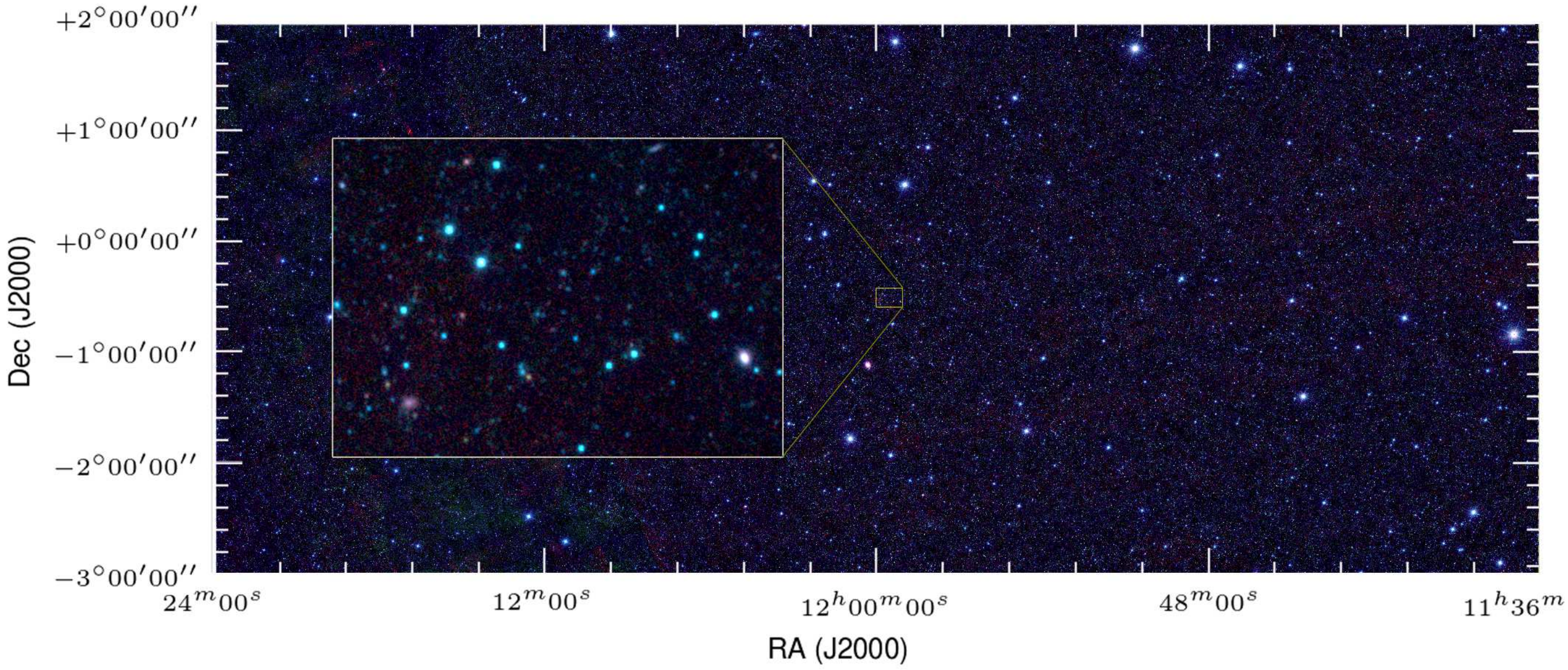

**Figure 1.** *WISE* equatorial view of the G12 field, covering 60 deg$^2$. The four bands of *WISE* are combined to create the color image; respectively, 3.4 $\mu$m (blue), 4.6 $\mu$m (green), 12 $\mu$m (orange) and 22 $\mu$m (red). The inset shows a zoomed view, $\sim$14 × 11 arcmin. In general, foreground stars appear blue in color, while background galaxies are red. There are nearly 1 million *WISE* sources in the G12 field.

The cosmology adopted throughout this paper is $H_0 = 70\,\mathrm{km\,s^{-1}\,Mpc^{-1}}$, $\Omega_M = 0.3$ and $\Omega_\Lambda = 0.7$. The conversions between luminosity distance and redshift use the analytic formalism of Wickramasinghe & Ukwatta (2010) for a flat, dark energy dominated Universe, assuming standard cosmological values noted above. Volumes length and size comparisons are all carried out within the co-moving reference frame. All magnitudes are in the Vega system (*WISE* photometric calibration described in Jarrett et al. 2011) unless indicated explicitly by an AB subscript. Photometric colors are indicated using band names; e.g., W1−W2 is the [3.4 $\mu$m]−[4.6 $\mu$m] color. Finally, for all four bands, the Vega magnitude to flux conversion factors are 309.68, 170.66, 29.05, 7.871 Jy, respectively, for W1, W2, W3 and W4. Here we have adopted the new W4 calibration from Brown et al. (2014b) , in which the central wavelength is 22.8 $\mu$m and the magnitude-to-flux conversion factor is 7.871 Jy. It follows that the conversion from Vega System to the monochromatic AB System conversions are 2.67, 3.32, 5.24 and 6.66 mag.

## 2. DATA AND METHODS

The primary data sets are derived from the *WISE* imaging and GAMA spectroscopic surveys. Detailed descriptions are given in Paper I and we refer the reader to this work. There are some differences in the data and methods, however, and below we provide the necessary information for this current study.

### 2.1. *WISE Imaging and Extracted Measurements*

Point sources and resolved galaxies are extracted from the *WISE* imaging in the four mid-infrared bands (Wright et al. 2010): 3.4 $\mu$m, 4.6 $\mu$m, 12 $\mu$m and 22 $\mu$m.[20] In the case of point sources, we use the ALLWISE public release archive (Cutri et al. 2012), served by the NASA/IPAC Infrared Science Archive (IRSA), updated from Paper I which used the AllSky public release data. Since the ALLWISE catalogs are optimized for point sources, in the case of resolved sources we re-sampled the image mosaics and extracted the information accordingly (see below).

The equatorial and North Galactic Cap G12 field, encompassing 60 deg$^2$, contains 803,457 *WISE* sources in total with $\geq$5-$\sigma$ sensitivity in W1, or 13,400 deg$^{-2}$. Many of which are detected at 4.6 $\mu$m, and many less detections in the 12 $\mu$m and 22 $\mu$m bands. To contrast with this impressive total, the total number of 2MASS Point Source Catalog (PSC) sources in the field is 1,600 deg$^{-2}$, and the 2MASS Extended Source Catalog (2MXSC; Jarrett et al. 2000) has far fewer, only 40 deg$^{-2}$. As we discuss in the next section, the resolved *WISE* sources are similar in number to the 2MXSC.

Confusion from Galactic stars is at a minimum in the Galactic caps, and based on our starcount model (see below), we expect no more than 3% of our extragalactic sample to have a star within 2 beam widths. Most of these stars are relatively faint and would only contribute a small percentage to the integrated flux. Blending with other galaxies, however, can be significant at the faint end where the source counts are at their peak.The bright end, represented by the GAMA selection, is expected to have a blending fraction of $\sim$ 1% (Cluver et al. 2014). As will be shown, the faint end, W1 > 17 mag, may have as many as $10^4$ galaxies per deg$^2$, which translates to 9% contamination for galaxies at the faint end (creating a well-known flux overbias). Bright galaxies will also have blending from faint galaxies, but the flux contamination is insignificant.

[20] Note however, the W4 filter response has a more redward sensitivity than first understood, its central wavelength is closer to 22.8 $\mu$m and has a color response similar to MIPS24; see Brown et al. (2014b).

Since the *WISE* mission did not give priority to extracting and properly measuring resolved sources, it is an absolute necessity to carefully do this using *WISE* imaging and appropriate photometry characterization tools. We carry out these tasks; first, we re-construct the image mosaics to recover the native resolution of *WISE* – which is not the case for the public-release mosaics –, and second we employ tools to extract and measure the extended sources.

Re-sampling with 1″ pixels using a 'drizzle' technique developed in the software package ICORE (Masci 2013) specifically designed for *WISE* single-frame images, we achieve a resolution of 5.9″, 6.5″, 7.0″ and 12.4″ at 3.4 $\mu$m, 4.6 $\mu$m, 12 $\mu$m and 22 $\mu$m, respectively, which is ∼30% improved from the public release "Atlas" imaging which is degraded to benefit point source detection; methods and performance are detailed in Jarrett et al. (2012). The resulting *WISE* imaging is showcased in the color panorama, Figure 1, where all of the mosaics have been combined to form one large view of the 60 deg$^2$ field. Inside are nearly a million *WISE* sources, including a few thousand resolved galaxies. The inset reveals the various kinds of sources, including stars, which appear blue, background galaxies (red-colored) and resolved galaxies, which are fuzzy and red, depending on the dust content and thermal properties.

As detailed in Paper I, resolved source extraction involves a number of steps. Candidate resolved sources are drawn from the ALLWISE catalog as follows: those sources with deviant, >2 W1 profile-fit reduced-$\chi^2$, and associated 2MASS resolved sources since resolved 2MASS galaxies are usually resolved by *WISE*; see Jarrett et al. 2013. Candidate sources are then carefully measured using the newly recast *WISE* mosaics and custom software that has heritage to the 2MASS XSC (Jarrett et al. 2000) and *WISE* photometry pipelines (Jarrett et al. 2011; Cutri et al. 2012; Jarrett et al. 2013). The automated pipeline extracts photometry, surface brightnesses, radial profiles and other attributes that are used to assess the degree of extended emission; i.e. beyond the expected point source profile of stars. Visual inspection and human-intervention are used for difficult cases, especially with source crowding, a major problem arising from the relatively large beam compared to, for example, *Spitzer*-IRAC or optical imaging, and added sensitivity of the 3.4 $\mu$m band.

Removal of foreground stars and other contaminants enables a clean and accurate characterization of the resolved *WISE* sources, including various combinations of resolved and unresolved bands – while W1 and W2 may be resolved, typically W3 and W4 are unresolved. With this identification and extraction method, we find 2,100 resolved *WISE* sources in the G12 field (35 deg$^{-2}$), which we refer to as the WXSC (*WISE* Extended Source Catalog).

We should caution that the WXSC is limited to sources that are clearly resolved in at least one *WISE* band; there are many more sources that are compact, but marginally resolved beyond the *WISE* PSF. These sources cannot be identified using the reduced-$\chi^2$ and will therefore not be in the initial WXSC selection. These cases will have systematically under-estimated profile-fit (WPRO) fluxes, and hence for extragalactic work, in which the target galaxies are local – for example using a sample such as SDSS/GAMA – it is better to use the ALLWISE Standard Aperture photometry or use your own circular aperture measurements that are appropriate to the size scales under consideration; more details can be found in Cluver et al. (in preparation) and Wright et al. (2016) .

### 2.2. *GAMA*

The spectroscopy and ancillary multi-wavelength photometry are drawn from the GAMA G12 field of the GAMA survey (Driver et al. 2009, 2011). The field is located at the boundary of the North Galactic Cap: (glon, glat) = 277, +60 deg, and equatorial R.A. between 174 and 186 deg, Dec between −3 and +2 deg; see Figure 1. There are approximately 60,000 sources with GAMA redshifts in the field, or 1000 deg$^{-2}$. It is important to note that pre-selection filtering using an optical-NIR color cut removed stars, QSOs and, in general, point sources (unresolved by SDSS) from the GAMA target list. Later we use these 'rejected' sources to help assess the stellar contamination in our *WISE*-selected catalogs. More details of the GAMA data, catalogs and derived parameters can be found in, for example, Baldry et al. (2010); Robotham et al. (2010); Taylor et al. (2011); Hopkins et al. (2013); Gunawardhana et al. (2013); Cluver et al. (2014). We expect SDSS point sources to also be unresolved by WISE. We will show that we are able to discern the unresolved extragalactic population from the Galactic stellar population, and hence recover distant galaxies, QSOs and the rich assortment of extragalactic objects.

Position cross-matching was carried out between the GAMA G12 redshift catalog and the *WISE* sources (ALLWISE + WXSC) using a 3″ cone search radius, which is generously large to capture source-blending cases. For each GAMA source, the match rate with *WISE* was well over 95%, forming a complete set from the GAMA view. From the viewpoint of *WISE*, only 1% of its sources have a GAMA counterpart. As will be shown in the next section, a fraction of *WISE* sources are Galactic stars and hence should not be in GAMA galaxy catalogs, although stars are used for flux calibration. But most are faint background galaxies, beyond the GAMA survey limit. In some cases, because of the large *WISE* beam and source blending, there can be more than one GAMA source per *WISE* counterpart; i.e., two separate optically-characterized galaxies are blended onto one *WISE* detected source. This problem is not widespread, however, as only 1.2% *WISE* sources have more than one GAMA cross-match within a 5″ radius, what is referred to as a 'catastrophic blend' in Paper 1, and does not adversely affect the GAMA-*WISE* statistics or analysis. More discussion of the GAMA-*WISE* blending is found in Paper I, but see also Wright et al. (2016) for a multi-wavelength deblending analysis of all GAMA photometry.

### 2.3. *Other data*

Radio-based observations are of interest to this and other multi-wavelength studies because of the SKA

pathfinders (e.g., JVLA) that are now coming online. Here we look at the number count and mid-IR color properties of galaxies detected in the Faint Images of the Radio Sky (FIRST) radio survey, as collated and classified in the Large Area Radio Galaxy Evolution Spectroscopic Survey (LARGESS; Ching et al. 2017), which covered 48 $deg^2$ of G12.

## 3. SOURCE CHARACTERIZATION RESULTS

In this section we present the photometry, cross-matching, source counts and statistics for the sources in the G12 field. Cross-matches between *WISE* and GAMA, as well as the resolved sources, provide the definitive extragalactic sample. Beyond the GAMA sensitivity limits lie most of the *WISE* sources, comprised of foreground stars and, >10× in number, background galaxies. We employ star-galaxy separation analysis to isolate a pure extragalactic catalog, which we then characterize using an infrared luminosity function of galaxies in the Local Universe.

### 3.1. *Observed flux properties*

*WISE* source detection sensitivity depends on the depth of coverage, which in turn depends on the ecliptic latitude of the field in question (see Jarrett et al. 2011). In the case of G12, the depth in the W1 (3.4 $\mu$m) band is about 25 coverages (i.e., 25 individual frames or epochs), and for the 800,000+ sources in the field, the S/N = (10/5) is $\sim$ 56/28 $\mu$Jy, in terms of Vega mags, 16.85 and 17.62, respectively. For W2 (4.6 $\mu$m) sensitivity, sources have S/N (10/5) limits of 118/57 $\mu$Jy,15.41 and 16.19 mag, respectively. Both W1 and W2 are sensitive to the evolved populations that dominate the near-infrared emission in galaxies, and hence are generally good tracers of the underlying stellar mass. These near/mid-infrared bands, however, are not without confusing elements that may arise from warm continuum and PAH emission produced by more extreme star formation (e.g., M82 has a relatively strong 3.3 $\mu$m PAH emission line) and active galactic nuclei, both of which would lead to an overestimate of the aggregate stellar mass (see e.g., Meidt et al. 2014).

The longer wavelength bands, tracing the star formation and ISM activity in galaxies, are not as sensitive as the W1 and W2 bands. In addition, their coverage is 2× lower (having not benefited from the second 'passive-warm' passage of *WISE*); W3 (12 $\mu$m) has S/N (10/5) limits of 1.44/0.67 mJy,10.76 and 11.59 mag, respectively, and W4 (22 $\mu$m) has S/N (10/5) limits of 10.6/5.0 mJy, 7.2 and 8.0 mag, respectively. The W1 S/N limits are close to the confusion maximum achieved by *WISE* (see Jarrett et al. 2011) and hence can detect $L_*$ galaxies to redshifts of $z \sim 0.5$. Conversely, the relatively poor sensitivity of the long wavelength channels means that only nearby galaxies are detected, and the rarer luminous infrared galaxies at greater distances (e.g., Tsai et al. 2015).

Our detection and extraction of resolved sources (see §2) draws $\sim$2,100 sources. These sources range from large – well resolved, multi-component galaxies – to small fuzzies reaching W1 depths of $\sim$0.5 mJy (14.5 mag in Vega). A representative sampling is shown in Figure 2. At the bright and large angular size end, it is computationally intensive to remove foreground stars and deblend other stars or galaxies, in general. Human 'expert' user intervention to the pipeline is particularly important when bright sources (stars or large galaxies) are in close proximity to the resolved target. Fortunately this number is relatively small. At the faint end, resolved sources are compact and can easily be confused with noise and complex, multi-component objects. For our resolved catalog, WXSC, we limit our study to clearly resolved, discrete objects (see e.g., Figure 2).

The GAMA survey covered the G12 field with high spectroscopic completeness ($\simeq$ 98.5%; Liske et al. 2015) to a limiting magnitude of $r_{\rm AB} = 19.8$ (Driver et al. 2009, 2011; Cluver et al. 2014) and a median redshift of $\sim$0.22. Assuming an r-W1 color of 0.5 mag, the corresponding W1(AB) is 19.3 mag, which is 16.6 mag in the Vega system, or 70 $\mu$Jy. Since *WISE* reaches much fainter depths, it means virtually every GAMA source has a *WISE* counterpart (see §2); while in some cases of blending, there are more GAMA sources than *WISE* sources (Paper I). The redshift range of GAMA is particularly suited to studying populations with $z < 0.5$, although much more distant, luminous, objects are cataloged in the survey. Cross-matching GAMA-G12 with the ALLWISE sources results in $\sim$58,000 sources, or 1000 $deg^{-2}$ (compared to 13,400 $deg^{-2}$, the cumulative number of *WISE* sources); see Figure 3. The few GAMA sources that are not in *WISE* are either *WISE* blends (i.e., two sources blended into one) or optically low surface brightness, as well as low-mass galaxies, of which infrared surveys tend to be insensitive to because of their low mass (hence, low surface brightness in the near-IR bands which are sensitive to the evolved populations) and those that are low opacity systems.

Differential W1 source counts for the three (ALLWISE total, WXSC, GAMA) lists are shown in Figure 3. Resolved sources perfectly track the bright end of GAMA galaxies, to a magnitude of $\sim$13.4, where they turn over, revealing the completeness limit of the WXSC: 1.35 mJy at 3.4 $\mu$m. The total number of resolved *WISE* galaxies is comparable to the number of resolved 2MASS galaxies (2MXSC), as can be seen in Fig. 3 and Table 1. The GAMA counts continue to rise to a limit of $\sim$15.6 mag (0.18 mJy) where they roll over with incompleteness, with the faintest GAMA redshifted sources reaching depths of $\sim$50 $\mu$Jy. Finally, the ALLWISE counts are much (10×) greater, although rising with a flatter slope. As we show below, this slope is being driven by Galactic stars, dominating the counts at the bright (W1 < 13th mag) end.

### 3.2. *Stars vs. Galaxies*

In this section we concentrate on separating the Galactic and extragalactic populations. The traditional methods for star-galaxy separation are employed, including the use of apparent mags and colors in conjunction with our knowledge of stellar and galaxy properties and their spatial distribution. Here we utilize a Galactic star-count model that yields both spatial and photometric information that we can expect to observe in the Galactic polar cap region. Finally, as discussed in the previous section, for the local universe, $z$ <0.2, we also identify galaxies by their resolved – low surface brightness – emission relative to point sources. However, this is only a small fraction of the total extragalatic population observed by WISE.

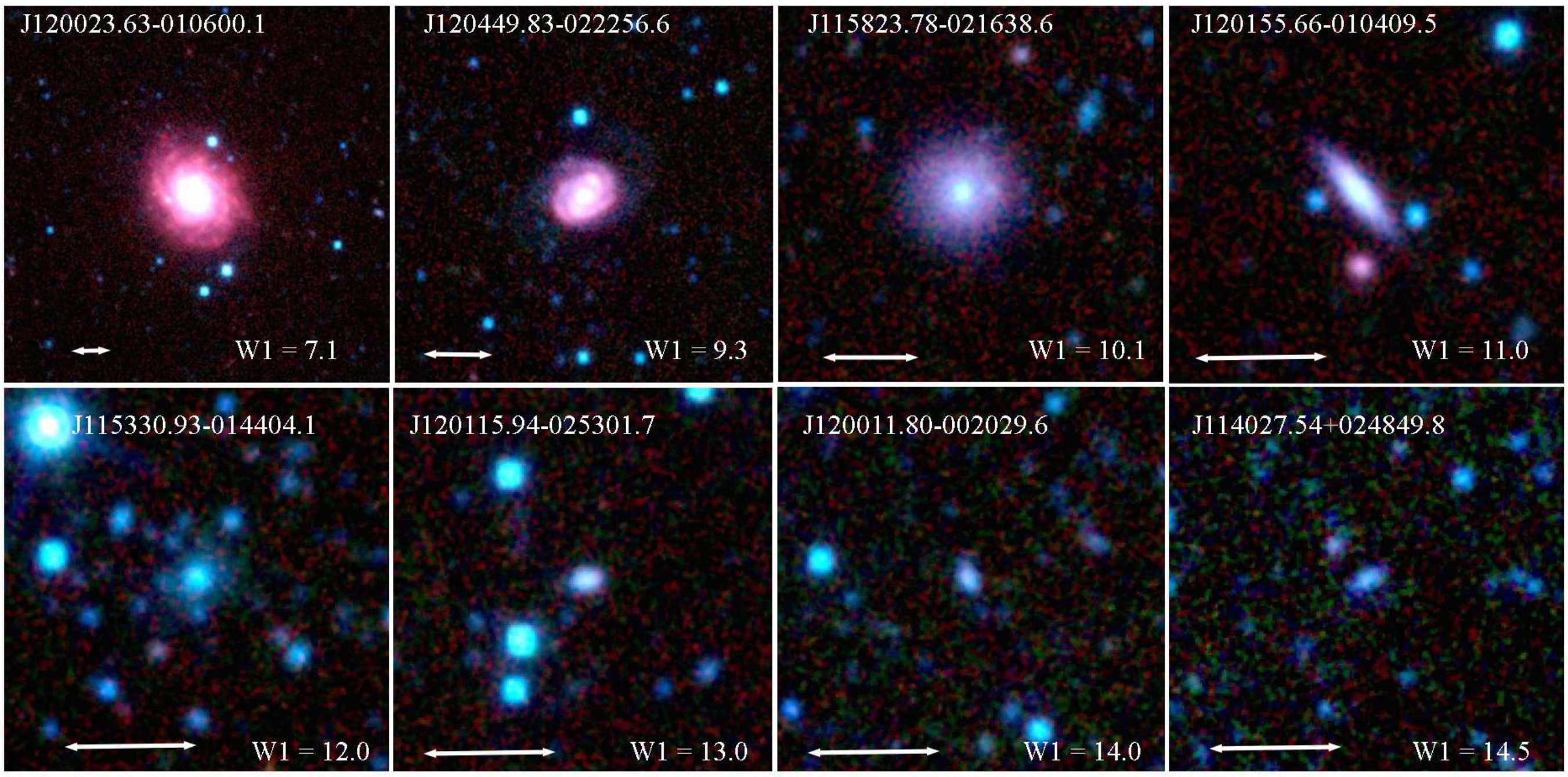

**Figure 2.** Examples of *WISE* resolved sources, ranging from bright (7.1 mag) to the faint (14.5 mag). Stars have not been removed in these examples. The faint end is limited by the angular resolution of the W1 imaging and, to a lesser extent, by the S/N. Image scale. 1 arcmin, is indicated by the arrow..

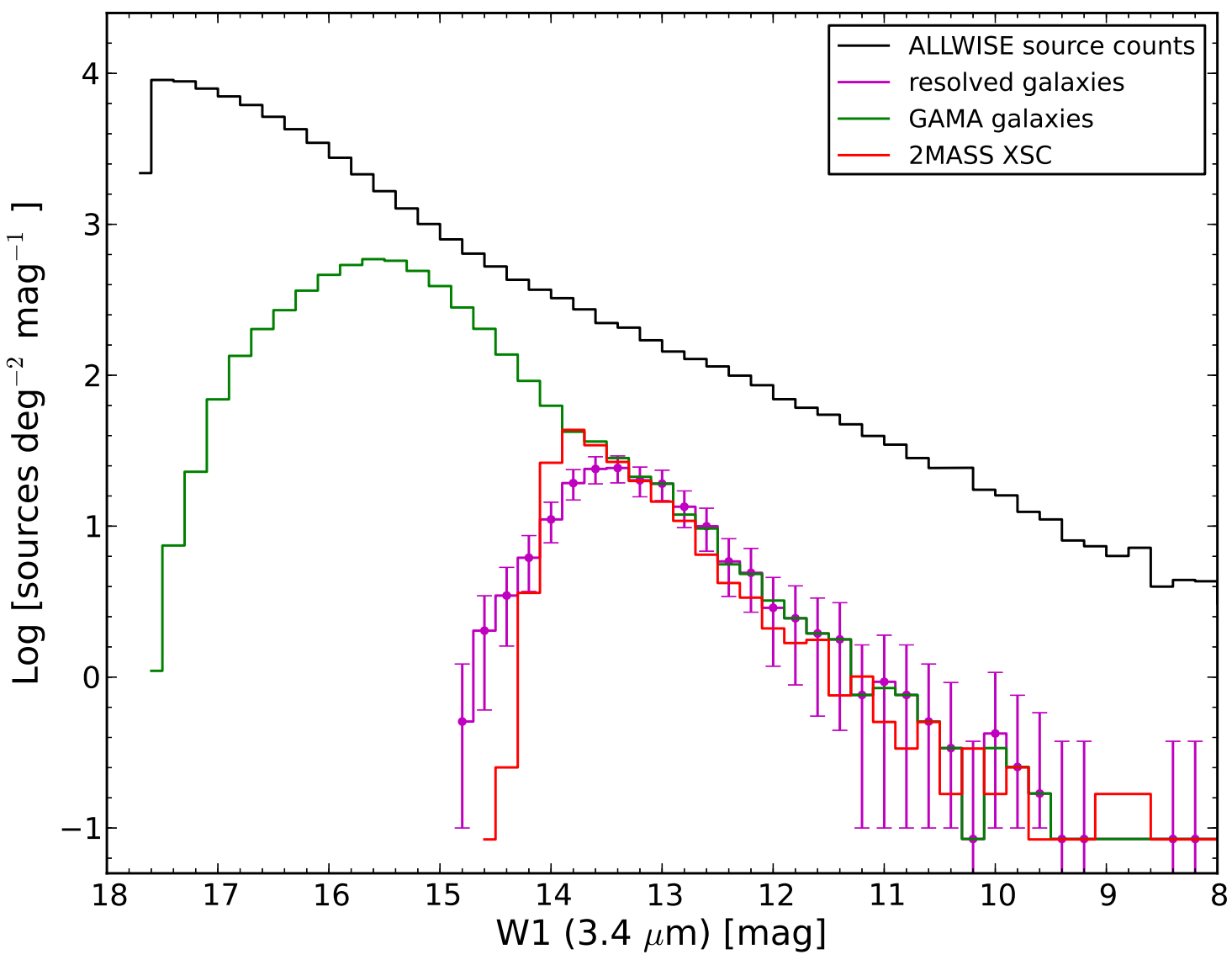

**Figure 3.** Differential W1 (3.4 $\mu$m) source counts in the G12 region; magnitudes are in the Vega system. The ALLWISE catalog of sources is shown in grey; cross-matched GAMA sources are delineated in green and resolved sources in black (with 1-$\sigma$ Poisson error bars). *WISE* detections are limited to S/N = 5, peaking and turning over at W1 $\sim$ 17.5 mag (31 $\mu$Jy). The total number of sources is $\sim$800,000, of which about 7.5% (60K) have GAMA redshifts, and about 2100 are resolved in the W1 channel. For comparison, the 2MASS XSC K-band galaxy counts for the G12 region are shown (red), where the constant color K-W1=0.15 mag has been applied.

We expect the bright sources in the *WISE* ensemble to be dominated by foreground Milky Way stars (see e.g., Jarrett et al. 2011). We demonstrate this using a 3-component (disk dwarf/giant, spheroidal) Galactic exponential star-count model, adapted from Jarrett et al. (1994) for optical-infrared applications. In addition to standard optical bands, the model incorporates the near-infrared (J, H, Ks) bands and the mid-infrared (L, M and N) bands, and was successfully applied to 2MASS, *Spitzer-SWIRE* and deep IRAC source counts (e.g., Jarrett et al. 2004, 2011). Here we estimate the L-band 3.5 $\mu$m counts corresponding to the Galactic coordinate location of the G12 field, and compare to the W1 (3.4 $\mu$m) counts. Note that we assume that the L-band (3.5 $\mu$m) and W1 (3.4 $\mu$m) band are equivalent for this exercise.

The resulting Galactic Cap star-counts are shown in Figure 4. Compared to the real *WISE* counts, the model suggests that stars dominate when W1 $<$ 14th mag. For the brightest *WISE* sources ($<$ 8th mag) evolved giants are the main contributor of the source population. Else for all other flux levels, it is main-sequence (M-S) dwarf stars that dominate the counts. K and M-dwarfs are the most challenging spectral type to separate from the extragalactic population because of their prodigious number density and colors that are similar to the evolved population in galaxies. At the faint end, W1 $>$ 17th mag, the star-counts become flat and the M-S population begins to decline in number, whereas the more distant Galactic spheroidal (halo and sub-dwarf) population is rising quickly, dominating the counts beyond the limits of the *WISE* survey. Compared to the *WISE* source counts, the star-counts contribute much less to the faint end, a factor of 2 less at W1=15th mag and a factor of 10 less at W1=17 mag. Nevertheless, this is still enough sources to render our galaxy catalogs unreliable, notably where GAMA sources drop off, and thus we require star-galaxy filtering to purify our catalogs.

Separating foreground stars from background galaxies is a challenging process, largely due to degeneracies in the parametric values of both populations. For instance, both are unresolved point sources – except, of course, for the tiny population of resolved extragalactic sources – and may share similar color properties in certain broad bands (see, for example, Yan et al. 2013; Kurcz et al. 2016. Kinematic information, which may be definitive, such as reliable radial and transverse motions, is difficult and expensive to acquire. For the most part, we are left with photometric information to delineate stars from galaxies. Here we use the near and mid-infrared information to study the photometric differences. Note that since we already have GAMA redshifts, shown in the next section to be complete to W1 $\simeq$ 15.5 mag, our aim is to separate stars from galaxies in the fainter population ensemble. Nevertheless, we consider the full observed flux range.

We first explore the W1 and W2 parameter space, the most sensitive *WISE* channels. We incorporate known populations to aid the analysis, including the GAMA cross-matches (confirmed galaxies), resolved (also confirmed galaxies) and *WISE* matches with SDSS QSO's. The latter were extracted from the SDSS DR12 based on their quasar classification (DR12, Alpaslan et al. 2015); we use this population in general as an AGN tracer. It should be noted that GAMA was not optimized to study AGN, and most QSOs and distant AGN are culled from the original GAMA selection. However, we do expect Seyferts and other low-power AGN to be in our sample. Finally, we have compiled a list of sources believed to be unresolved, including rejects of the original GAMA color selection (Baldry et al. 2010), either known SDSS stars or unresolved sources which may be distant galaxies, although less likely in the brighter magnitudes. We call this group "SDSS stars or rejects", but it is not an exhaustive list to any degree, and is only used as a qualitative guide as to where some stars may be located in the diagrams to follow.

The color-magnitude diagram (CMD) results are shown in Figure 5a. The large – nature unknown – ensemble of *WISE* sources are shown in greyscale, and the known populations are labeled accordingly. We denote an S0-type galaxy track (dashed line), allowing it to change (curving redward) its W1−W2 color with increasing redshift. Classic QSOs are expected to be located above the W1−W2=0.8 mag threshold (Stern et al. 2012; see also Assef et al. 2013), although lower power AGN and Seyferts may have much bluer colors due to their host galaxy dominating the mid-IR emission (Jarrett et al. 2011). As expected, the QSO (cyan contours) population is faint (W1 $>$ 15 mag) and red in the W1−W2 color. GAMA (green contours) galaxies span the entire range, but generally have W1−W2 colors less than 0.5 mag. Nearby resolved galaxies (blue contours) are bright, and relatively blue in W1−W2 color. The reason that nearby galaxies are relatively blue compared to their fainter counterparts is because of cosmological band-shifting – *WISE* galaxies become redder in the W1−W2 color (illustrated later in this paper). For this parameter space, the only obvious separation is that foreground stars are brighter in W1, as is expected from Figure 4), and bluer than most galaxies. There is a clear degeneracy at the fainter magnitudes where low S/N halo dwarfs confuse the CMD and redder stellar populations (e.g., M and L-dwarfs) become important. Finally, and as will be clearly demonstrated in the next section, stars tend to dominate the total source counts for W1 $<$ 14.5 mag (0.5 mJy), while galaxies become the dominant population for magnitudes fainter than this threshold.

The separation of populations appears clearer in the W2−W3 CMD, Figure 5b, where stars are considerably brighter and bluer than galaxies. Unfortunately W3 is less sensitive in flux compared to W1 and W2 and far fewer sources are detected in this band. Note that stars have very faint W3 fluxes because the Rayleigh-Jeans (R-J) tail for evolved giants is dropping fast at mid-infrared wavelengths. Hence, if W3 is detected at all, and W1 $>$ 12th mag, it means the source is very likely a galaxy with some star formation (SF) activity. There is a small grouping of rejects at faint magnitudes, which are plausibly unresolved galaxies or those hosting AGN.

Exploring this SF aspect further, we now look at the *WISE* color-color diagram, Figure 6a, often used for morphological classification (see §4.1, below). There is now clear separation of the QSOs, which fill the AGN box proposed by Jarrett et al. (2011). GAMA galaxies span the disk and spiral galaxy zone, as do the resolved sources; i.e., they are likely very similar galaxy types. Stars tend to have ∼zero color, well-separated from the extragalactic population. There is a grouping of unknown sources at the blue end, just above the rejects and to the left (blueward) of the nearby galaxies. What are these sources? Too blue to be distant galaxies, and slightly too blue for resolved galaxies, which should catch all the nearby ellipticals, lenticulars and other quiescent or quenched galaxies. It is possible these are blends, likely a combination of a blue foreground star and a fainter and redder background galaxy. We know that about 1% of the *WISE* galaxies have blended pairs (Cluver et al. 2014), while some 1-3% may have blends or confusion from faint foreground stars. Visual inspection of a random sampling of these odd color sources reveal only nominal galaxies whose emission is dominated by luminous-evolved populations; i.e., early type galaxies. It is possible these are relatively early-type, large galaxies with minor blending contamination.

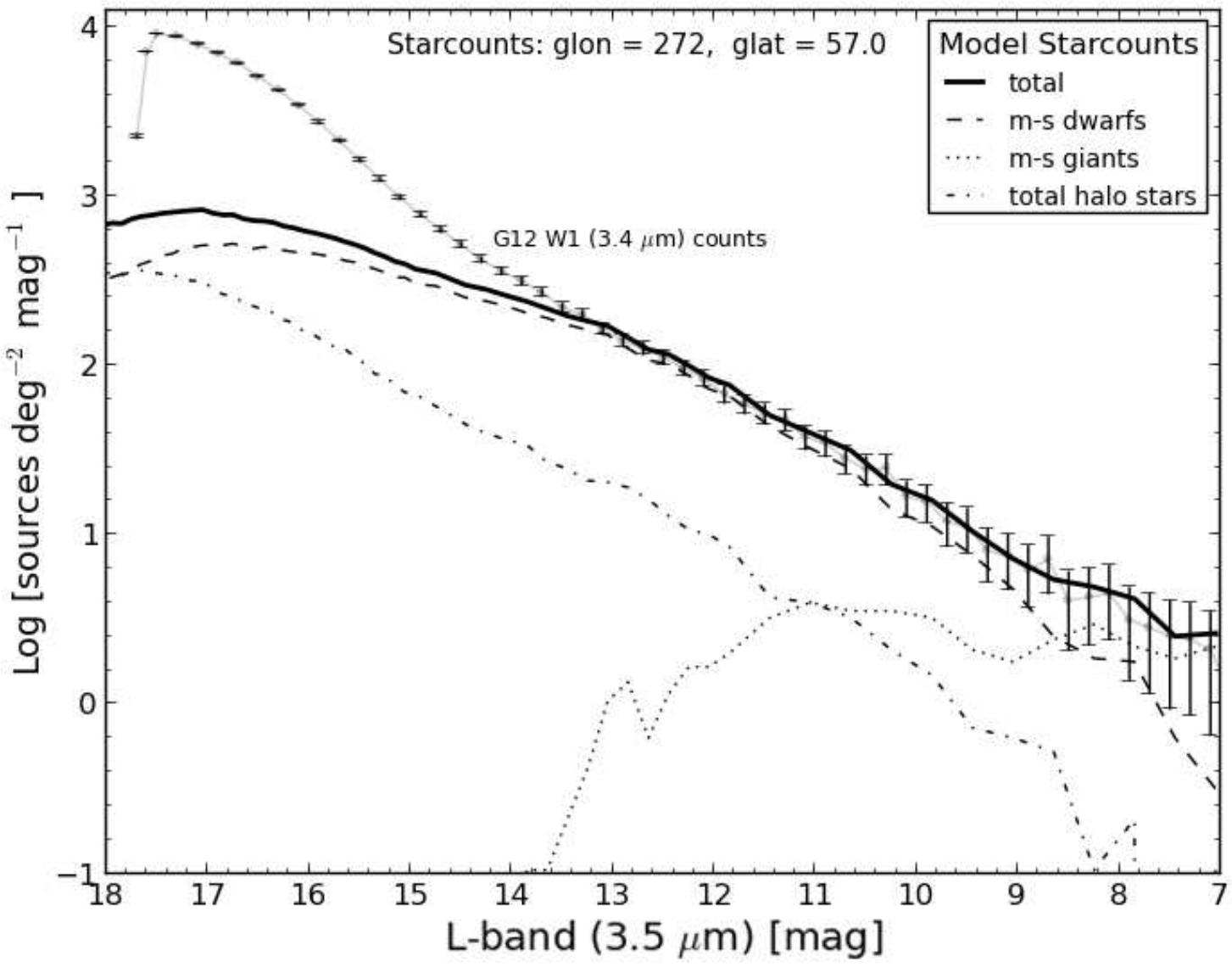

**Figure 4.** Expected L-band (3.5 $\mu$m) star-counts for the direction of the sky that contains the G12 field, the polar cap region. For comparison, the ALLWISE W1 (3.4 $\mu$m) counts are indicated (with Poisson error bars) and connected using a faint grey line. Stars dominate the source counts for W1 < 13.5 mag (1.2 mJy). We assume a negligible difference between L-band (3.5 $\mu$m) and W1 (3.4 $\mu$m).

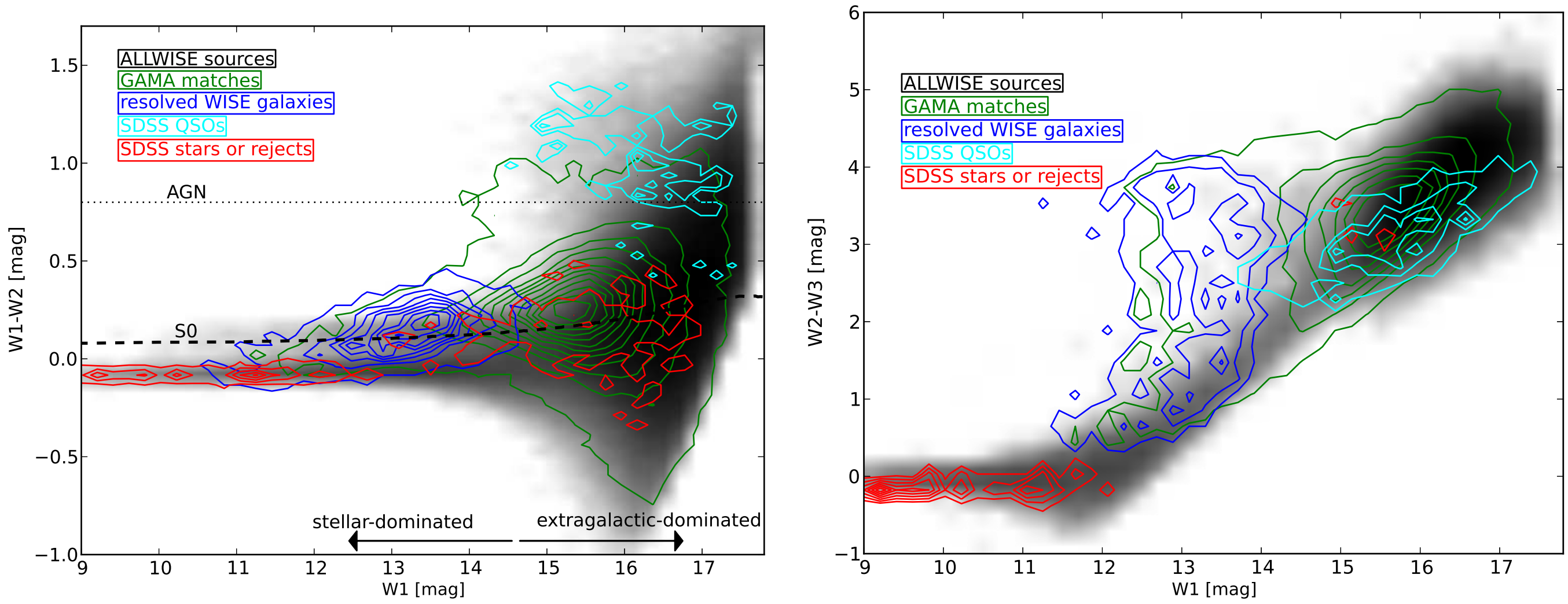

(a) W1−W2 vs W1

(b) W2−W3 vs. W1

**Figure 5.** Color-magnitude diagram for the G12 detections: the left panel (a) shows W1−W2 vs. W1, and the right panel (b) W2−W3 vs W1. Different populations are indicated: the greyscale shows all *WISE* sources; GAMA matches are shown with green contours; resolved galaxies with blue contours; SDSS QSOs with cyan; select stars or otherwise rejected sources are in red. The contour levels have log steps from 1-90%. The expected classical QSO populations lie above the dotted line W1−W2 = 0.8. We denote an S0-type galaxy track (dashed line), allowing it to change (redden) its W1−W2 color with increasing redshift from zero to 1.5.

Finally, we note that adding another band to create a new color, in this case the J-band 1.2 $\mu$m photometry, Figure 6b, can significantly help to break degeneracies. Bilicki et al. (2014) exploited this property by using SuperCOSMOS + 2MASS + *WISE* to create photometric redshifts by virtue of machine learning algorithms. Unfortunately, the whole-sky 2MASS PSC is not nearly as sensitive as *WISE*, and hence only useful for W1 brighter than 15.5 mag (0.2 mJy). What the figure does show for this magnitude range is that galaxies are much redder in J-W1 than most stars; J−W1 colors greater than 1.0 mag are most likley galaxies. The only possible contamination comes from Galactic M-dwarfs, which are highlighted in the figure as the magenta-hatched track. As constructed from our Galactic star-count model, the M-dwarf track is wide since it incorporates the range from M0 types (lower end of the track) to M6 types (upper end of the track), and some degeneracy with galaxies which occur at low S/N detections. Fortunately, the M-S population declines relative to the extragalactic population where the degeneracy is at its worst. Clearly, the near-infrared J, H and K bands are valuable toward separating stars from galaxies. With the maturing deep and wide surveys (e.g., VISTA-VHS; McMahon et al. 2013), and the optical southern surveys (ANU's SkyMapper; Keller et al. 2007), it will be possible to combine much more effectively with the *WISE* catalogs.

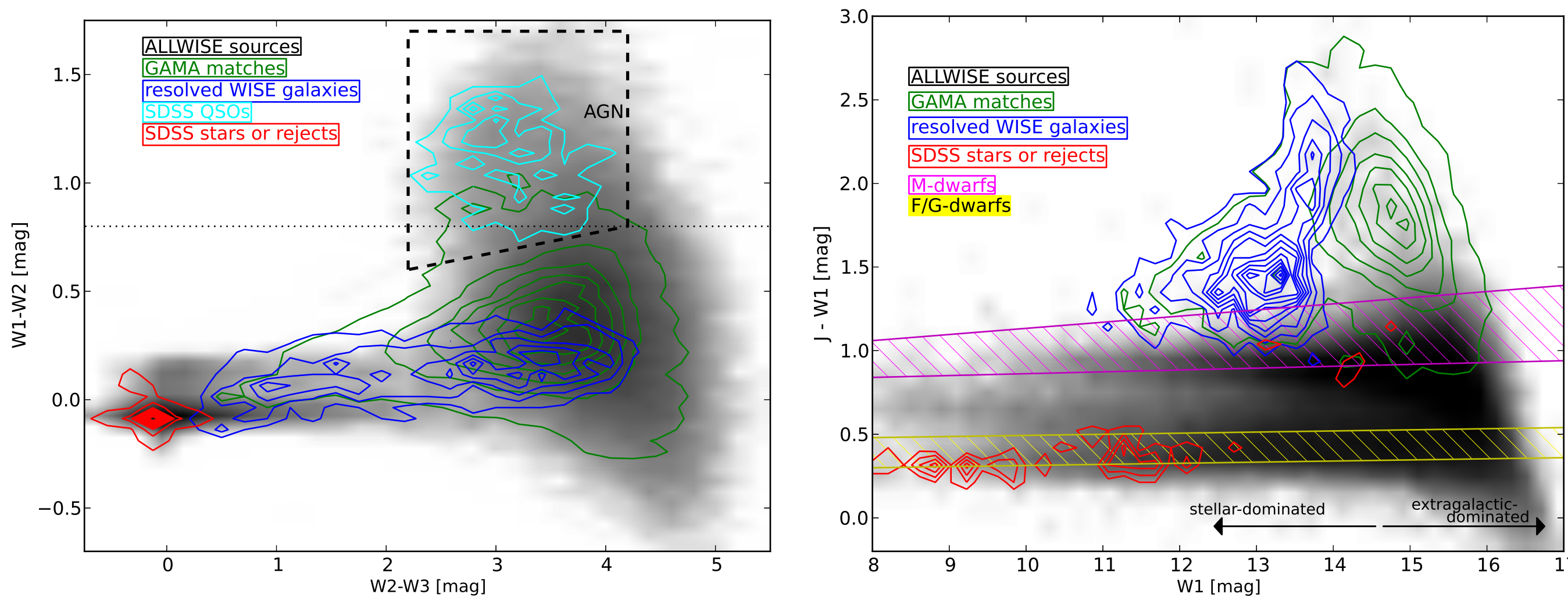

(a) W1−W2 vs. W2−W3

(b) J−W1 vs W1

**Figure 6.** The power of colors: (a) W1−W2 vs W2−W3, and (b) near-infrared J−W1 colors. See Figure 5 for a description of the contouring. The AGN box is from (Jarrett et al. 2011). Note that the J−W1 color is limited by the 2MASS J-band sensitivity; hence, only the bright W1 sources are shown. The expected track for main-sequence M-dwarfs is shown with the magenta shading, and the brighter F/G dwarfs with the yellow shading. K-dwarfs are located between these two tracks.

Based on the color-mag and color-color diagrams, we apply the following filters to remove likely stars:

- W1 $<$ 10.7 and W1−W2 $<$ 0.3
- W1 $<$ 11.3 and W1−W2 $<$ 0.05
- W1 $<$ 12.4 and W1−W2 $< -0.05$
- W1 $<$ 14 and W1−W2 $< -0.12$
- W2−W3 $<$ 0.35 and W1−W2 $<$ 0.30
- W1 $<$ 11.75 and J−W1 $<$ 1.05
- W1 $<$ 14.25 and J−W1 $<$ 0.97
- W1 $<$ 17.2 and J−W1 $<$ 0.75

These represent hard thresholds, so any one of these can eliminate a source, and are mostly applicable to bright sources. For all remaining sources, however, we use a weighting scheme where the proximity in the color-color and CMD diagrams, in combination, determine a star-galaxy likelihood – as presented in the next section.

### 3.3. *Extragalactic sample*

To create an uncontaminated galaxy sample we use the color-magnitude diagrams, applying color/magnitude cuts as noted above and the relative distributions, in combination with our star-count model to produce a likelihood – or put more simply, weighted – measure of its nature: galaxy or Galactic star. The final probability that a source is stellar, and hence rejected from the galaxy catalog, is driven by the expected distribution, Figure 4. In this sense, the faint sources in the extragalactic sample are in all likelihood to be real galaxies, although some may actually be masquerading foreground stars or even (rare, but not impossible) slow-moving Solar System bodies.

This selection is applicable to high Galactic latitude fields where the stellar number density is relatively minimal. In thie case, the North Galactic Cap, it is reassuring that stars rapidly diminish in importance for W1 magnitudes fainter than ∼14.5 mag. It is not straight forward to assess the reliability of the classification using WISE-only colors (see e.g., Krakowski et al. 2016 ). However, as noted earlier, the stellar contamination and blending is expected to be minimal in this field. We caution that the same cannot be said for fields closer to the Galactic Plane, where exponentially increasing numbers of stars completely overwhelm the relatively clean star-galaxy separation presented here. Photometric error scatters stars across the CMDs, notably with K/M-dwarfs (e.g., Fig. 6b), creating degeneracies that are very difficult to break without additional optical and near-IR color phase space information – see Bilicki et al. 2016 and Kurcz et al. 2016 for an all-sky analysis of star-galaxy separation using optical, near-IR and mid-IR colors. Below and in Section 3.5 we consider the completeness of the counts.

The final extragalactic sample is presented in Figure 7, showing the W1 differential counts for the northern Galactic cap. Statistics for the sample and the principal components are listed in Table 1. Of the total number of ALLWISE sources, about 74% comprise the final galaxy sample, ∼591,400 sources. Most (90%) of the sources do not have redshifts, but do have properties consistent with being extragalactic. A small percentage, $<$ 1%, are resolved in the W1 channel, and only a few percent are in the relatively shallow 2MASS near-IR catalogs (PSC

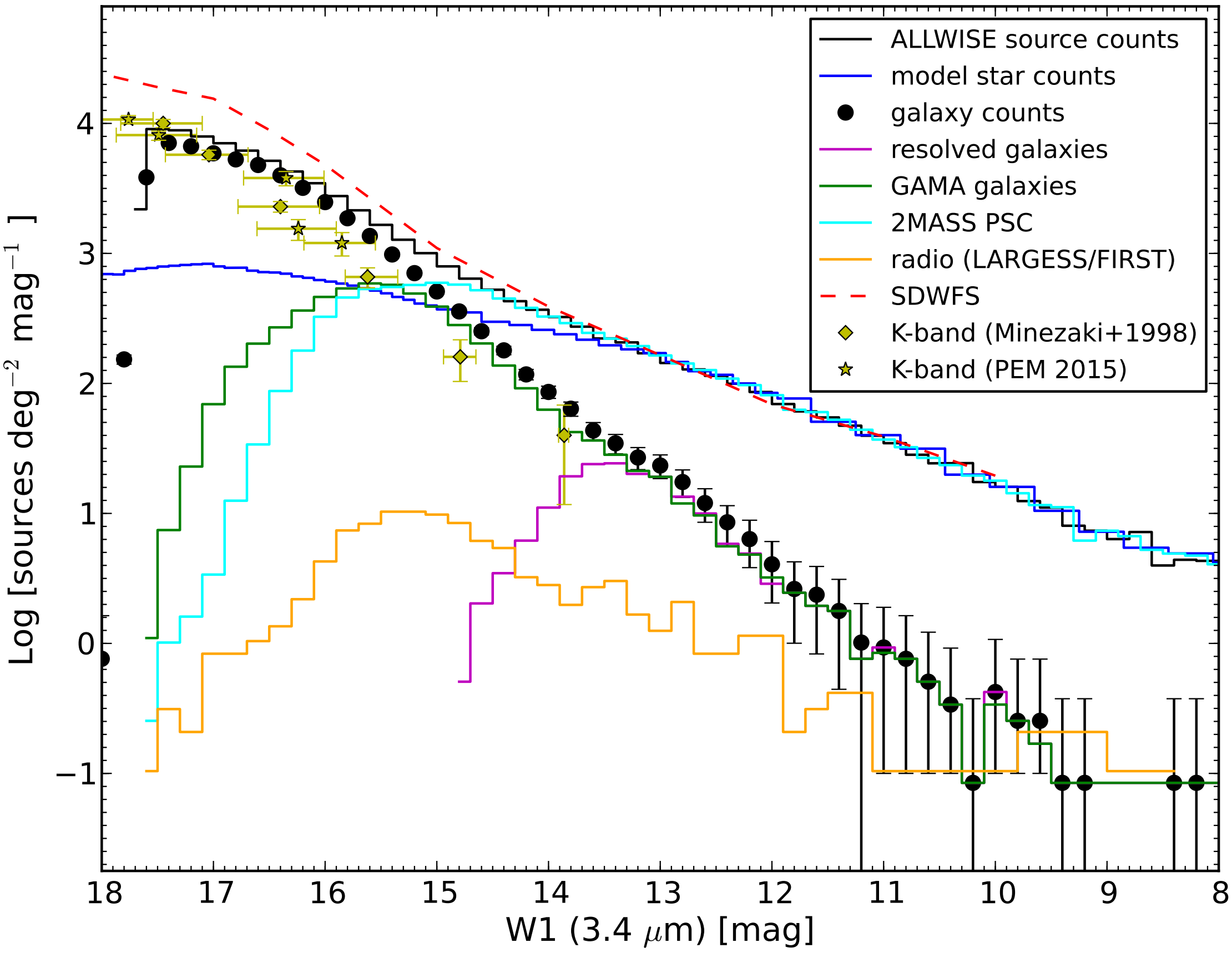

**Figure 7.** Final differential W1 (3.4 $\mu$m) source counts in the G12 region. The total galaxy counts are denoted with solid black circles and Poisson error bars. *WISE* sources that are also GAMA (green), resolved (magenta), 2MASS PSC (cyan) and LARGESS radio galaxies(orange) are indicated (see §4.2). For comparison, we show deep IRAC-1 counts from the *Spitzer* Deep-Wide Field Survey, deep K-band galaxy counts, rest frame-corrected to the W1 channel, from the Minezaki et al. (1998) and Prieto & Eliche-Moral (2015) (PEM) studies.

**Table 1**
*WISE* Cross-Match Statistics

| *WISE* G12 Detections (803,457 in total with W1 $\geq$5-$\sigma$, 60 deg$^2$) | | |
|---|---|---|
| Type | number | Percentage (%) |
| Extragalactic Population | 591,366 | 74% of all *WISE* sources |
| GAMA redshifts | 58,126 | 9.8% of galaxies |
| 2MPSC | 26,210 | 4.4% of galaxies |
| WXSC | 2110 | 0.4% of galaxies |
| 2MXSC | 2430 | 0.4% of galaxies |
| SDSS QSOs | 1167 | 0.2% of galaxies |
| LARGESS radio galaxies | 986 | 0.2% of galaxies in 48 deg$^2$ |

2MPSC and 2MXSC are the 2MASS point and resolved sources;
WXSC is the resolved *WISE* sources;
QSOs are from SDSS identifications (see text)
LARGESS discussed in Section 4.2.

and XSC).

By definition, the extragalactic sample matches the complete and reliable bright end distributions of W1-resolved and GAMA galaxies; see Figure 7. At the faintest GAMA magnitudes, 15 to 15.5 mag, there are a few percent more total extragalactic sources than 2MASS PSC or GAMA galaxies, likely due to incompleteness in these surveys, while slight contamination from foreground stars is possible. Extrapolating to the faintest bins where GAMA is highly incomplete, 15.5 - 17.5 mag (e.g., at 16.5 mag the GAMA counts are 90% incomplete compare to the *WISE* galaxy counts), the *WISE* galaxy counts continue to rise, with a slight upward increase in the slope, before slowly flattening beyond 16.5 mag (78 $\mu$Jy), with incompleteness beginning at 17.5 mag (31 $\mu$Jy), peaking at 7,900 sources per deg$^2$. There is no obvious signature of Eddington bias in the shape of the curve, which may be a clue that incompleteness is entering more than just the last magnitude bin.

Finally, the radio continuum sources from the LARGESS survey, discussed in Section 4.2, have a shallower slope compared to all extragalactic sources, becoming increasingly incomplete at fainter magnitudes. This is expected given the relatively shallow continuum survey ($\sim$1 mJy) the sources are drawn from (FIRST/NVSS). We further discuss the radio properties of our extragalactic sources in Section 4.2.

### 3.4. *Comparing source counts with previous work*

We perform two separate external comparisons. The first comes from the *Spitzer* Deep-Wide Field Survey (SDWFS), which focused on 10 deg$^2$ in Bootes (Ashby et al. 2009). The SDWFS counts reach impressively faint levels, $\sim$3.5 $\mu$Jy in IRAC-1, and are shown by the red dashed line in Figure 7. For comparison, here we assume the 3.6 $\mu$m IRAC-1 band is equivalent to the *WISE* 3.4 $\mu$m band (they are within $< 4\%$ for low redshifts, and up to 10% for high redshifts). At the bright end, W1 $< 13$th mag, the SDWFS agrees very well with the *WISE* source counts, where the counts are completely dominated by stars. At fainter magnitudes, where galaxies become the dominant population, the SDWFS grows slightly faster than the *WISE* counts; e.g., at 17th mag (50 $\mu$Jy), the SDWFS counts are nearly a factor of two larger than the *WISE* counts.

Either this difference is a real cosmic variance effect (plausible, there is large scale structure in both fields), or *WISE* is becoming incomplete due to confusion and source blending at those depths, consistent with a lack of strong upturn expected with flux over-bias. We should note, the Bootes field has more Galactic stars than the G12 field (because it is closer to the Galactic Bulge); our star-count model predicts 30% more stars in the Bootes region compared to the polar cap. Hence, at least a few percent of the SDWFS excess is due to stars. A better comparison would be to remove the expected star-counts from the SDWFS sources, as follows: at 17th mag, the SDWFS counts are 15,500 deg$^{-2}$, while the star-count density is 1,100 deg$^{-2}$; hence, the expected extragalactic counts should be 14,400 deg$^{-2}$ at 17th mag, which is still larger than the observed *WISE* W1 extragalactic counts at this flux level.

A second external comparison uses small-area, yet deep, K-band (2.2 $\mu$m) galaxy counts from the Minezaki et al. (1998) survey of the South Galactic Pole (SGP), and the Prieto et al. (2013) near-IR study of a field in the Groth Strip (GS). The SGP galaxy counts reached a limiting K magnitude of 19.0 in the 181 arcmin$^2$ field, and similarly the GS observations reached 19.5 mag (90%) in a 155 arcmin$^2$ area. The 2.2 $\mu$m and the 3.4 $\mu$m bands are sensitive to the same stellar populations for galaxies in the local universe. However, this is in fact a far more challenging comparison because the bands are sufficiently different that at faint magnitudes, or high redshifts, there is a large color difference due to cosmic redshifting. We can determine the rest frame-corrected (k-correction) behaviour using galaxy templates (e.g., early to late-types) and our knowledge of the source distribution with redshift. We present in the next section (3.5) a detailed analysis that elucidates the expected color differences.

At rest wavelengths, the K-band and the W1 (or IRAC-1) bands are sensitive to the same light-emitting population; i.e. evolved giants, and the color difference is $\sim$ 0.15 mag. The band-shifting due to redshift, or cosmic reddening, is small and roughly constant for both bands in the Local Universe ($z < 0.2$) and is generally not a concern for nearby galaxies. At intermediate redshifts, however, there is an abrupt transition and the $K_s - W1$ color rapidly reddens because K-band is no longer benefiting from the H-band stellar bump. By $z = 1$, the color is nearly 1 magnitude for an S0-type galaxy, and by $z = 2$ it is 1.5 magnitudes. Consequently, to perform a comparison between W1 and K, we need to derive the mean $K - W1$ color for each W1 magnitude bin, using our expected galaxy distribution model; see next section and Fig. 8 for details.

Applying the expected mean $K - W1$ colors (Section 3.5) and their associated expected scatter represented by the horizontal errorbars, we arrive at the W1-converted deep K-band galaxy counts shown in Figure 7. Except at the very faint end, W1 $> 16.5$ mag, the SGP and GS counts are slightly lower than the W1 counts, which is either a cosmic variance difference – this is plausible given that the K-band surveys have very small areas –, incompleteness in the K-band counts, or that the $K - W1$ color is even redder than expected at lower redshifts, relevant to these intermediate flux levels. The large spread in K-W1 colors, $> 0.3$ mag (see Fig 8b), functions as a limitation to comparing between 2.2 and 3.4 $\mu$m counts.

Finally, one interesting feature of note: there is a kink or slope change at W1 $\sim 16.5$ mag (78 $\mu$Jy) , which is readily apparent in the *WISE* counts, SDWFS counts and the GS K-band counts, as well as other deep K-band surveys (see e.g., Vaisanen et al. 2000). The follow-up Prieto & Eliche-Moral (2015) study of the GS highlighted this slope change – at an observed K-band $\sim 17.5$ mag, corresponding to W1$\sim$16.5. The flattening they attribute to a sudden population change from early-type (S0) galaxies to late-type disks dominating at redshifts greater than 1. Our *WISE* extragalactic sources are consistent with this scenario. As we see in the next sub-section, attempting to model the faint (>17th mag) source counts is complicated by the mix of galaxy types spread across a large range in redshift, and hence k-correction and LF evolutionary effects.

### 3.5. *Expected faint galaxy counts*

In this section, we characterize the faint extragalactic counts in the 3.4 $\mu$m bandpass, notably the redshift distribution of the *WISE* galaxy population detected in W1. Although a more detailed and sophisticated treatment is beyond the scope of this paper, we apply an infrared-based luminosity function (LF) method to help understand what may be happening at these faint flux levels. The major caveat with the following analysis is that we have incomplete knowledge of LF evolution at redshifts $> 0.6$, hence we caution interpretation of the counts at the faintest levels that *WISE* can detect.

Our approach is to characterise the galaxy population using the 3.6 $\mu$m (IRAC 1) luminosity functions derived by Dai et al. (2009), which employed a non-parametric stepwise maximum-likelihood (SWML) method to characterize populations up to $z = 0.6$. Two variations, and a combination of the two, are investigated – the first is a single LF, but includes redshift-evolution of $M_*$, and the second fits Schechter functions to three redshift shells, and hence evolutionary and normalization differences that may arise. There is no change or difference in the slope, $\alpha$, for the LFs, which stretches to an absolute magnitude of $-18$. We find that a combination of the two LFs give the closest fit to the WISE number counts: where the first LF is used for redshifts $< 0.5$, and the second LF with the deepest redshift shell, 0.35 to 0.6, is used for all high redshift sources, $z > 0.5$.

With these LF combinations, we explore the resulting expected source counts that arise from different mixing of early and late-type galaxies, thereby exploring the range in k-corrections that are plausible. For example, in one trial we employ a 50/50 mix of early (E-type) and late (Sc-type) galaxies, which have slightly different k-correction responses at high redshifts (early types tend to result in 10% higher counts in the faint source counts compared to late-types). Fractions with relatively more late-types are explored and motivated given the results of Prieto & Eliche-Moral (2015) discussed in the previous section (3.4). With this stochastic mixing technique, we find 5 to 20% differences in the model source counts, where the best (data matching) results appear to be higher (2:1) fractions of late-types. Given the uncertainties in the LF for high redshifts, the exact fractions cannot be determined with any fidelity.

For k-corrections, we use the Brown et al. (2014) and Spitzer-SWIRE/GRASIL (Polletta et al. 2006, 2007; Silva et al. 1998) SED templates to redshift and measure synthetic photometry of the *WISE* filter response functions (Jarrett et al. 2011) and in this way derive the flux ratios between rest and redshifted, $(1+z)\lambda$, spectra in the W1-band, or IRAC-1, band. The standard k-correction magnitude is then $-2.5\,Log\,[\text{flux ratio} * (1+z)]$. Further, we carry out trials using the k-corrections in Dai et al. (2009), which are slightly smaller, ∼5-10%, compared to our k-correction SED families, but which only make a small, ∼few %, difference in the resulting counts, and are duly reflected in the spread in model counts presented in Fig. 8a.

To help understand the faint end of the *WISE* source count diagram, the volumes are sampled to high redshifts, limited to $z = 2$. This limit was chosen to make sure that we probe deep enough to see how – qualitatively – the faint bins are populated by luminous high redshift galaxies. Finally, and to emphasize, we assume the resulting IRAC-1 counts are equivalent to the W1 counts, although as noted earlier, real differences may arise in the faintest mag bins where distant galaxies dominate. The difference between the IRAC-1 and WISE W1 bands can be assessed by their k-correction response; e.g., at redshift zero for a late-type galaxy SED, W1 is brighter by 4% compared to IRAC-1, whereas by redshift 1.5 it rises to a 10% difference. Future work will employ LFs purely derived using WISE measurements, removing this potential complication.

Following Dai et al. (2009), we account for evolution by parametrising the LF as a Schechter function using the best-fit values from the 3.6 $\mu$m (IRAC 1) determination of Dai et al. (2009). In the first case, using a single Schechter function with $M_*$ brightening with redshift by a factor of 1.2, and in the second case, jointly fitting in three redshift bins: $z \leq 0.2$, $0.2 \leq z \leq 0.35$ and $0.35 \leq z \leq 0.6$. For the later case, in all redshift bins the faint end slope is fixed to $\alpha = -1.12$, whilst $M_* - 5\log h$ and the normalisation $\phi_*$ ($10^{-2}$ h$^3$ Mpc$^{-3}$) are fitted. For the lowest redshift bin ($z \leq 0.2$) the characteristic magnitude is $M_* - 5\log h = -24.09$ and $\phi_* = 1.45$; the middle bin ($0.2 \leq z \leq 0.35$) has $M_* - 5\log h = -24.34$ and $\phi_*$ =1.01 in the higher redshift bin ($0.35 \leq z \leq 0.6$) is $M_* - 5\log h = -24.63$ and $\phi_*$ =0.85. Our sample contains sources with redshifts higher than $z > 0.6$, hence we extrapolate the LF derived in the $0.35 \leq z \leq 0.6$ bin to higher redshifts (out to $z < 2$). The impact of this assumption is discussed in more detail below, but clearly it introduces a serious limitation to the analysis at the faint end. We find that for the first case, the $M_*$ evolution with redshift is far too strong for redshifts greater than 0.6, and note it was never designed to be applied here, and hence we do not employ this LF for redshifts beyond 0.5.

We then proceed as follows: the particular case-1, case-2, or a combination thereof, LF model is used to predict the number density of sources in bins of absolute magnitude ($\Delta M = 0.02$ mag) from −28 to −18 (at 3.6 $\mu$m). We apply the luminosity distance modulus and the appropriate redshift band-shifting to the magnitudes associated with these sources, employing a pre-defined mix of elliptical (E) and spiral (Sc) galaxies and their associated k-corrections. The key assumption of using one, two (or more) types of galaxies provides a straightforward, albeit simplistic, modelling of the morphological diversity of the G12 sample, which tends to impact the high redshift galaxies.

We estimate the final source counts from this magnitude-selected LF distribution by sampling redshift *shells* of $\Delta z = 0.0025$ out to a maximum redshift of $z_{\rm max} = 2$. These are scaled by the co-moving volume of each shell in an area of 5000 square degrees for statistical stability. The differential source counts of redshift shells are then computed, directly comparable to our *WISE* galaxy counts. The results are shown in Figure 8a, highlighting representative redshift shells and the corresponding accumulative source counts (grey shaded-region), which is compared to the actual galaxy counts (black filled points) in G12. The shaded curve reflects the spread in values that arise from using different population mixes and LF combinations which are all plausible.

The model-to-observed correspondence is particularly good at magnitudes brighter than ∼15.5 mag (0.2 mJy), which suggests the simplifications are robust and the Dai et al. (2009) LF is representative of the G12 volume, $z < 0.5$ toward the Galactic polar region. The dominant redshift distribution appears to be 0.1−0.3 for this intermediate magnitude range – the green, yellow and orange curves in Figure 8a – not unlike the GAMA redshift distribution, and hence is quite realistic. Conversely, at fainter magnitudes the model counts are systematically larger than the data, with a steeper slope at W1 > 17 mag arising from high-redshift sources, $z > 0.75$ to 1.5, however diminishing rapidly beyond that limit as these galaxies are far too faint to detect with WISE. At face value, this robust (in spread) result suggest the W1 source counts are incomplete at the faint end, notably for the moderate to high-redshift ($z > 0.5$) populations, consistent with Yan et al. (2013; see their Fig. 6). We expect with Malmquist Bias, the high-redshift detections to be luminous in nature, which likely means they are dominated by early-type, quenched and clustered galaxies; we discuss population clustering in the Section 4.4. A few interesting statistics follow from the LF modeling results: integrated to a limiting magnitude of 17.5 (31$\mu$Jy) for all redshifts, the total number density is 15603 deg$^{-2}$, of which 72% have redshifts >0.5, and 48% have redshifts >0.75, and fully 27% are beyond a redshift of 1.

We caution however, extrapolating the Dai et al. (2009) LF to high redshifts is uncertain – the luminosity evolution correction term is only designed to redshifts <= 0.6 – meaning that large and potentially-systematic uncertainties are in play at these faint magnitudes. Moreover, as Prieto & Eliche-Moral (2015) conjecture, there may be strong effects happening at high redshift (z∼1) that significantly alter the LF since the counts should flatten, not increase. We note that UV-LF studies at such high redshifts, and beyond $z > 4$, indi-cate strong evolution in the slope ($\alpha$) and $\phi_*$, even as the functional form remains Schechter-like, which clearly highlights the importance of using the appropriate LF for the given source population (e.g., see Bouwens et al. 2015). Nevertheless, the model counts suggest that the measured *WISE* counts are not complete for these faint bins, due in part to the large (6 arcsec) beam of W1 exacerbating the blending of faint sources, which include both stars and background galaxies, and losses from increased noise around brighter foreground sources. Using our starcount model, we have estimated that 3% of the extragalactic sky is lost to foreground stars brighter than 18th mag for a masking diameter of 2×FWHM, which is exacerbated at the faint, high redshift end.

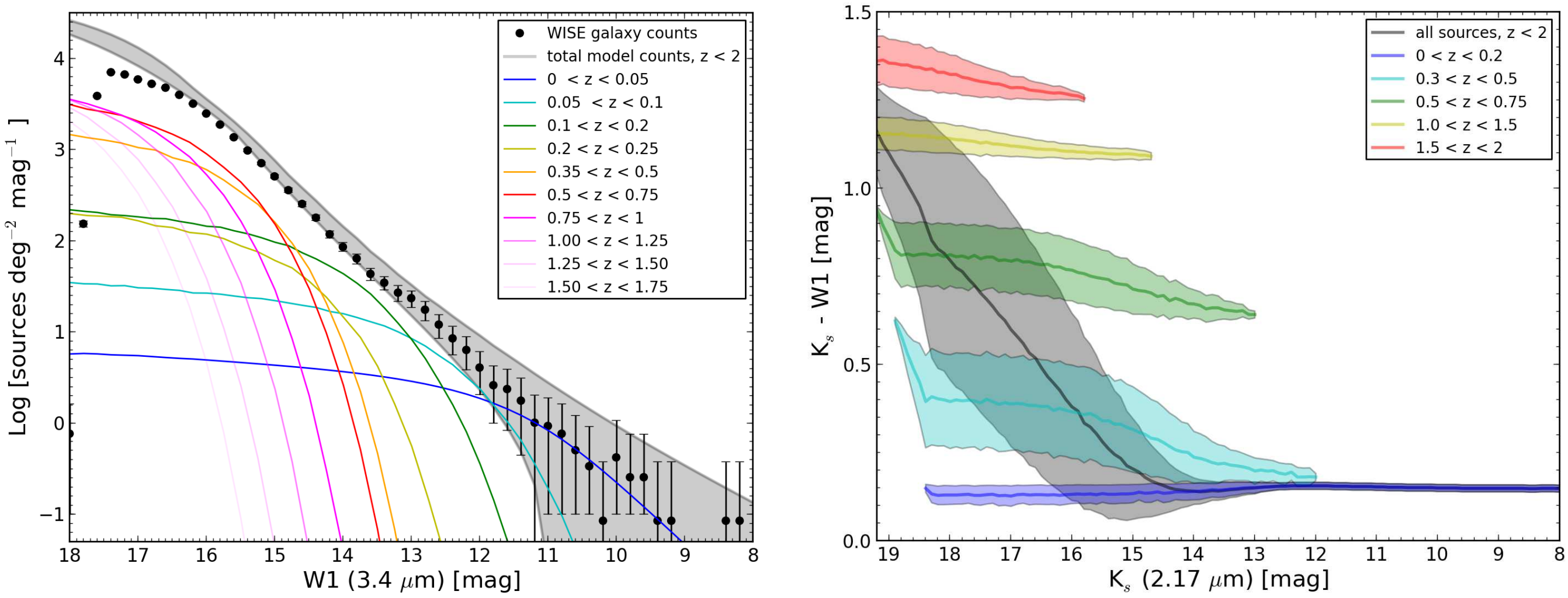

(a) Source Count Models

(b) Expected near-IR color vs. $z$

**Figure 8.** Modeling the extragalactic source counts. (a) Expected extragalactic source counts: differential source counts in comparison to the measured *WISE* values (solid filled points), highlighting a series of redshift shells. The shaded curve represents the spread in values using a mixture of k-corrections and two different infrared LFs of Dai et al. (2009). (b) The expected near-infrared K-W1 color as a function of the apparent K-band (Vega) magnitude, for the extragalactic population; the grey-shaded region corresponds to all redshifts (up to 2); the other shadings represent redshift shells, demonstrating the significant band-shifting differences between 2.2 $\mu$m and 3.4 $\mu$m at redshifts $> 0.2$.

As part of the LF modeling, we track the K−W1 color variation across redshift since it is relevant to our comparison of the W1 source counts with the more prevalent and deep K-band source count studies (previous section, 3.4). The method is straight forward – the zero redshift Vega color, K−W1=0.15 mag, changes due to the differing k-corrections in the 2MASS 2.17 $\mu$m and WISE 3.4 $\mu$m bands. The results are shown in Figure 8b, which depicts the spread in K−W1 color as a function of the K Vega magnitude. Here we have accounted for the density of sources at a given magnitude bin. For example, at K=17 mag there is a wide range in sources at different redshifts (and hence, k-corrections) – from local sources, which may be late types, to $z = 2$ distant galaxies, which are early-type luminous galaxies. As expected, the color spread is consequently large, in some cases 0.5 mag or more. Depending on the redshift, the color can range from 0.15 (low-z) to 1.5 mags (high-z); redshift shells have approximately the same color, but vastly different values between redshift shells. This means converting from the near-infrared to the mid-infrared is only straight forward at low redshifts, $z$ <0.2, and far more problematic at fainter magnitudes where higher redshifts populations are dominant.

### 3.6. *Redshift Distribution*

The LF modeling aids the determination of the redshift distribution of sources that are detected by WISE in the 3.4 $\mu$m band. We now compare the GAMA distribution with the LF source count model; see Figure 9. At this juncture we remind the reader that SDSS/GAMA is an optically-selected sample, while *WISE* is a mid-infrared survey. At higher redshifts, the former is sensitive to optically 'blue' galaxies, currently star forming, while *WISE* W1 and W2 are sensitive to the evolved stellar population, i.e., past generation of star formation. Consequently, sample differences, as well as GAMA incompleteness at high-z, mean that we should not be surprised to see significant differences at increasingly higher redshifts; we are pushing the limits of our respective data sets.

In panel Figure 9(a) we plot the N($z$) vs $z$ distribution of *WISE* resolved sources (grey line), GAMA sources (solid line), and the model distributions (mean of the spread; dotted lines) for two different magnitude limits,16 and 17 mag, respectively). As noted above, the model mean performs well for magnitudes brighter than 16th mag, which suggests that GAMA is highly discrepant – both incomplete and divergent, by comparison – for redshifts $> 0.35$, as shown by the red dashed line. Panel (b) presents a different view of this redshift incompleteness, clearly showing that the missing sources in GAMA for W1 $> 15.5$ mag (see Fig. 3 and 7), are intrinsically red – i.e. dusty and SF. At the fainter limits, beyond the sensitivity of GAMA (panel b), the incompleteness extends to all redshifts, even Local Universe, which is either due to a paucity of low-luminosity sources; i.e. dwarfs that are too faint to be captured by the GAMA selection function, or the model is over-estimated. We emphasize the favorable redshift k-correction with deep mid-IR counts that is driving the behavior seen in Figure 9; namely, the high redshift tails of *WISE* flux-limited samples and the high amplitude of $N(z)$ for W1<17 mag, even at low z, clearly show the utility of *WISE* to see far and deep.

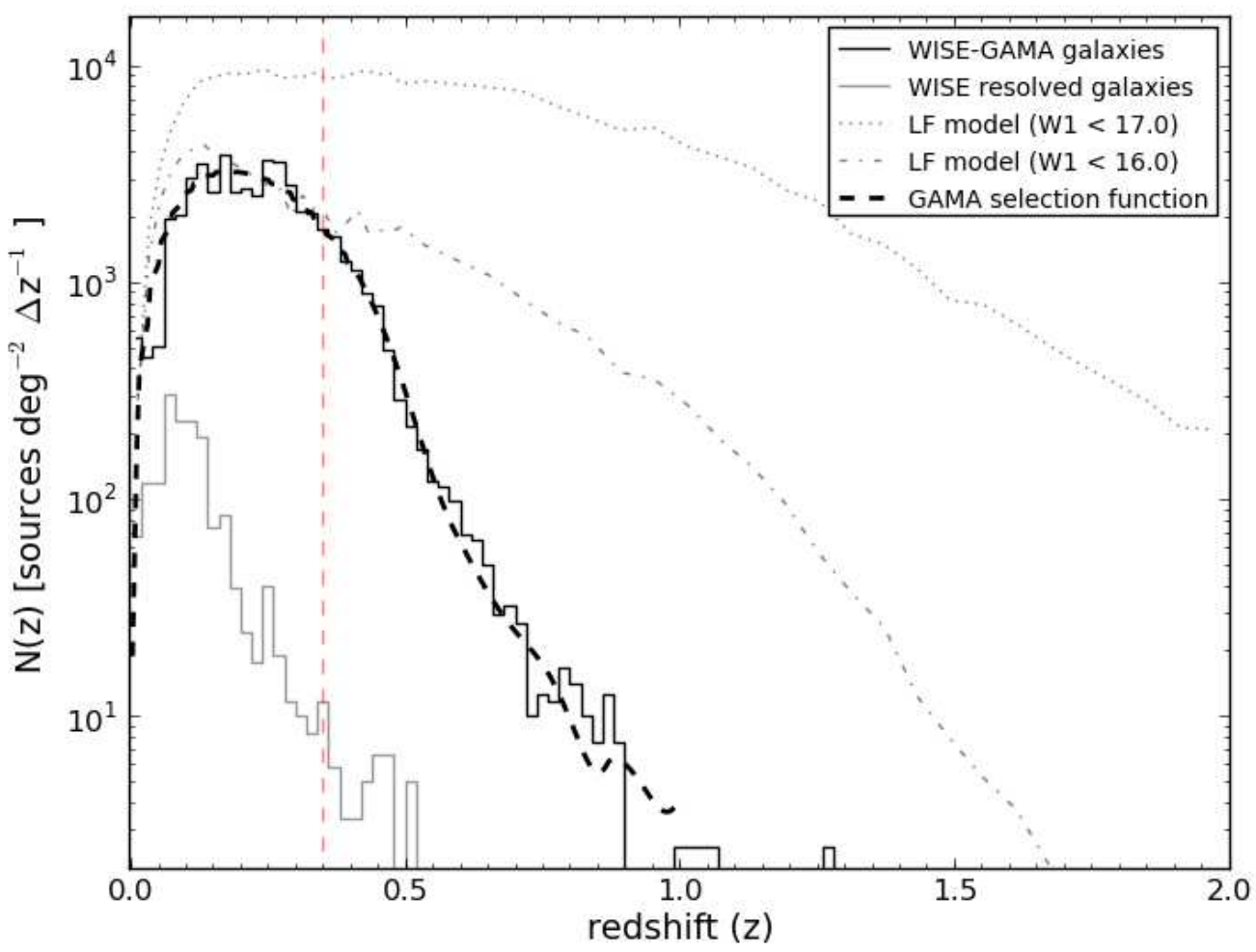

(a) redshift distribution

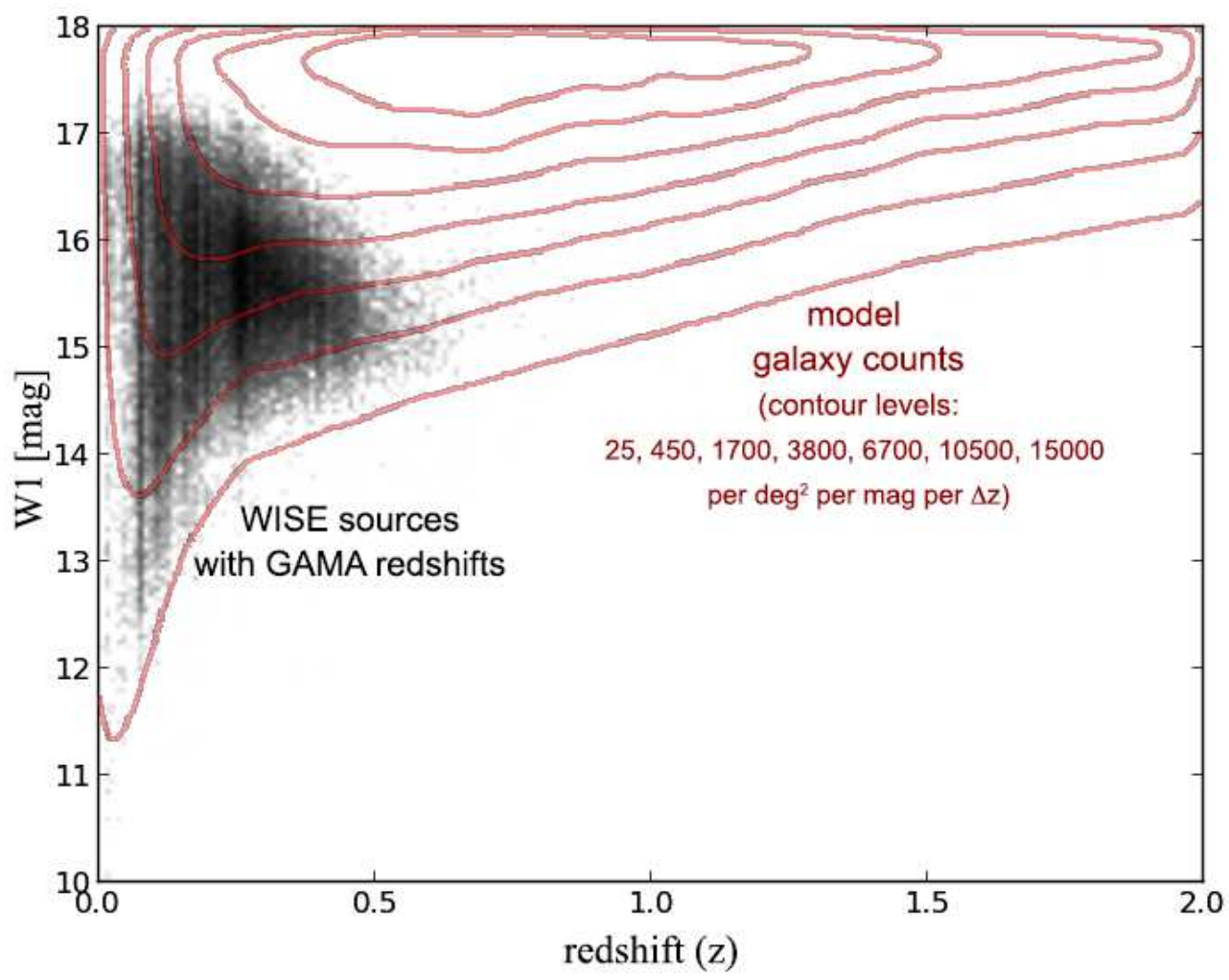

(b) Mag versus $z$

**Figure 9.** Redshift distributions of the real and expected G12 sources; (a) Denoting all GAMA matches (solid line) and those that are resolved in *WISE* (grey line). The derived $N(z)$ function is shown with a dashed line, and used later as the selection function in the 2PCF analysis; also shown are the expected (model-averaged) distributions at magnitudes (W1 = 16 & 17 mag) fainter than the GAMA detection limit. (b) Expected redshift distribution of *WISE* sources as a function of the observed W1 magnitude (red contours); real *WISE*-GAMA matches are shown in grey-scale.

Finally, we draw attention to the dashed line in Figure 9a, which corresponds to the GAMA variance-free distribution, which we later use as the GAMA $N(z)$ selection function, equivalent to the distribution of GAMA galaxies if there were no clustering signature. This was derived using a hybrid method that combines the source-cloning method of Cole (2011) and the LF model in which we attempt to apply GAMA-like W1 magnitude incompleteness to the counts (Figure 7, note the green curve). We refer the reader to the Farrow et al. (2015) study of clustering in the GAMA fields, in which they describe in detail the source-cloning method. Indeed the resulting hybrid selection function is entirely consistent with the function derived by Farrow et al. (2015) who used a much larger GAMA sample. In the next section, we use the GAMA-G12 selection function to estimate the degree of structure and clustering in the northern Galactic cap field.

### 3.7. *Equatorial projection maps*

The projected 2D distribution of *WISE*-GAMA extragalactic sources is presented in Figure 10. All sources are plotted in the top (a) panel, mixing a wide range of redshifts, but dominated by the high redshift volumes (see Figure 9a), hence blurring the large scale structure. The grey scale distribution is heavily Gaussian-smoothed to reveal correlated structures, sometimes washing out the smallest scale features. The middle (b) panel of Figure 9 is restricted to the GAMA redshifts, and although it is also a mix of redshifts ($z < 0.5$), some structure is readily apparent. Finally, the bottom (c) panel shows the resolved *WISE* sources, which have relatively low redshifts ($< 0.1$), but also show some diagonal structure central to the G12 field. In the next section, we attempt to identify overdensities and structures that comprise the cosmic web in the volume studied.

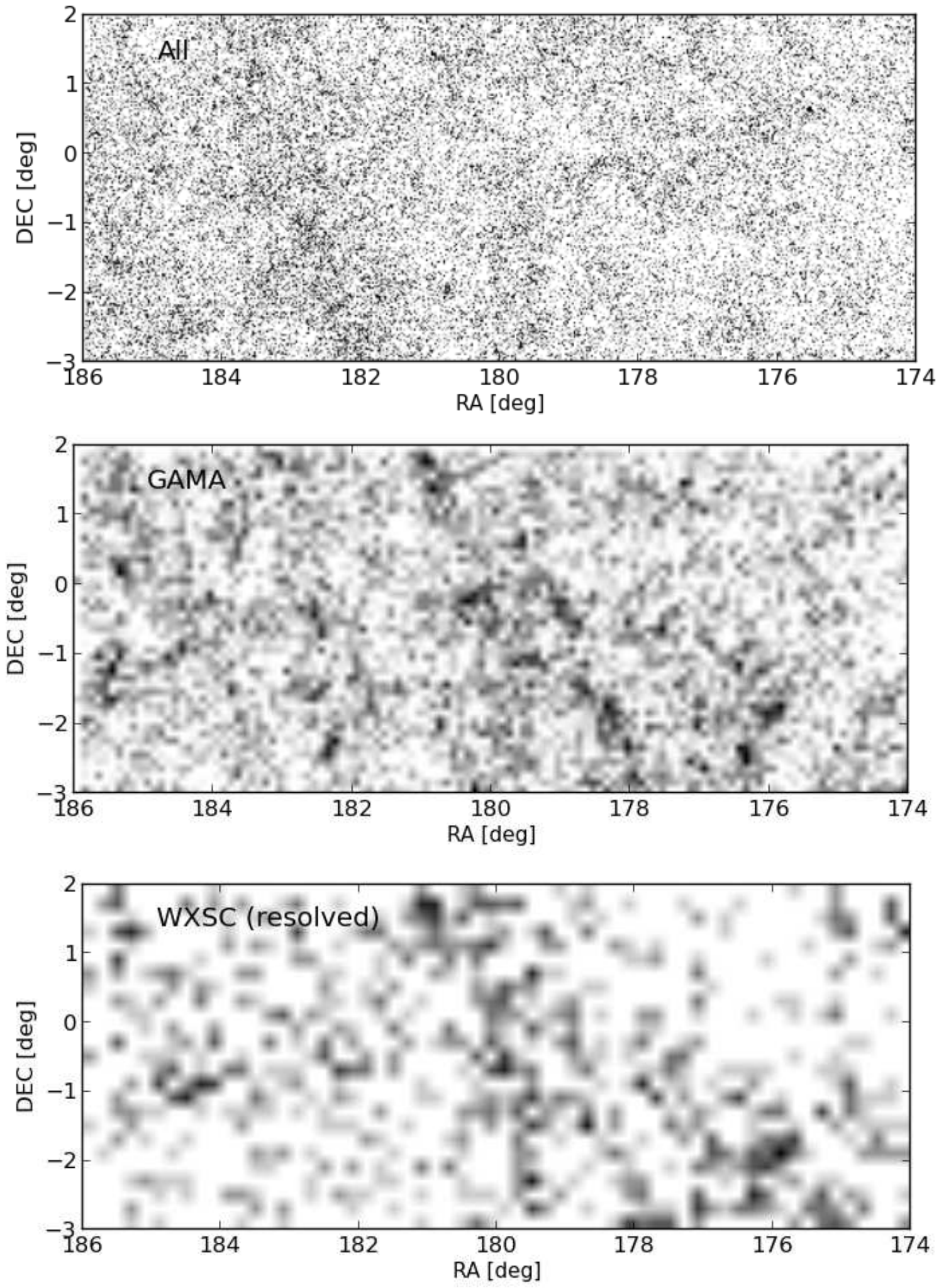

**Figure 10.** Projected distribution of *WISE*-GAMA galaxies in the G12 field: top panel is all ∼600,000 sources; middle panel shows sources with GAMA redshifts; bottom panel shows all sources resolved by *WISE*. The grey scale is logarithmic, and 15 arcsec gaussian smoothing has been applied. Over-densities and filamentary structures are evident even with a large range in volume.

## 4. THE GALAXY POPULATION AND ITS SPACE DISTRIBUTION TO $Z < 0.3$

Thus far the main objectives were to cross-correlate the *WISE* and GAMA data sets for G12, characterize the resulting catalogs using basic statistical measures, producing detailed source counts and redshift selection functions, and pushing the *WISE* galaxy counts beyond the GAMA detection limits. In this section, we advance to characterizing and mapping the galaxies using our GAMA redshifts, aiming to construct a 3D mapping of the space distribution that extends to $z < 0.3$. We start with the basic host properties of color, stellar mass and star formation (SF) activity. Next we search for spatial overdensities, and quantify this using a two-point correlation function analysis, and finally we show 3D constructions of the G12 field.

### 4.1. *Past to present star formation history*

In this section we consider the derived stellar mass and SFR properties of the *WISE*-GAMA sample in G12. The redshift range is limited to a maximum of $z < 0.5$ to mitigate incomplete selections, k-correction modeling accounts for spectral redshifting, and we consider effects that are redshift-dependent, for example, Malmquist bias. For luminosity calculations, we use the redshift to estimate the luminosity distance, corrected to the Local Group frame of reference, in Mpc.

Combining the optical, near-infrared and mid-infrared photometry, we construct spectral energy distributions (SED) for each galaxy. We then used extragalactic population templates from Brown et al. (2014a) and SWIRE/GRASIL models (Polletta et al. 2006, 2007; Silva et al. 1998), to find the best-fit template to the measurements, thereby characterizing the source, based on the template type, as well as correcting the source for spectral shifting in the bands; see also Paper I, where this technique was applied to the GAMA fields G9, G12 and G15. The resulting rest frame-corrected *WISE* colors are shown in Figure 11a.

One can see that with spectral shifting, the apparent (observed) colors shift blueward in W2−W3, and redward in W1−W2, i.e., shifting the ensemble to the upper left. With redshift, the observed W1 magnitude gets brighter relative to the rest value[21]. For interpretation purposes, Figure 11b shows where various types of galaxies are located in this color-color diagram (adapted from Wright et al. 2010; Jarrett et al. 2011; see also Paper I) and is used below when describing the clustering and spatial distributions. We reiterate that W3 and W4 are considerably less sensitive than W1, and correspondingly fewer sources have the W2−W3 color available – this is illustrated in the next section. Consequently, early-type galaxies, with R-J dominated emission, are only detected in the local universe. To first order, if a galaxy has a W3 detection it is likely to have star formation activity. More so, W4 detections are usually associated with starbursting systems and luminous infrared galaxies ( e.g., Tsai et al. 2015).

We estimate the stellar mass ($M_\star$) and the dust-obscured star formation rate (SFR) using the rest-frame-corrected flux densities. For the stellar mass, the procedure is to compute the 3.4 $\mu$m in-band luminosity, $L_{W1}$ and apply it to the M/L relation that employs the W1−W2 color relation; see the description in Jarrett et al. (2013), and Paper I. In this work, we use the 'nearby galaxy' M/L relations from Paper I, in which the *WISE* stellar masses, derived from the W1 in-band luminosity, were calibrated with the GAMA stellar masses derived by SDSS colors (Taylor et al. 2011):

$$\log_{10} M_\star/L_{\rm W1} = -2.54({\rm W1} - {\rm W2}) - 0.17, \qquad (1)$$

with $L_{\rm W1}$ $(L_\odot) = 10^{-0.4(M - M_{\rm sun})}$
where M is the absolute magnitude of the source in W1 and $M_{\rm sun} = 3.24$ is the in-band solar value; see Jarrett et al. 2013.

For this $M_\star/L_{\rm W1}$ relation, we place floor/ceiling limits on the W1−W2 color: −0.2 to 0.6 mag, to minimize the contaminating effects of AGN light which tends to drive W1−W2 color redward, and to minimize the S/N effects from W2 being less sensitive to W1, which can induce unphysically blue colors. For galaxies with only W1 detections or colors with S/N $\leq$ 3, we apply a single M/L = 0.68.

Not surprisingly, this M/L relation is similar to the IRAC version derived for S4G IRAC-1 (3.6 $\mu$m) imaging; e.g., at zero W1−W2 color, the M/L is about 0.68, which may be compared with the general value of 0.6 recommended for Spitzer S4G measurements in Meidt et al. 2014, but see also their Figure 4, showing the M/L color dependence, and the work of Eskew, Zarisky & Meidt (2012). It is worth repeating that W1, as well as IRAC-1, is susceptible to relatively short SF history (SFH) phases in which the mid-infrared emission is enhanced beyond these standard relations due to starburst, AGN and thermally-pulsating, warm and dusty, AGB (TP-AGB) populations (e.g., Chisari & Kelson 2012). The 3.3 $\mu$m PAH emission line is generally an insignificant contributor to the integrated W1 (or IRAC-1) band flux, but may be important for starbursting systems – for example this line is detected in M82 – which would lead to a stellar mass over-estimation.

The resulting stellar mass distribution, ranging from $10^{7.5}$ to $10^{12} M_\odot$ is given in Figure 12, which shows how the stellar mass changes with redshift shells. Because of the W1 dependency, and unlike the SF metrics, the stellar mass may be estimated for galaxies at large redshifts; nevertheless, Malmquist bias will favor the most massive galaxies, generally spheroidal, dispersion-dominated systems, at great distances. As expected, the most massive sources are also the most luminous and thus are detected to all depths although being relatively rare in the Universe. Conversely, the lowest mass galaxies are under-luminous ($< L_*$) and thus only detected in the nearby Universe, and only the Local Volume (D < 30 Mpc) for the lowest mass galaxies. As we see later, many of these dwarf galaxies have early-type colors, indicative of dwarf spheroids that are lacking any SF activity. For the GAMA sample, the peak in the distribution of stellar mass is at $\log_{10}$ $M_\star \sim 10.3 - 10.6$, and for the WXSC the mode of the distribution is much higher at ∼11.0. This reflects the fact that the resolved sources are low redshift galaxies which are large in angular size, translating to massive hosts.

Assuming the contribution from AGN emission is small compared to SF processes, the obscured – dust absorbed, re-radiated – SFR can be estimated from the 12 $\mu$m and 22 $\mu$m photometry, where the former is dominated by the 11.3 $\mu$m PAH and 12.8 $\mu$m [NeII] emission features, both sensitive to SF activity. The latter measures the warm, T∼150 K, dust continuum and is generally a more robust SF tracer, while the former is sensitive to metallicity and radiation field intensity (e.g., Draine et al. 2007; Seok et al. 2014). Heavily dust-obscured galaxies, such as Arp220, will also have a significant 10 $\mu$m silicate absorption feature in the W3 band.

Because of the decreased sensitivity in these two *WISE* channels, SF activity can only be measured for (1) relatively nearby galaxies, (2) relatively dusty galaxies and (3) luminous infrared galaxies, which can be seen at any redshift. For example, relatively quiescent galaxies such as early-type spirals are only detected in W3 in the Local Universe. This implies a SF activity and dust-content bias with redshift, as noted in Paper I. Moreover, we caution that even though the mid-infrared provides very convenient metrics for SF, they are rather large extrapolations of the dominant bolometric emission that arises from the cold dust (T≃25 K) in the far-infrared, and hence have large uncertainties and potentially discrepant excursions – a prime example is the HI-massive galaxy, HIZOA J0836-43, which has under-luminous mid-infrared compared to its far-infrared emission; see Cluver et al. 2010. Nevertheless, for "normal" metallicity galaxies, and typical stellar mass ranges, $10^9 - 10^{11} M_\odot$, they have proven to be a powerful tool to study galaxies (see e.g., Calzetti et al. 2007; Farrah et al. 2007; Paper I).

Here we employ an updated SFR calibration based on the total infrared luminosity of typical, nearby systems correlated to the corresponding mid-infrared luminosities (Cluver et al., in prep). Both the 12 and 22 $\mu$m SFRs follow from the spectral luminosities: $\nu L_\nu$, where $\nu$ is the bandpass central frequency and is normalized by the bolometric luminosity of the Sun. It is important not to confuse the spectral luminosity with the in-band luminosity, as they are very different in value due to the bolometric vs. in-band normalization; see Jarrett et al. 2013).

[21] This favorable k-correction is one of the reasons why *WISE* is useful for working with high redshift samples, and is even capable of finding some of the most distance QSOs (see e.g., Blain et al. 2013) and hyper-luminous infrared galaxies (Tsai et al. 2015).

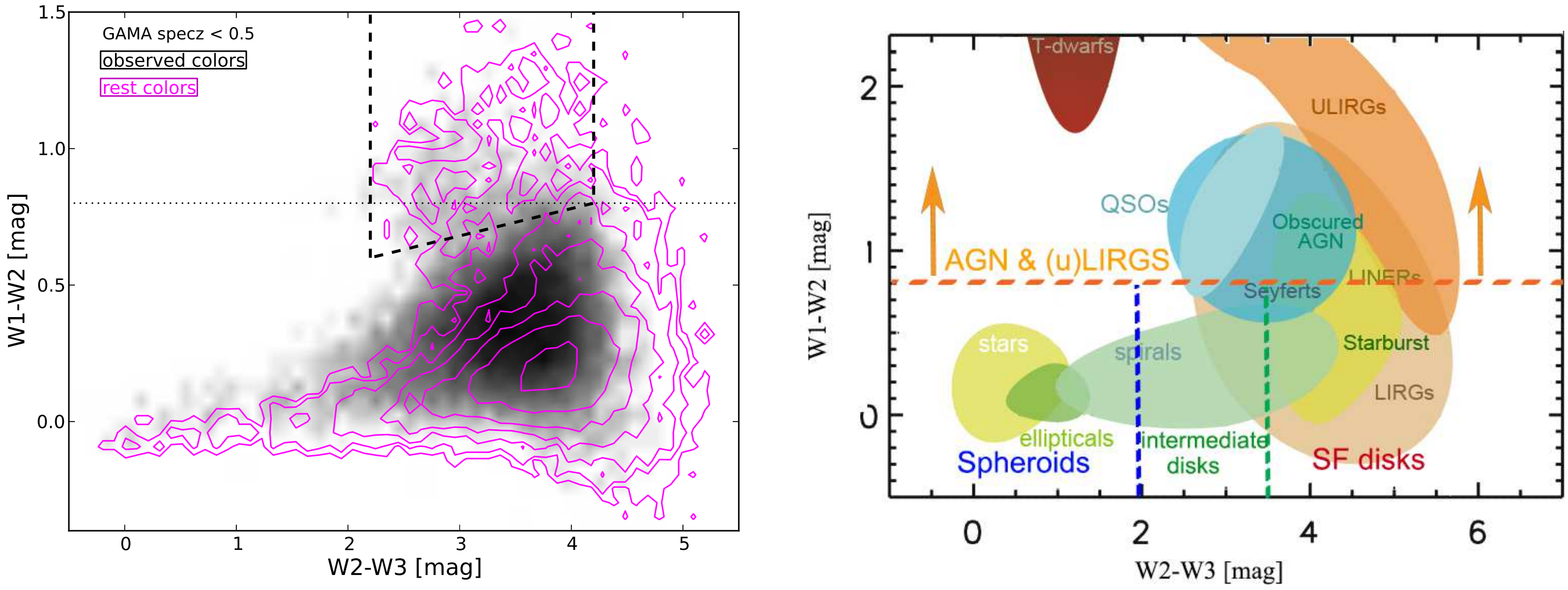

(a) Mid-infrared colors (b) Color diagram

**Figure 11.** *WISE* mid-infrared colors. (a) Color-color distribution of galaxies, comparing their observed and rest frame-corrected rest measurements. The corrections are such that the *WISE* W1 and W2 magnitudes appear brighter with redshift, shifting the ensemble towards the upper left. The horizontal dashed line is the AGN threshold from Stern et al. (2012), and the dashed lines denote the QSO/AGN zone from Jarrett et al. (2011). (b) color-color diagram that illustrates how galaxies separate by type; showing the simple divisions for early (spheroidal), intermediate (disk) and late-type disk galaxies.

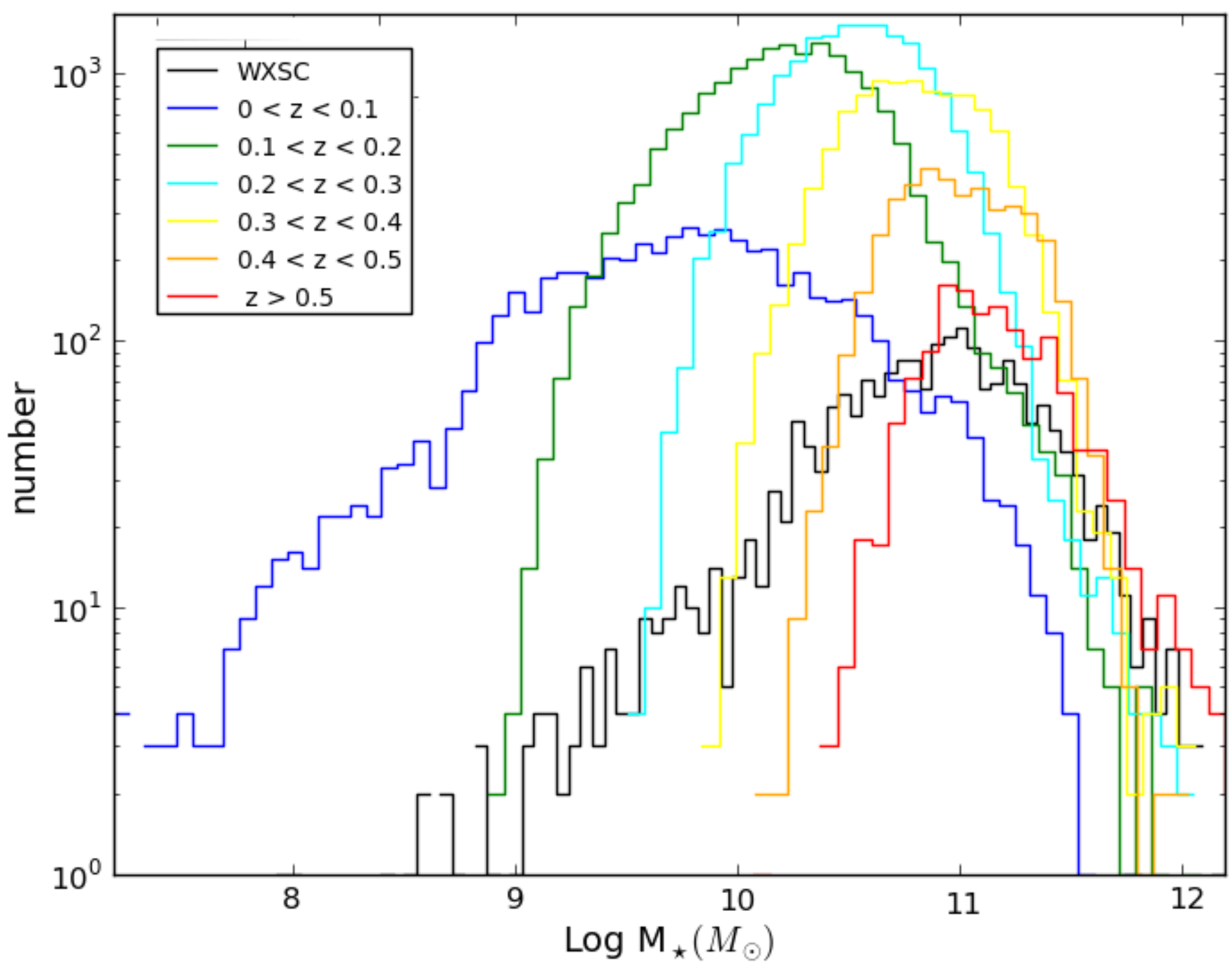

**Figure 12.** Stellar mass distribution in redshift shells. The lowest mass galaxies detected in GAMA, $M_\star < 10^9 M_\odot$, are only seen at low redshifts ($z < 0.1$) because of the faint surface brightness, while the highest mass galaxies, $> 10^{11} M_\odot$, are seen well beyond the Local Universe. For redshift where GAMA is relatively complete, $z < 0.3$, the mass distribution peaks around $> 10^{10.5} M_\odot$.

The resulting ensemble SFRs are shown in Figure 13, where we use the SFR(W4) when it is available, otherwise we revert to the SFR(W3). To help make sense of the SFR distribution, we relate the resulting SFRs with the corresponding host stellar masses, effectively the past-to-present SFH, shown in Figure 13a. This SFR-$M_*$ approach has been described as a kind of evolutionary or galaxy star formation "sequence", where ever larger SFRs track with ever larger stellar masses. This linear trend holds even for galaxies at high-z (see for example Noeske et al. 2007; Elbaz et al. 2007; Rodighiero et al. 2010; Bouché et al. 2010). Deviations occur for galaxies that are no longer forming stars – falling to the bottom right corner of the diagram– and those that are forming at a prodigious rate – starbursts rise to the upper left, but eventually return to the "sequence" after a relatively rapid period of disk building and continue their passive evolution.

Employing Figure 11b, we have divided our sample by the mid-infrared color, which is a proxy for the morphology i.e. galaxy type. Spheroidals and early-type spirals unsurprisingly fall off the "sequence", while the intermediates (green curves) appear to be in transition toward quenched, or decreased SF activity, with substantial bulges in place. Spiral/disk galaxies define the main-sequence, while infrared luminous galaxies have excess SFRs and move upwards relative to the main ensemble. Those hosting AGN will have over-estimated SFRs, – the prevailing problem is therefore separating the accretion-driven and SF-emitting components.

The slope in the sequence is linear for a large mass range, $10^{7.5}$ to $10^{10}$ ($M_\odot$), where it appears to turn over and become flatter. There may be two different populations creating the kink/inflection at this critical stellar mass threshold. It is interesting to note that the apparent slope in the lower-mass SF population, $M < 10^{10}$ ($M_\odot$), is steeper then what is seen with optically or UV-selected samples. For example, the relation of Grootes et al. (2013) used a volume-limited sample, to $z < 0.13$ and $M_* > 10^{9.5}$, of morphologically-selected GAMA spirals detected in GALEX, resulting in a SFR-$M_*$ relation (see dashed magenta line in Figure 13) that is much flatter than what is seen in our infrared-selected sample, except for the high-mass range, $M > 10^{10}$ ($M_\odot$). The difference in our two samples may be fundamental to the wavelength bands – we demand that W3, or W4, be detected, and thereby select dusty and more massive, possibly starbursting systems, compared to UV or optically-selected samples. GALEX, for example, is far more sensitive to dwarf and transition/quenched, low SF systems, but insensitive to dusty systems. The sequence we see in our G12 ensemble is very similar in slope to the high-z result of Elbaz et al. (2007), which would also be selecting higher-mass, as well as higher gas mass, SF systems. In any event, we seem to have a forked sequence of three tracks: galaxies on the sequence at lower masses, and high-mass systems which are either SF-active due to either tidal or major merger interaction, or transitioning to passive and ultimately quenched states.

Another way to view the galaxy star forming history is to normalize the SFRs by the host mass, thus forming the specific star formation rate (sSFR). In this way the sSFR mitigates the mass-dependent slope seen in the sequence, and provides a more straight-forward view of how galaxies build. The resulting sSFR is presented in Figure 13b. SFRs range from fully quenched to active >100 $M_\odot yr^{-1}$, but with the bulk of galaxies forming stars at >1 $M_\odot yr^{-1}$, consistent with the GAMA survey selection of SF blue galaxies, as well as the *WISE* sensitivity to SF galaxies in the W3 and W4 bands.

As expected, the sSFR diagram exhibits a flatter distribution than the SFR sequence, although there is an inverse trend in sSFR with stellar mass. Galaxies with high sSFR tend to be lower mass galaxies – that is, they are actively building their disks (e.g., NGC 3265) and some, such as M82, are doing so in rapid, starburst fashion, populating the upper envelope in sSFR which may be part of a fork or 'track' of enhanced SF extending to larger baryonic masses. At the extreme SF locus, the ultra-luminous NGC6240 is an example of a hybrid starburst+AGN merging system. Most nearby spiral galaxies fall into the center of the ensemble, including the large nuclear starburst, NGC253, and the barred grand-design spiral M83 (see Lucero et al. 2015; Jarrett et al. 2013, and Heald et al. 2016). Relatively quiescent disk galaxies, such as M81, and elliptical galaxies (e.g., NGC 4486; M87) have large stellar hosts and diminishing SF activity – they fall to the right corner of the diagram.

Indeed, the nearby M81 with its visually-stunning spiral arms is still producing new stars, but it has such a large and old bulge that its sSFR is relatively small in comparison. In the grand scheme of its lifetime, it has built most of its stars and is now gently evolving to retirement. In the case of the giant elliptical galaxy, NGC 4486 (M87), emission from old stars dominates all *WISE* bands – M87 is your classic "massive, red and dead" galaxy – although there is an infrared excess due to hot accretion from a supermassive black hole lurking in the center; see (Jarrett et al. 2013) for a detailed SED of M87.

To summarize: (1) lower mass galaxies are actively building, even while their global SFRs are relatively small, average < 1 $M_\odot$ $yr^{-1}$, (2) intermediate-massed galaxies have typical ensemble or evolutionary-sequence building, SFRs ∼ few, but may also be in their starburst phase and populate an upper-level track in the sSFR diagram, and (3) massive galaxies, >$10^{11}$ $M_\odot$, have consumed their gas reservoirs and for the most part completed building their super-structure, existing in a quiescent, passive, quenched or 'dead' state. That is not to say that massive galaxies cannot be re-activated to some degree with gas-rich, dissipative merging and major accretion events.

Finally, we caution that infrared-based SFRs for low-mass dwarf galaxies gives an incomplete census of the SF activity since much of the UV light produced by the young and massive populations, which trace the overall SF activity, escapes the galaxy. Optically-thin systems require both UV/optical spectro-imaging and infrared imaging to estimate the total SF activity. Hence, the *WISE* mid-infrared estimated SFRs are lower limits for the total SF activity in dwarfs and low opacity systems.

### 4.2. *Radio galaxy population*

The Large Area Radio Galaxy Evolution Spectroscopic Survey (LARGESS; Ching et al. 2017) is a spectroscopic catalogue of radio sources drawn from the FIRST radio survey, chosen to span the full range of radio AGN populations to ∼ 0.8. As part of this study, optical spectra of radio-selected objects were obtained in the GAMA fields, including G12. Cross-matching using a spatial cone radius of 5 arcsec with the *WISE* galaxy catalogue has a > 90% match rate (see Figure 7). The total sample, however, is relatively small in number, less than 1000 total sources; see Table 1. Their classification scheme delineates the sources into four general classes, the first three are AGN-dominated: strong or high-excitation lines (HERGs), weak or low-excitation (LERGs), and broad emission lines (AeBs), while the fourth are those sources dominated by dominant star SF activity. AeBs are similar to the classic Type-I QSOs

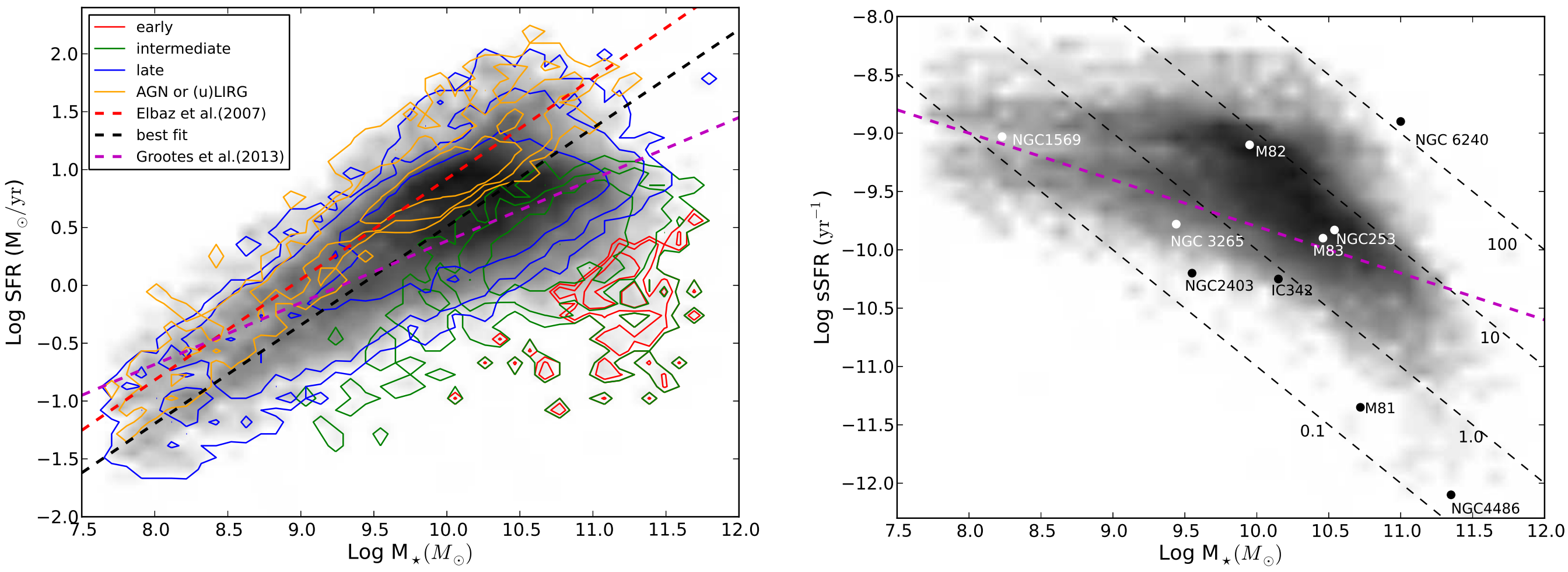

(a) Star Formation Rate Sequence (b) Specific Star formation Rate

**Figure 13.** Star formation rate (SFR) relative to the host stellar mass, $M_*$ ($M_\odot$). The grey scale represents all sources with $z < 0.3$, including photometric redshifts. The left panel (a) shows how the rate changes with mass, delineated by *WISE* color: early, intermediate, late and AGN and infrared-luminous types. The black dashed line represents the average 'sequence' for the sample; the red dashed line the relation at high redshift (Elbaz et al. 2007), and the dashed magenta line is for a nearby GALEX-GAMA sample (Grootes et al. 2013). The right panel (b) shows the equivalent sSFR distribution, with dashed lines representing lines of constant SFR (0.1, 1, 10 and 100 $M_\odot$ yr$^{-1}$), and the magenta dashed line the relation from GALEX-GAMA. For comparison, a few nearby galaxies values are indicated, from massive spheroidal (NGC4486; M87) to star-forming (NGC 253) spirals and starbursts (M82), and to dwarfs (NGC1569).

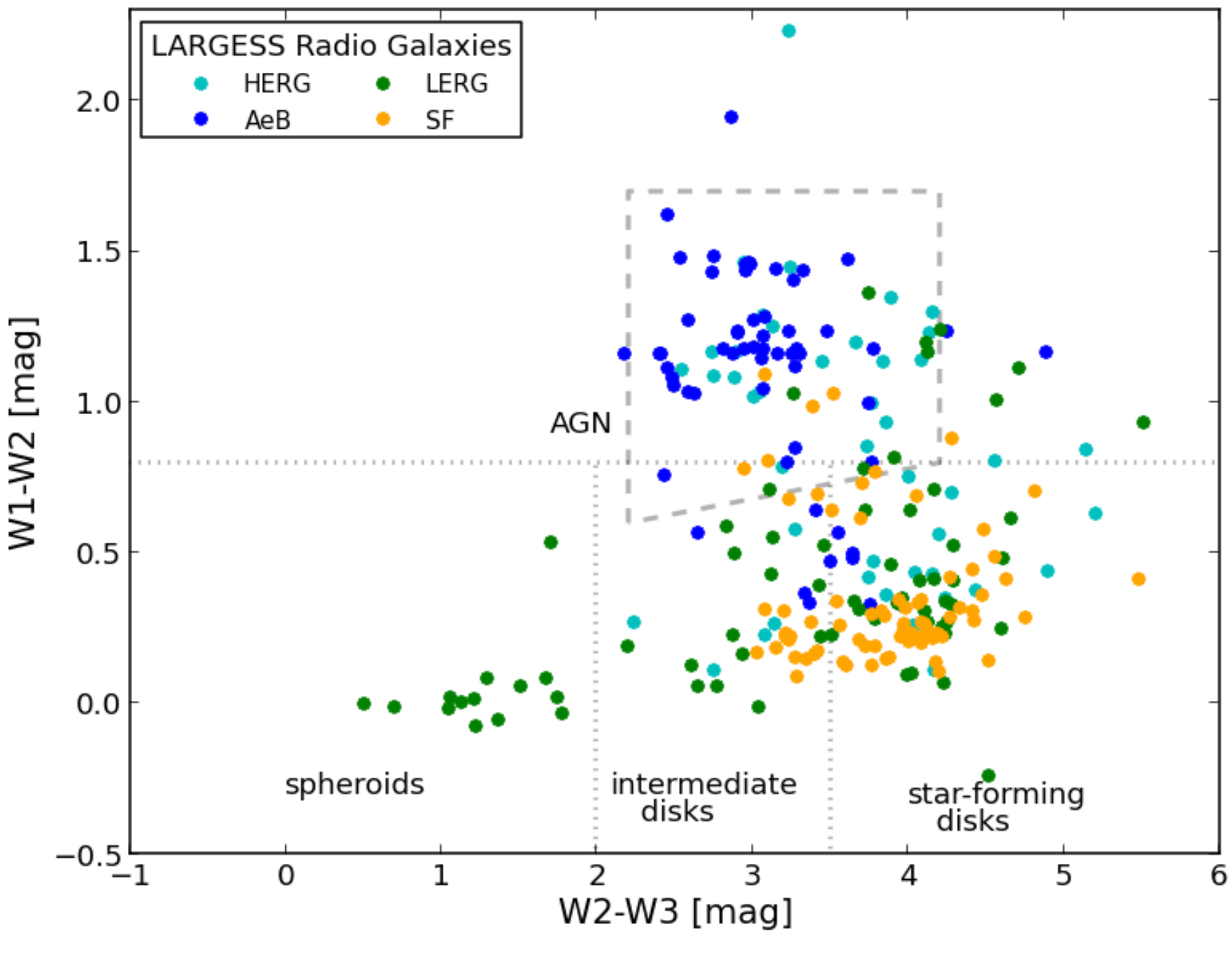

**Figure 14.** Radio galaxy colors in the *WISE* W1−W2-W3 diagram. Radio galaxies from the LARGESS (Ching et al. 2017) are delineated by their spectroscopic properties, including those with high-excitation lines (HERGs), weak or low-excitation (LERGs), those specifically with broad emission lines (AeBs), and with dominant star formation properties (SF).

– the rarest type in the G12 galaxy sample since these were selected against in the original GAMA selection of SDSS galaxies, eliminating sources that are not resolved in the optical imaging. For less extreme-power galaxies, the difference between LERG and HERG is thought to be driven by the accretion mechanism: jet-mode and radiative-mode, respectively (see Heckman & Best 2014, and the discussion in Ching et al. 2017).

Here we investigate how the infrared colors discriminate between these classes; Figure 14 shows the *WISE* colors of the LARGESS radio galaxies, which may be compared with the larger-sample results of Ching et al.

(2017; see their Figure 15) and Yang et al. (2015). Not surprisingly, the AeBs populate the *WISE* QSO region of the diagram (defined in Jarrett et al. 2011), displaying very warm, accretion-dominated, W1−W2 colors. At the other end of the spectrum, galaxies whose host emission is dominated by SF populate the disk/spiral galaxy region of the diagram; ie. cool W2−W3 colors. Nevertheless, some galaxies have much warmer colors, suggesting AGN activity, a reminder that these spectroscopic classification schemes are not always reliable or robust to degeneracies. Finally, the LERG and HERG populations have diverse infrared colors, ranging across the *WISE* diagram. Interestingly, only LERGs – the most common radio source – are located in the early-type/spheroidal region of the diagram; i.e., stellar emission dominated hosts, consistent with high-mass halos driving jet/hot-mode accretion. LERGS can also exhibit strong ISM emission – the hosts are either undergoing a 'wet' merger event or undergoing a recent starburst trigger. HERGs populate both the ISM and AGN-dominated regions of the diagram with about equal numbers, signifying both SF hosts and AGNs in lower-mass halos, driven by radiative or cold-mode accretion. Hence HERGS are likely those with hybrid (SF/AGN) mechanisms. We conclude that *WISE* colors, although crude in fidelity, may be used to study extragalactic radio sources and their evolutionary state, which should be notably helpful with the SKA-era now underway with ASKAP, MeerKAT and APERTIF..

### 4.3. *Source density maps*

Quantifying the spatial and density distribution of galaxies has a number of applications, for example, improving photometric redshifts by using the properties of the cosmic web as a prior input to statistical (e.g., neural network) assessment of photometric information, as well as investigating the environmental influence upon evolution (see e.g., Aragon-Calvo et al. 2015). Moreover, since future surveys (e.g., ASKAP-EMU) will be combined with *WISE* and other multi-wavelength datasets to study the cosmic web, our goal is to understand how infrared-selected samples characterize the clustering in large scale structure, including the nature – stellar mass and SF activity – of the host galaxies. To this end, we investigate the number density distribution, angular and radial correlations, and the 3D structures in the G12 field.

Here we consider a straightforward method to locate overdensities and coherent structure in the field using the *WISE*-GAMA redshift sample. The goal is to highlight clustering on scales of a few Mpc to tens of Mpc. More exhaustive methods have been applied to GAMA data, for example, using cylinders, nth nearest neighbor, and friends-of-friends to construct clustered catalogs and to study their environmental effects Alpaslan et al. 2015; Brough et al. 2013; Robotham et al. 2011. In this study, we simply count the number of sources in 5 Mpc-diameter spheres and catalog the largest overdensities. The sphere size is chosen to be large enough to include clusters and their associated redshift distortions, but also likely too large to discern sub-Mpc environmental conditions; we do not correct for the incompleteness ($z > 0.2$). To search for correlations between individual clusters, we use a large, 20 Mpc-diameter sphere to identify superclusters or larger, possibly connected structures, and compare to previous 'skeleton' constructions. Note that given the relatively small volume at low redshifts for the G12 field, the larger sphere has little meaning for $z < 0.1$. Lastly, we compare with a galaxy groups catalog that shows linked-structures.

The method uses a similar approach to source identification, i.e. find distinct local maxima representing overdensities. For each source, we derive the spherical coordinates, co-moving XYZ in Mpc, luminosity distance, co-moving distance, and finally the co-moving radius. In spherical coordinates it is trivial to find proximal neighbors, although redshift distortions are still in play, although negligible here – the relatively large sphere minimizes this complication. Boundaries and edges are corrected for by computing the effective volume for the spheres centered on the overdensities; clusters near boundaries may have a smaller number or incompleteness, whereas the space density – number per volume – corrects for this using the correspondingly smaller volume.

To identify local maxima, we use a brute force method in which for each source in the sample, with redshifts, we count the number of nearby sources within a 2.5 Mpc radius, for example, identifying the cluster with the highest count. The centroid is then computed, thus refining the central location of the grouping. The overdensity is cataloged and all sources within 2.5 Mpc of the location are then removed. The process is then repeated. In each iteration, one maximal density is identified, cataloged and sources removed. In this way we build a top-down, or maximal, density catalog which will later be used to interpret the 3D cone diagrams.

A sampling of the largest overdensities (with N > 15), based on the 5 Mpc-diameter sphere and sorted by redshift is given in Table 2. The 'density' metric is simply the log number per $\mathrm{Mpc}^3$. The centroid locations of the over-densities is given by the equatorial coordinates (J2000) and the spherical coordinates (X, Y, Z), which geometrically follow from the Galactic coordinates. Also indicated are the central luminosity distance, the mean host galaxy mass and the mean W1 absolute mag for the group ensemble. As expected, the nearest 'groups' are sensitive to the sub-luminous and lowest mass systems, while the most distant grouping in the GAMA volume have mean stellar masses and luminosities greater than $M^*$.

The densest clustering is located at a distance of ∼90 Mpc ($z \sim 0.02$), which appears to be a filament (see next section) of small groups, none of which are true clusters. This nearby grouping is not remarkable, but rather a consequence of detecting more lower-mass objects nearby, thus inflating the density metric. Instead, the most striking clustering occurs in the next redshift shell – this is discussed further in the next section. Using a larger search area, sphere of 20 Mpc in diameter (Table 3), the outstanding overdensity is at 500 Mpc, $z \sim 0.1$, comprised of smaller clusterings that appear to make a more complex super-structure suggesting merging super-groups.

A graphic illustration of the 5 Mpc over-densities is shown in Figure 15, which contains the projected maps, but now separated by three redshift ranges. The nearby shell, $z < 0.1$, has a log (co-moving) volume = 5.67 $\mathrm{Mpc}^3$, the intermediate shell, $0.1 < z < 0.2$, has a log volume = 6.47, and the distant shell, $0.2 < z < 0.3$, has log volume = 6.86; all together, the cone has a total co-moving volume of $10^{7.03}$ Mpc$^3$. The locations of dense clusters are depicted using red circles, whose radius corresponds to 2.5 Mpc, therefore apparent size differences are a depth effect. The larger spheres, 20 Mpc in diameter, are color coded green and identify groupings of clusters – note that for the first shell, the larger sphere is ignored since the volume is too small.

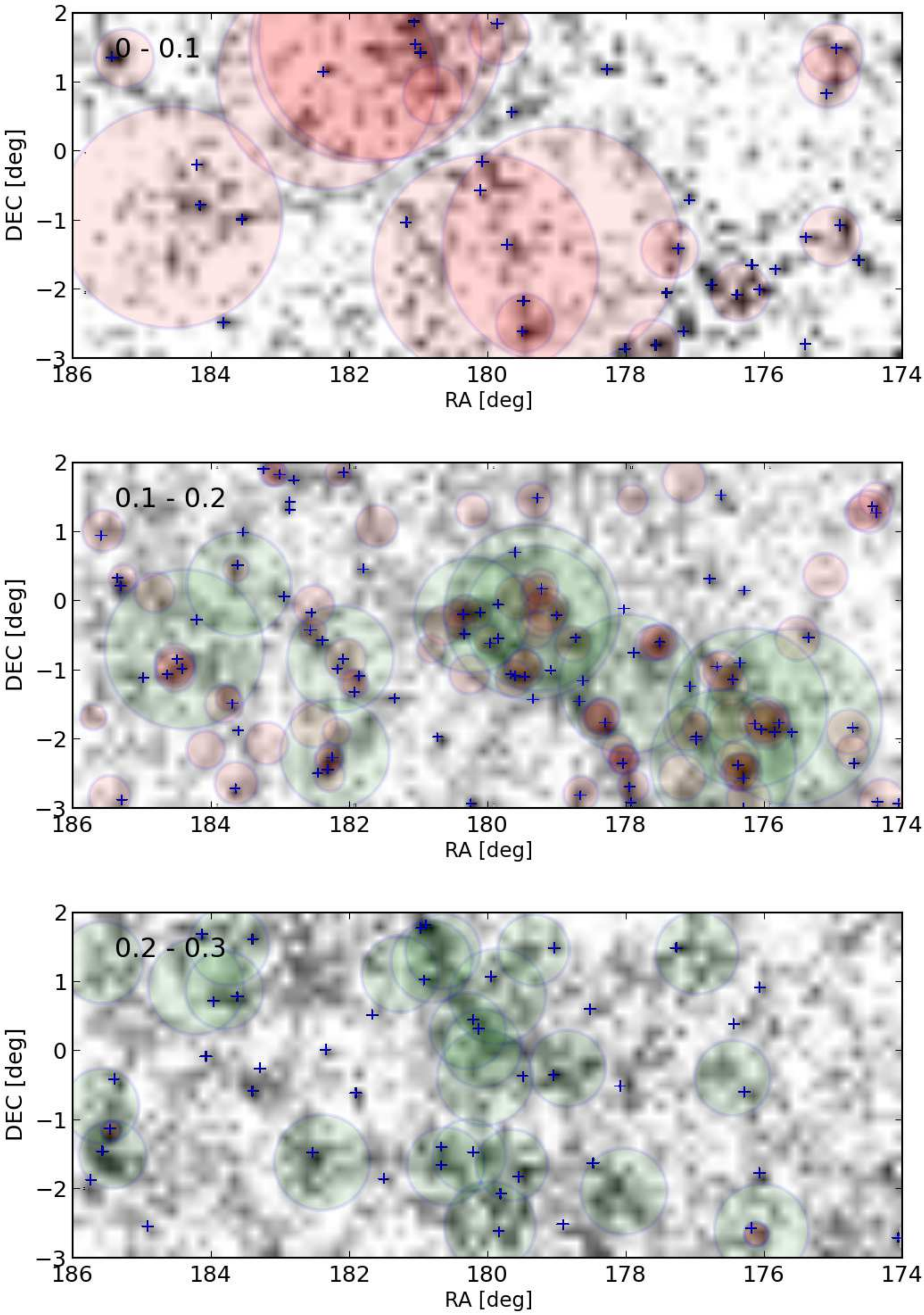

**Figure 15.** Projected distribution of *WISE*-GAMA galaxies in three redshift shells: $z < 0.1$, $0.1 < z < 0.2$, $0.2 < z < 0.3$. Overdensities are indicated with filled red circles, with diameter 5 Mpc, and by the green spheres (20 Mpc diameter). Apparent sizes vary due to depth effects in the shells. For comparison, blue crosses demark the locations of groups and clusters as given in the GAMA Catalog of Galaxy Groups (Robotham et al. 2011).

For comparison, we show the friends-of-friends GAMA Galaxy Groups Catalogue (G3C; Robotham et al. 2011), whose discrete locations are denoted with blue crosses. By eye, there is generally good agreement between the two methods, although a few clear differences can be located, chiefly in the diffuse regions. A more rigorous cross-match between the overdensity maxima and the G3C reveals match rates of over 50%; for example, in the redshift range between 0.1 and 0.2, there are ∼150 identified G3C groups whose membership number is greater than 7 'friends of friends', and from this study ∼140 density peaks (N >= 10) of which 92 (66%) match spatially within 2 Mpc of a G3C group. The mean positional offset between the two catalogues is 1.4 ± 0.5 Mpc, which reflects the different methods for computing the group centers and the blending/confusion between groups using a large spherical diameter (5 Mpc) filter. An an example of a relatively large group, $\sigma = 508$ km/s, G3C ID = 200009 (ra,dec,redshift) = 176.3816, -2.5257, 0.13159) is located within 1.0 Mpc of density peak (176.3770, -2.4768, 0.13169), LogDensity = -0.436 Mpc$^{-3}$, with the mean host mass of Log M/M$_\odot$ = 10.59 ± 0.44, and mean absolute magnitude of -23.0 ± 1.4 mag (see Table 2).

It is interesting that in the last redshift shell ($z$ >0.2), where incompleteness sets in, the GAMA groups (blue crosses) and large 20 Mpc regions have spatial correspondence, whereas there are very few 5 Mpc overdensities. That is to say, larger volumes; i.e. spheres, are needed to identify clustering at higher redshifts because of the GAMA incompleteness (see Figure 9a) coupled with the increasing bias toward higher mass yet rarer systems.

Most of the clustering appears in the middle panel, $0.1 < z < 0.2$, which for GAMA is optimal in terms of detection completeness and spatial volume – see the selection function, Figure 9a). Two prominent overdensities at $z \sim 0.17$ in redshift, appear to be centered within grouping complexes (see middle panel, Figure 15). Here we attempt to estimate their respective velocity dispersions and cluster masses; Figure 16 shows the peculiar velocity distribution for each overdensity, where clustering peculiar motions are described by (in the non-relativistic case)

$$v_i = c\,\frac{z_i - \bar{z}}{1+\bar{z}} \qquad (2)$$

and the dispersion follows as the root mean square (RMS) of the distribution. The virial radius and the R200, relative to the cosmic critical density, in combination with the velocity dispersion are then used to compute the virial and M200 masses, respectively (see e.g., Navaro et al. 1995).

The first grouping, $z = 0.1649$, has a mean radial velocity of 45,400 km s$^{-1}$ and a corresponding dispersion of 200 km s$^{-1}$ in a 2.5 Mpc radius. The equivalent virial mass is $4{\times}10^{13}M_\odot$, which is more typical of a galaxy group, while the *WISE*-GAMA detections are of relatively massive galaxies, with mean Log M/M$_\odot$ = 10.71 ± 0.34. The small velocity dispersion implies a modest R200 ∼0.7 Mpc, and a corresponding M200 mass of $9{\times}10^{12}M_\odot$. It is reasonable to conclude that it is part of a filamentary web of galaxy groups, likely to still be in a dynamic phase and unlikely to be virialized, not unlike the merging Eridanus Supergroup, Brough et al. 2006. Moreover, the G3C does not match with the density peak, but does have two groups that are adjacent and within 10 Mpc, also suggesting that this is a filamentary complex. The angular extent of the *WISE*-GAMA members is 22×17 arcmin (Δra vs. Δdec), which would be a relatively large area to cover for future high-z galaxy lensing studies (e.g., JWST; Euclid).

The second is larger in mass, with a mean radial velocity of 47,800 km s$^{-1}$ and a corresponding dispersion of 350 km s$^{-1}$ in a generous 3.3 Mpc radius. The equivalent binding or virial mass is $1.5{\times}10^{14}M_\odot$, which implies it is a modest-sized galaxy cluster, but still unlikely to be relaxed. The implied R200 is ∼1.1 Mpc, and $4.9{\times}10^{13}M_\odot$, which is still group-sized. The nearest G3C object is 4 Mpc in radius (ID=200121) and has a similar velocity dispersion, 390 km s$^{-1}$. Consequently, this greater web-like structure may indicate dynamical assembly, although considerably smaller in size and mass than, e.g., the still-forming Abell 1882 complex in the GAMA G15 field (see Owers et al. 2013). The corresponding angular extent is similarly large,14×23 arcmin on the sky, for lensing consideration.

Ultimately these projections are rather limited for interpretation because of crowding and the two-dimensional projection. Exploration is best suited using 3D visualization tools – in the next section we use more sophisticated tools to explore the 3D-spatial structures.

### 4.4. *Two-point correlation functions*

Correlation functions are a way to quantitatively describe clustering, or structure, in the spatial distribution of galaxies, effectively used to study the baryonic acoustic oscillations imprinted in the matter distribution (cf. Blake et al. 2011; Jeong et al. 2015). Methods employed focus on the angular ($\Delta\theta$) relationship between galaxies, and when redshift or radial information is available, the line-of-sight ($\Delta z$) clustering. The most straight forward statistical method is the galaxy two-point correlation functions (2PCF), which includes the redshift space component, $\xi(r)$, and the angular component, w($\theta$), which quantifies the amplitude of clustering relative to a random (non-clustered) distribution. Very simply, this is done by tracking pairs of galaxies across all scales – i.e. their separations in space – and hence named the two-point correlation. For G12, we use our *WISE*-GAMA galaxies, i.e. redshifts are known, to define the sample that will be characterized with the 2PCF method.

#### 4.4.1. *Galaxy angular clustering*

We begin with the angular correlation function. The galaxy sample is divided into stellar mass and color ranges in order to explore how the clustering depends on the host properties. We limit the analysis to sources with redshifts $z < 0.5$, and note that although density evolution is a factor, the intrinsic properties of the galaxies dominates the clustering analysis. For each sub-sample, we construct a random, simulated non-clustered distribution which will be used to compare to the real distribution. The random sample has 50× the size of the real data for statistical robustness, but we ignore the small 'lost' spaces around bright stars; i.e., the mask is uniform, which is a reasonable approximation for the clean *WISE* imaging of G12.

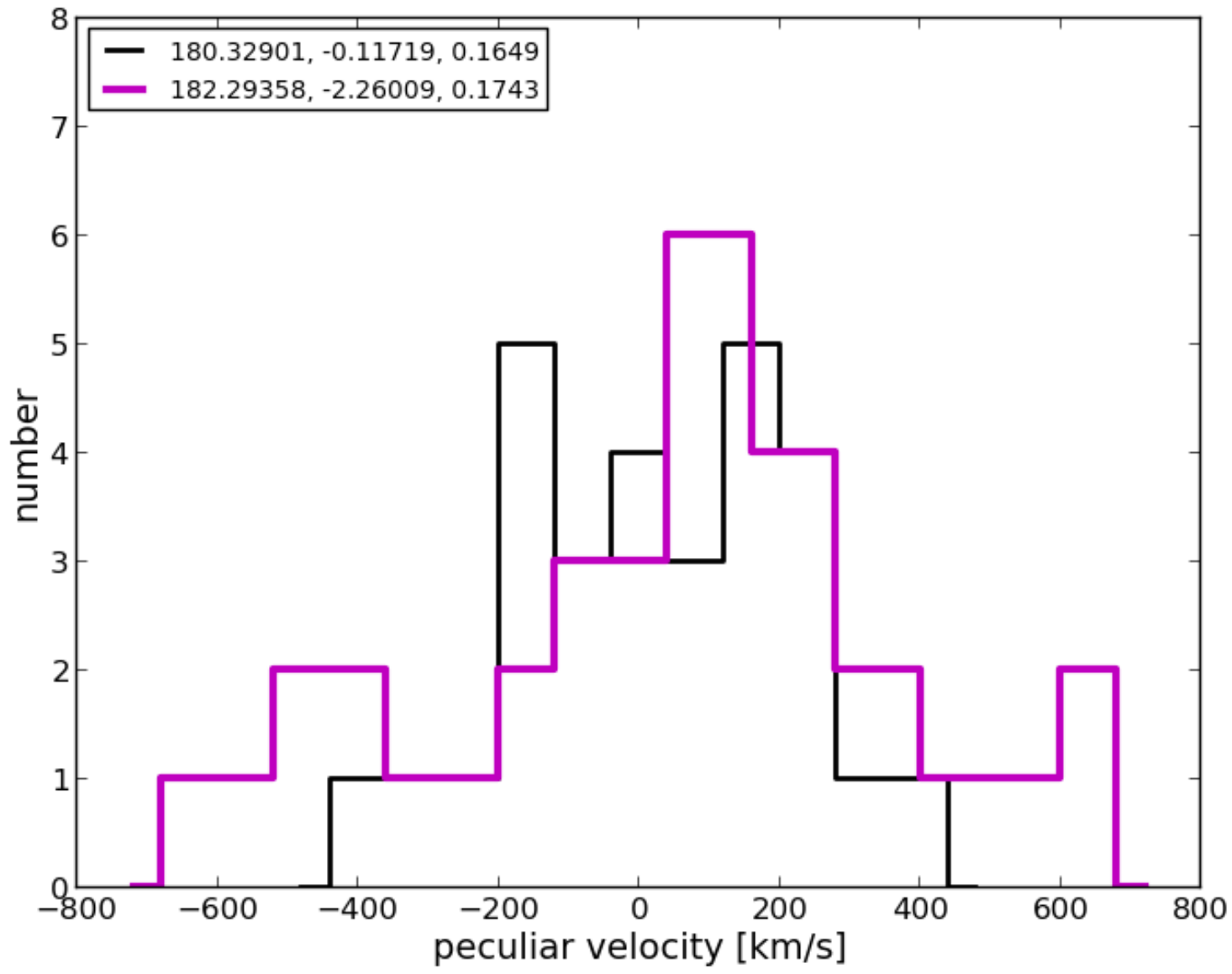

**Figure 16.** Peculiar velocity distribution of two small clusters near $z \sim 0.17$. The central locations (ra, dec, redshift) of the two objects are noted in the figure legend.

This statistical comparison – between pairs of real galaxies, fake galaxies and combinations thereof – is carried out with the Landy & Szalay (1993) estimator, and we employ the free open-source code Correlation Utilities and Two-point Estimation (CUTE ; see Alonso 2013) to perform the calculations over the $12 \times 5$ degree field. Because of the field size constraints, we are unable to probe $\theta$ scales larger than ∼2 degrees. The results are shown in Figure 17 for a volume extending to z < 0.5, with mass ranges in panel (a) and color in panel (b). In addition, and for comparitive purposes, we show the angular correlation for resolved *WISE* galaxies (WXSC), which represent a wide range of galaxy type, but with typical redshifts less than 0.2.

Figure 17a, comparing between stellar mass ranges and the WXSC, in all cases w($\theta$) follows a power-law trajectory, $\sim \theta^{-0.8}$. The angular clustering trend has been noted in many studies going back decades (e.g., Groth & Peebles 1977; Lidman & Peterson 1996; Wang et al. 2013) and it is a scaling property of the cosmic web through a combination of the real-space clustering and the redshift distribution of the sources $N(z)$. What is different is the amplitude of the clustering, with a consistent decrease in the clustering at lower stellar masses. The most massive galaxies, which tend to be morphological ellipticals and S0's, have the strongest clustering, while low mass field isolated and dwarf galaxies have the least amount of clustering, which is consistent with results from other large surveys (e.g., Zehavi et al. 2016). It should be noted that the massive galaxies (red and magenta curves) are seen at all redshifts because they are also the most luminous ($> L_*$) in the *WISE* 3.4 $\mu$m band. Conversely, the lowest mass galaxies (blue and cyan curves) are only seen nearby, $z < 0.1$; see also Figure 12. The observable fact that $N(z)$ peaks at different redshifts for a given stellar mass range means that interpretation of the real-space clustering through the angular clustering is not straightforward and should be tempered accordingly.

Similar results are obtained if the galaxy sample is delineated by apparent W1 magnitude: brighter magnitudes cluster more strongly than fainter magnitudes, although because of mixing across redshifts, and hence, mass range, the signal is muted. And likewise for SDSS studies, clustering amplitude is dependent on the apparent optical magnitude (see Wang et al. 2013, Figure 15); but ultimately, it the intrinsic properties, such as host mass, that reveal clustering behavior (e.g., Norberg et al. 2001; Connolly et al. 2002).

To explore the clustering to host-type connection, in Figure 17b we separate our sample according to the mid-infrared colors, using the simple divisions described previously (Figure 11b). This method demands that all three bands (W1, W2 and W3) be detected with adequate S/N; as a consequence, spheroidals tend to be only detected in the Local Universe ($z < 0.1$) because W3 is very weak for these types, while star-forming disk galaxies tend to be biased to the high-mass systems. As noted in the SFR-Mass relation, Figure 13, the steepness of the trend is likely due to *WISE* W3 selecting large opaque/dusty systems as opposed to SF dwarf systems.

The resulting curves show two clear differences: spheroidal galaxies have the strongest clustering, while disk/spiral galaxies have the lowest amplitudes, consistent with the stellar mass results. Spheroidals are the most massive galaxies and tend to live in galaxy clus-ters, while spirals tend to be filamentary and field distributed (e.g., Zehavi et al. 2016). Presumably the SF spirals may be even less clustered than what is shown here given that *WISE* W3 is selecting higher mass and hence more clustered systems; see Brown et al. (2000). Galaxies with intermediate colors, which are a mix of early and late-type disks/spirals, have a clustering similar to the WXSC, a mix between field and cluster (i.e., the dominant component of both). It is interesting to note that the sources with warm W1−W2 colors, including galaxies that may be harboring AGN, have relatively low clustering amplitudes, and possibly a scaling distribution that is flatter than the $\sim \theta^{-0.8}$ trend for all other samples. As noted earlier, the GAMA point-source exclusion eliminates most high-z QSOs and those in which the AGN is much brighter than the host; consequently, the AGNs in this study will have hosts that are detected by *WISE* and are primarily nearby, low-power AGN and Seyferts. There are many studies of AGN and QSOs at high redshifts –including using *WISE* (e.g. Donoso et al. 2014) – that suggest powerful AGN preferentially exist within over-dense environments, and would therefore exhibit strong angular clustering; see for example the recent AGN clustering work from Jones et al. (2015); Assef et al. (2015); Chehade et al. (2016); Mendez et al. (2016). Given that GAMA is not optimal for studying AGN and the completeness is therefore poor, the clustering results should be interpreted with caution.

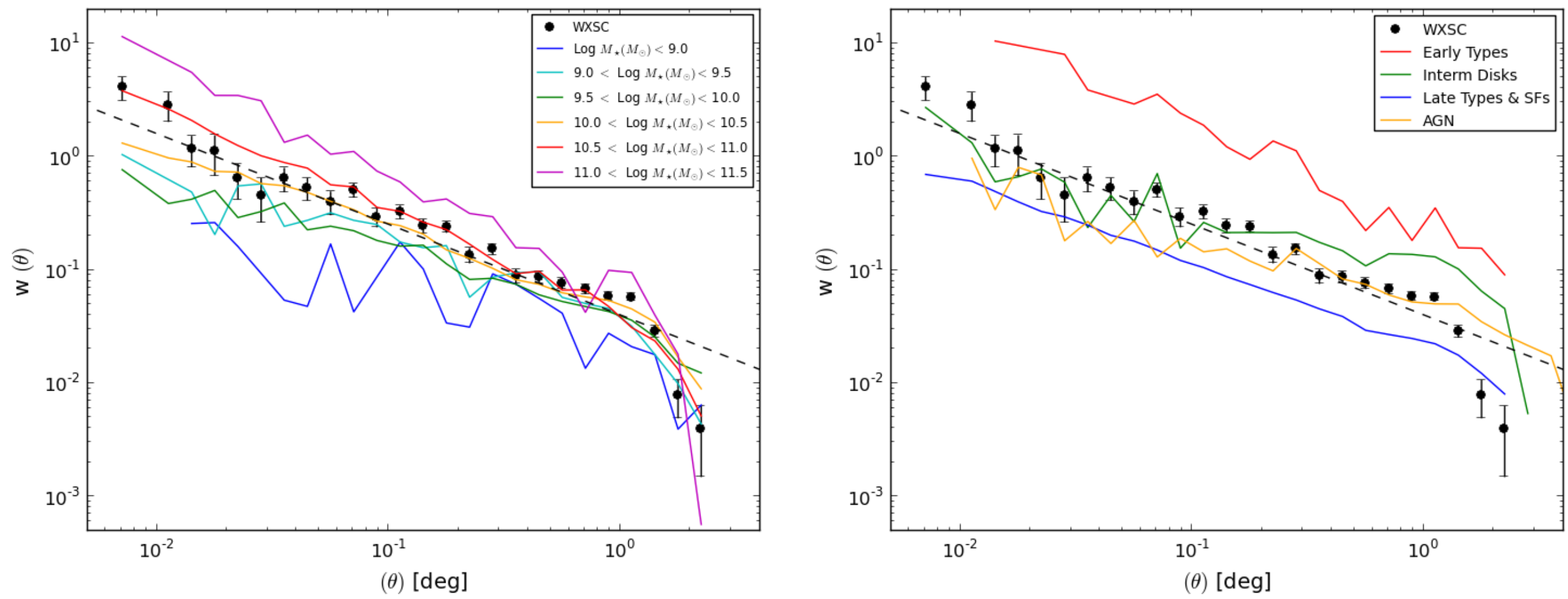

(a) Angular Correlation by Stellar Mass

(b) Angular Correlation by Color Type

**Figure 17.** Galaxy two-point angular correlation function for (a) stellar mass and (b) *WISE* color ranges for $z < 0.5$. For comparison, we also show the result for *WISE* resolved sources (WXSC) including their 1-sigma uncertainties (filled points and bars). The dashed line represents a power law of index $-0.8$. For color separation, the plane W2−W3 vs. W1−W2 is used to separate galaxy type (see Figure 11 and text).

#### 4.4.2. *Radial and transverse clustering*

Combining the angular and redshift information, it is possible to probe the spatial clustering of galaxies. Here we focus on the radial 2PCF correlation, $\xi(\Delta z)$, the 2-dimensional parallel-to-transverse correlation, $\xi(\pi, \sigma)$ and the projected radial correlation, $w_p(\sigma)$. Since redshift distortion renders radial 2PCF correlation $\xi(\Delta z)$ difficult to interpret, the usual procedure is to integrate $\xi(\pi, \sigma)$ along the radial axis ($\pi$) to arrive at the projected relation, as follows (see also Farrow et al. 2015 for further details and analysis):

$$w_p(\sigma) = 2\int_0^{\pi_m} \xi(\pi, \sigma) d\pi \qquad (3)$$

As with angular correlations, a random sample must be drawn that can be used to compare with the real data. We employ the GAMA galaxy $N(z)$ function as the selection function, described in Section 3.5 (shown in Figure 9a) to construct a random sampling that covers the 60 deg$^2$ G12 field to a redshift of 0.5. Here again we use CUTE to carry out the 2PCF calculations, setting the maximum angular aperture to be 1 degree – the assumption for co-alignment along the z-axis. This choice of diameter is a balance between collecting enough sources to be statistically meaningful, while avoiding combining structures that are not actually correlated i.e. blending the signal.

The correlation results are shown in Figure 18. In the first panel (a) we show the radial correlation delineated by three coarse redshift shells: the complete shell (0 to 0.5), the lower end (0.1 to 0.3) and the higher end (0.3 to 0.5). Recall that the most distant GAMA redshift shells are dominated by luminous, massive galaxies, which are also the most strongly clustered. The largest amplitudes occur near $\Delta z \sim 0$, i.e., very small separations, and then floor at higher separations. There are various upward wiggles, likely noise or artifacts from the small volumes, and conservatively we can conclude that there are no correlations for $\Delta z > 0.01$. The sample is likely too small, and likewise, the volume too small, while the aperture diameter is too large to cleanly delineate radial structure. We do, however, highlight an interesting repeatable feature at $\Delta z \sim 0.06$. This separation corresponds to roughly $\sim$25 Mpc h$^{-1}$ at $z = 0.4$, and is visually seen in Figure 18b, which shows the radial-transverse correlation for the *WISE*-GAMA source distribution, as a ver-

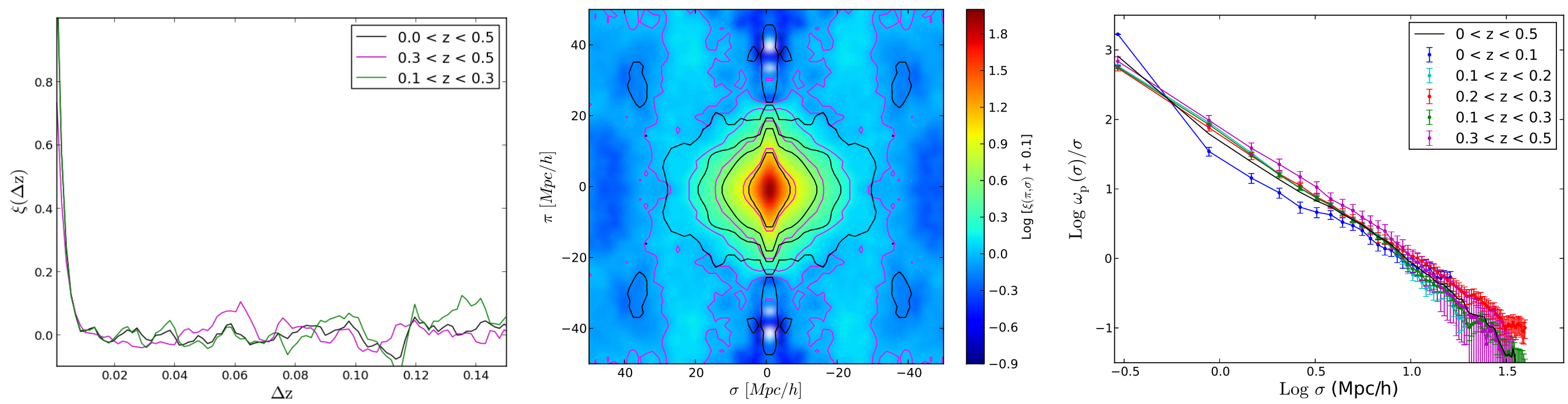

(a) radial $\xi(\Delta z)$ correlation (b) 2D $\xi(\pi, \sigma)$ correlation (c) projected $w_p(\sigma)$ correlation

**Figure 18.** Radial and 2D two-point correlation function of the G12 galaxy distribution. Left panel (a) is the radial (or line-of-sight) correlation, $\xi$ vs $\Delta z$, where sources with angular separations less than 1 degree are considered co-aligned. Three coarse redshift families are compared,; note that $\xi(r)$ is subject to distortion along the $\Delta z$ axis. The middle panel (b) shows the 2PCF in the $\pi$ (parallel) vs. $\sigma$ (perpendicular) plane. Here we compare two redshift families: $0.1 < z < 0.3$ (black contours) and $0.3 < z < 0.5$ (background grey-scale and magenta contours). The contour levels are the same for both; Log $\xi(\pi, \sigma) = 0.1$, 0.3, 0.5 and 1.0. The elongation along the $\sigma = 0$ axis is due to 'finger of god' redshift distortion, inducing a negative correlation (white space). The right panel (c) shows the projected radial correlation, $w_p(\sigma)$, for narrower redshift shells to clearly delineate the nearby and more distant clustering.

tical cyan band feature (20 − 25 Mpc $h^{-1}$).

In the middle panel, Figure 18b, the redshift distortion is readily apparent, creating the vertical elongated inner structure – the distortion works on scales of galaxy clusters, < 7 Mpc $h^{-1}$. The diagram also shows the 0.3 to 0.5 shell in color-scale, with contours in magenta to guide the eye, while also showing (black) contours for the closer redshift shell, 0.1 to 0.3. Although they are very similar, the distant shell has a more extended distribution, and more power in the $\xi(\Delta z)$ feature at ∼25 Mpc $h^{-1}$, most apparent along the $\sigma$ (transverse) axis. The location of this feature is consistent with a dynamical flattening distortion from the Kaiser Effect (Kaiser 1987). Alternatively it could be a small-sample statistical anomaly, perhaps associated with the small spatial extent of the G12 field, and should be investigated using the full GAMA repository.

Our attempt to minimize the effect of redshift distortion is presented in the last panel, Figure 18c, where $\xi(\pi, \sigma)$ is integrated along the $\pi$ axis (Eq. 5) to a radial limit of 40 Mpc $h^{-1}$ (after which noise overwhelms the signal). Now we further divide into smaller redshift shells to reveal any clustering differences, similar to Figure 17. The clustering power and linear trend, in log-log space, are similar to the GAMA results of Farrow et al. (2015); it is interesting to note that the strongest clustering is seen at the highest redshifts (0.3 to 0.5) and the weakest in the nearest shell (0 to 0.1), fully consistent with the angular correlation results showing the strongest clustering with massive spheroidal galaxies and the weakest for low-mass, field disk galaxies (Figure 17). Finally, we note that the 20−25 Mpc $h^{-1}$ feature seen in $\xi(\Delta z)$ (Figure 18a and possibly in the 2D correlation, panel b), is mostly washed out in the projected radial correlation, suggesting its origin may be distortion related; nevertheless, there is a hint of something irregular at these distances – between 1.3 and 1.4, in the log – in the 0.1 to 0.3 redshift shell.

### 4.5. *3D distributions*

The overarching goal is to explore the *WISE*-GAMA galaxy catalog using tools that better visualize the 3D structures, which may be used in future studies that further explore the cosmic web, e.g., using the underlying LSS to improve photometric redshift estimates. We start with a simple, pseudo-3D method that is often utilized to represent multi-wavelength imaging: assign RGB colors to different layers, in this case, redshifts between $z < 0.1$ (blue), $0.1 < z < 0.2$ (green) and $0.2 < z < 0.3$ (red). Although these ranges are relatively large and blunt, Figure 19 does yield a crude projection map as to where structure is located. Dense clustering is seen for all layers, but specifically linked, filamentary structures are crossing the blue to green layers – overlap is revealed by cyan composite – toward the southern end (180 to 178 degree R.A.). It is interesting to note the projection of nearby (blue) and distant (red) structures around 181 degrees R.A. Bare in mind that the physical scale differences are significant for the three redshift shells; for example, the volume is much smaller in the first shell, only a small piece of the cosmic web is probed in the 'blue' shell.

Breaking free of projection and volume effects, we now utilize 3D visualization to explore the data, notably the Partiview system (Levy 2001)[22]. Partiview was developed for scientific research of complex data sets, and was applied to astronomical data sets through the Digital Universe effort (Brian Abbot of the AMNH), including the 2MASS XSCz (Jarrett et al. 2004). It is perfectly suited for exploring the *WISE*-GAMA catalogs; we have translated our galaxy catalog to spherical coordinates and the Partiview format. Not only do we examine our data looking for particle over-densities, but also color-code our galaxies (or points, in Partiview) according to some of their outstanding attributes – notably their *WISE* W2−W3 color, and their stellar mass.

Figures 20 and 21 show a spatial view of the galaxy sample as visualized in 3D by Partiview. The data appear as a circular cone, extending from the origin to a redshift of 0.3 (to the right), but it is in fact a rectangular cone constrained by the G12 equatorial limits (see Fig-ure 1). The total volume is about 0.011 $\mathrm{Gpc}^3$, which is adequate for discerning marginal large-scale structures. The plots are, of course, limited by the 2D nature of flat projections; it can only be fully appreciated using 3D analysis tools. Nevertheless, we can point out some interesting features. In projection the cosmic web is readily apparent: small clusters, filaments, walls and voids are apparent, notably in the central 0.1 to 0.2 regions. The lower panel of Figure 20 features the Alpaslan et al. (2015) filamentary catalog, constructed using minimal spanning trees, that form a 'skeleton' with connecting 'bones' of the underlying structure. This is, reassuringly, tracked closely by the 20 Mpc overdensity spheres (in green; see also Table 3). Redshift distortion is apparent with the more densely packed groupings; indeed, linear features are a telling signature for a galaxy cluster. For the last redshift shell ($z > 0.2$), the points are noticeably spread out and less dense, which is partly an illusion of the increasing volume; i.e., spatial scale is diminishing, in conjunction with the decreasing completeness at these depths: Malmquist bias favors the luminous, rare members of clusters in these diagrams.

[22] http://virdir.ncsa.illinois.edu/partiview/

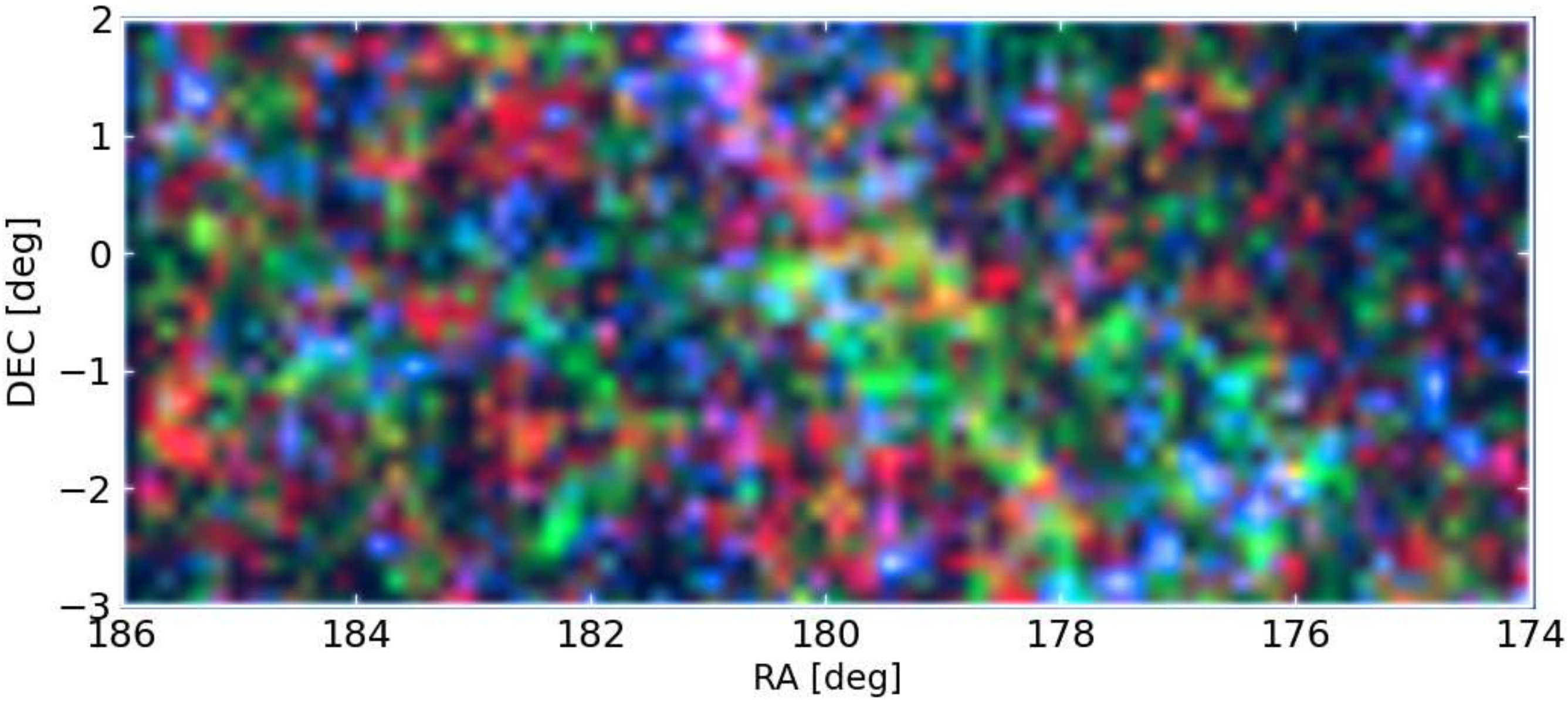

**Figure 19.** 3D multi-color view of *WISE* galaxies in the G12 region to a redshift limit of 0.3. Here blue represents sources with $z < 0.1$, green is $0.1 < z < 0.2$, and red is $0.2 < z < 0.3$. The horizontal and vertical features are an artifact of the 1-D smoothing process. Be aware that the physical scales between the shells are significantly different, see e.g., Figure 15.

A compelling demonstration of this kind of selection, redshift-dependent bias is seen in the top panel of Figure 21, which shows the same distribution, but now color-coded by host stellar mass, ranging from low mass in blue, to high mass in red. As expected, low mass galaxies can only be detected nearby since they are faint, while massive galaxies are seen all throughout, but notably at large distances since they are relatively rare and a larger volume is needed to see them in appreciable numbers. Moreover, they appear very clustered at large distances, fully consistent with the 2PCF clustering results presented in the previous section.

The more interesting result is shown in the lower two panels of Figure 21, now coded by the rest frame-corrected *WISE* W2−W3 color, which as we have demonstrated is, roughly, a proxy for galaxy Hubble Type (see Figure 11) and the sSFR. It can be seen that the early and intermediate types (red and green points) are more strongly clustered than the late-types (blue points, which make up the majority of field galaxies in this diagram). However, note that dust-free and early-types (red points) appear more in the Local Universe ($z < 0.1$) and are nearly invisible at high redshifts – this is because the diagram is coded by the *WISE* W2-W3 color, which is highly insensitive to galaxies whose light is dominated by old stars, the R-J tail in the mid-IR. There are plenty of spheroidals in the region (see Fig 21a), but they are not generally detected in the *WISE* W3 and W4 bands; hence, only the clustering from the SF disk galaxies is seen thoughout the diagram. Moreover, the GAMA selection introduces a bias to SF, disk galaxies.

In terms of sSFR, quenched and quiescent galaxies are mostly seen at low redshifts and are highly clustered, while the more actively SF galaxies, which are building their stellar mass, are more broadly distributed, as well as filling the entire volume, consistent with the *WISE* W3 sensitivity and the GAMA selections. These cone diagrams do not do full justice to the rich detail seen in the cosmic web, because of the 2D limitations; we refer the reader to ancillary 3D animations that accompany this paper.

The final figure, Figure 22, zooms into one of the most interesting complexes, in this case, at $z \sim 0.17$ (see also Figure 16), which demonstrates both the Mpc-scale clustering and the larger, connected super-structures as illustrated by the GAMA Groups skeleton. This redshift distance is still close enough that *WISE*-GAMA is still sampling a large range in stellar mass and galaxy types. However, as with Fig 21b, this diagram is coded by the *WISE* W2-W3 color, which is highly insensitive to galaxies whose light is dominated by old stars, the R-J tail in the mid-IR. There are plenty of spheroidals in the region (see Fig 21a), but they are not generally detected in the *WISE* W3 and W4 bands; hence, only the clustering of the SF galaxies is apparent. The tight clustering in the center of the diagram is dominated by the intermediate spiral galaxies, depicted in green, while the more randomly distributed field galaxies are mostly late-type and lower mass disks – Figs 12 & 13. This apparent clustering pattern is consistent with the 2PCF, highlighted in Fig 17b, showing the strongest clustering from spheroidals, not visible in this diagram, and the intermediate disk galaxies. Many studies have shown the environmental effects on galaxy evolution, with the most recent SDSS and *Spitzer* work pointing toward more rapid evolution and quenching for denser environments (see e.g., Walker et al. 2010; Cluver et al. 2013), which could explain the presence of lenticular and intermediate 'green valley' galaxies populating the groups and structures in the *WISE*-GAMA field. The intricate cosmic web of galaxies reveals some of its secrets with these diagrams, but clearly this is only scratching the surface, and powerfully demonstrates the need for redshift and multi-wavelength surveys to study the extragalactic Universe.

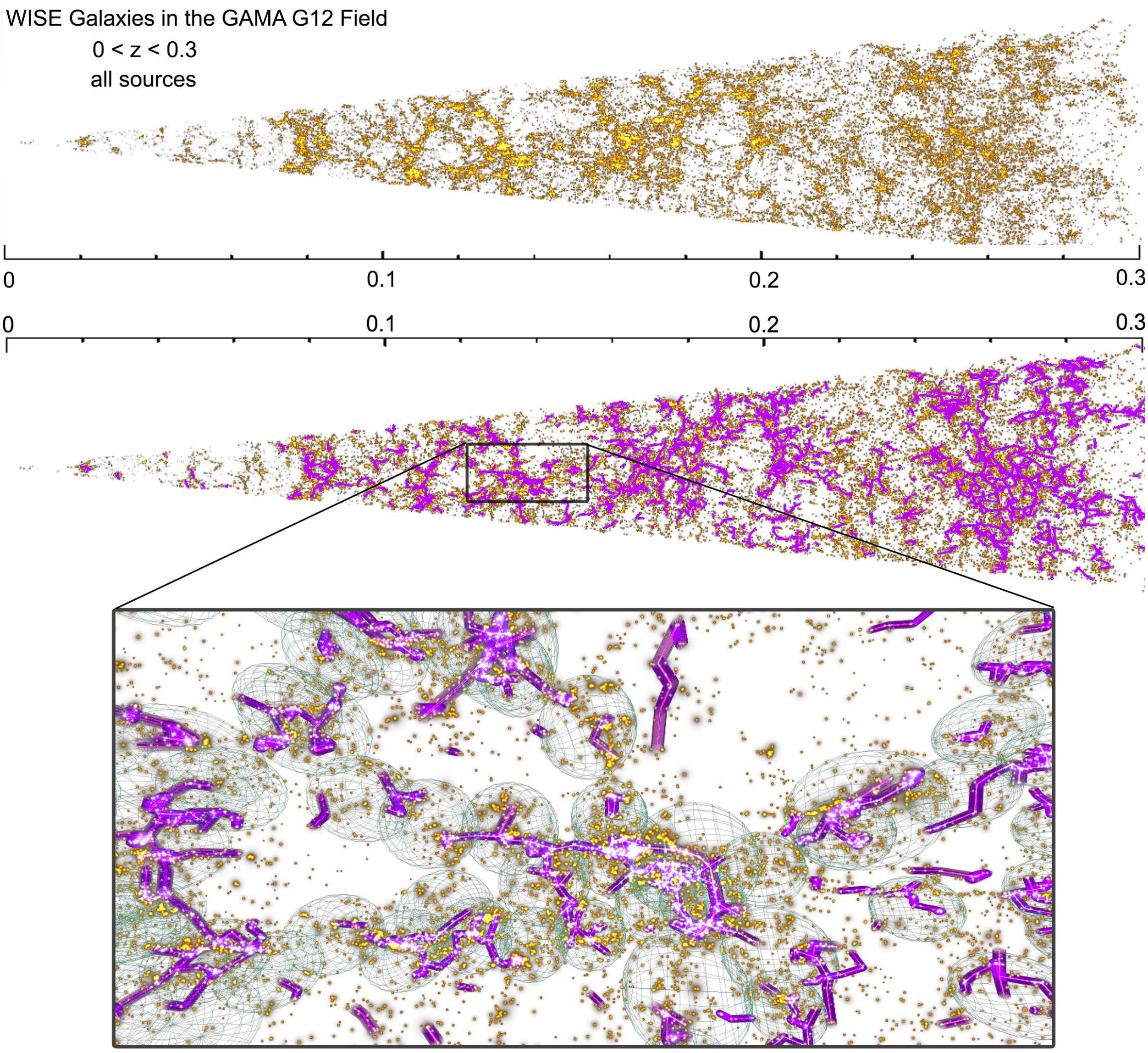

**Figure 20.** *WISE*-GAMA galaxies in G12, displayed using a 3D cone that extends to a redshift of 0.3, or a luminosity distance of 1564 Mpc (1200 Mpc in co-moving frame), and a total co-moving volume of 0.011 Gpc$^3$. The lower panel shows the same view, but with the GAMA galaxy groups filament catalog (Alpaslan et al. 2015) overlayed in magenta, illustrating the underlying 'skeleton' of the web of galaxies. The inset image shows a zoomed view of an overdensity region near z $\sim$ 0.14 to gain a better view of the 3D distribution of filaments and clusters. The green spheres are the over-dense regions determined in a 20 Mpc diameter region (see Table 3 and Figure 15). The graphics were made using the Partiview visualization system.

A set of fly-through animations have been prepared that provide a 3-D experience and fully render the information in this section. They are available online.

**Table 2**
Clustering in the G12 Field: The Largest overdensities in 5 Mpc spheres. The complete list is available online.

| R.A. deg | Dec deg | $z_{\rm spec}$ — | $D_L$ Mpc | N — | density Log N Mpc$^{-3}$ | X Mpc | Y Mpc | Z Mpc | Log $M_\star \pm$ err $M_\odot$ | $M_{\rm W1} \pm$ err mag |
|---|---|---|---|---|---|---|---|---|---|---|
| 181.55313 | 1.64476 | 0.01892 | 82.399 | 25 | -0.3169 | 5.225 | -37.219 | 71.605 | 9.872 0.488 | -19.10 0.94 |
| 178.93069 | -1.36402 | 0.01929 | 84.317 | 25 | -0.4180 | 4.142 | -43.012 | 70.538 | 10.354 0.583 | -19.17 0.83 |
| 181.76935 | 1.51087 | 0.01993 | 87.035 | 50 | -0.0729 | 5.891 | -39.278 | 75.528 | 10.429 0.732 | -19.07 0.90 |
| 180.03653 | -1.67075 | 0.02026 | 89.336 | 41 | -0.2031 | 6.087 | -45.129 | 74.789 | 10.309 0.555 | -19.09 0.83 |
| 184.57907 | -0.95518 | 0.02090 | 91.073 | 25 | -0.4180 | 11.818 | -41.875 | 77.878 | 9.543 0.592 | -19.11 3.58 |
| 182.32007 | 1.05400 | 0.02097 | 91.340 | 43 | -0.1824 | 7.272 | -41.322 | 79.016 | 9.995 0.608 | -19.07 0.79 |
| 179.46967 | -1.43977 | 0.02139 | 93.916 | 17 | -0.5855 | 5.418 | -47.520 | 78.531 | 9.873 0.572 | -19.47 0.79 |
| 176.58344 | -2.14046 | 0.02814 | 122.299 | 18 | -0.5606 | 2.474 | -65.099 | 99.527 | 10.137 0.437 | -20.23 1.01 |
| 175.27148 | -0.40431 | 0.02843 | 125.157 | 16 | -0.6118 | -1.691 | -65.030 | 102.852 | 9.845 0.523 | -19.87 0.83 |
| 184.88451 | -0.79698 | 0.04004 | 177.670 | 18 | -0.5606 | 23.193 | -79.399 | 149.468 | 10.380 0.456 | -20.62 0.82 |
| 174.27612 | -0.52319 | 0.04603 | 205.962 | 16 | -0.5192 | -5.530 | -106.877 | 165.275 | 10.243 0.646 | -20.72 0.83 |
| 177.86795 | -0.12341 | 0.05958 | 266.943 | 16 | -0.6118 | 5.893 | -128.969 | 216.339 | 10.183 0.483 | -21.38 0.85 |
| 185.37892 | -2.32491 | 0.06772 | 305.810 | 16 | -0.6118 | 44.690 | -137.525 | 247.230 | 10.237 0.468 | -21.73 0.77 |
| 184.23335 | -0.49195 | 0.07105 | 322.783 | 22 | -0.4735 | 37.176 | -140.536 | 263.992 | 10.586 0.511 | -22.25 0.77 |
| 174.28540 | 1.38668 | 0.07400 | 336.320 | 22 | -0.4368 | -13.800 | -162.122 | 267.558 | 10.490 0.391 | -21.82 0.89 |
| 175.11880 | 0.93215 | 0.07485 | 340.899 | 21 | -0.4937 | -8.725 | -164.271 | 271.163 | 10.353 0.436 | -21.93 0.88 |
| 184.17326 | 1.32090 | 0.07496 | 340.140 | 19 | -0.5372 | 33.893 | -140.204 | 281.632 | 10.414 0.390 | -21.70 0.83 |
| 175.06786 | 1.64995 | 0.07521 | 341.340 | 22 | -0.4569 | -10.902 | -161.523 | 273.083 | 10.561 0.391 | -22.25 0.89 |
| 183.76874 | -0.80118 | 0.07528 | 342.547 | 19 | -0.5372 | 37.886 | -151.064 | 277.900 | 10.265 0.442 | -22.00 0.76 |
| 175.01956 | 1.41704 | 0.07667 | 348.207 | 35 | -0.2718 | -10.710 | -165.659 | 277.555 | 10.536 0.394 | -22.02 0.82 |
| 174.42085 | -1.63919 | 0.07724 | 352.856 | 17 | -0.5841 | -5.386 | -182.174 | 272.169 | 10.137 0.392 | -21.59 7.57 |
| 177.87677 | -2.70261 | 0.07727 | 351.440 | 22 | -0.4444 | 14.770 | -177.765 | 273.146 | 10.600 0.480 | -22.05 0.85 |
| 179.83263 | 1.68845 | 0.07734 | 352.230 | 27 | -0.3597 | 12.438 | -154.703 | 287.758 | 10.451 0.371 | -22.08 0.87 |
| 176.27466 | -2.03264 | 0.07757 | 352.250 | 22 | -0.4735 | 4.960 | -179.181 | 273.365 | 10.507 0.504 | -21.77 0.78 |
| 175.03677 | -1.23212 | 0.07765 | 353.026 | 37 | -0.2477 | -3.435 | -179.086 | 274.283 | 10.426 0.460 | -22.08 5.93 |
| . | . | . | . | . | . | . | . | . | . | . |
| . | . | . | . | . | . | . | . | . | . | . |
| . | . | . | . | . | . | . | . | . | . | . |
| 176.34511 | -2.45364 | 0.12960 | 610.872 | 32 | -0.3108 | 10.708 | -299.033 | 450.460 | 10.683 0.461 | -22.96 2.60 |
| 179.62001 | -0.97884 | 0.12992 | 611.714 | 18 | -0.5606 | 31.045 | -275.964 | 464.726 | 10.752 0.519 | -23.16 2.55 |
| 179.30650 | -0.41568 | 0.13042 | 615.673 | 17 | -0.5855 | 26.039 | -274.966 | 469.414 | 10.406 0.366 | -22.95 3.55 |
| 179.62994 | -1.11922 | 0.13062 | 617.519 | 19 | -0.5372 | 32.046 | -279.351 | 468.237 | 10.591 0.456 | -22.95 1.55 |
| 179.24068 | 0.20777 | 0.13141 | 619.659 | 22 | -0.4735 | 22.753 | -272.339 | 474.631 | 10.535 0.439 | -22.84 2.45 |
| 176.37701 | -2.47675 | 0.13169 | 620.117 | 24 | -0.4357 | 11.227 | -303.029 | 456.403 | 10.586 0.435 | -22.98 1.36 |
| 178.77011 | -0.61900 | 0.13170 | 620.931 | 21 | -0.4937 | 22.685 | -280.689 | 470.891 | 10.678 0.459 | -22.88 0.78 |
| 179.47545 | -0.94969 | 0.13178 | 623.706 | 16 | -0.6118 | 30.251 | -281.312 | 472.908 | 10.436 0.259 | -23.18 2.57 |
| 179.33405 | 1.45814 | 0.13328 | 629.577 | 22 | -0.4735 | 17.998 | -266.695 | 487.000 | 10.540 0.334 | -22.91 0.79 |
| 174.58043 | 1.27971 | 0.13375 | 633.179 | 25 | -0.4180 | -21.589 | -288.789 | 477.532 | 10.759 0.397 | -23.28 1.96 |
| 178.34377 | -1.65366 | 0.13415 | 634.961 | 19 | -0.5372 | 24.386 | -295.660 | 474.795 | 10.411 0.343 | -23.26 2.19 |
| 177.51169 | -0.59069 | 0.13454 | 637.224 | 18 | -0.5606 | 12.296 | -292.419 | 479.375 | 10.380 0.243 | -22.93 2.73 |
| 174.37976 | 1.46601 | 0.13488 | 639.237 | 18 | -0.5606 | -24.387 | -290.651 | 481.865 | 10.595 0.372 | -23.00 0.77 |
| 178.65443 | -2.78462 | 0.13591 | 642.356 | 21 | -0.4849 | 32.697 | -305.433 | 474.796 | 10.532 0.444 | -23.14 1.79 |
| 177.56216 | -0.58711 | 0.13608 | 644.198 | 21 | -0.4937 | 12.835 | -294.978 | 484.099 | 10.774 0.588 | -23.08 1.55 |
| 178.43216 | -1.73963 | 0.13612 | 644.388 | 21 | -0.4937 | 25.885 | -299.782 | 480.788 | 10.674 0.537 | -22.73 1.86 |
| 178.36096 | -1.66714 | 0.13769 | 651.799 | 17 | -0.5855 | 25.171 | -302.582 | 485.841 | 10.417 0.348 | -23.07 3.09 |
| 177.62721 | -0.61205 | 0.13811 | 654.623 | 18 | -0.5606 | 13.712 | -299.125 | 491.094 | 10.759 0.418 | -23.53 0.77 |
| 180.22279 | 1.30882 | 0.13943 | 660.291 | 16 | -0.6118 | 27.357 | -275.366 | 509.153 | 10.698 0.439 | -23.29 0.79 |
| 179.46954 | -0.92528 | 0.14524 | 691.864 | 16 | -0.6118 | 32.983 | -308.229 | 518.526 | 10.627 0.445 | -23.22 2.36 |
| 178.12276 | -2.14124 | 0.15045 | 718.693 | 16 | -0.6118 | 27.674 | -334.828 | 526.670 | 10.599 0.450 | -23.36 0.82 |
| 177.91139 | 1.47029 | 0.15788 | 759.225 | 18 | -0.5606 | 6.940 | -321.721 | 571.308 | 10.648 0.454 | -23.31 0.78 |
| 182.28485 | -2.53203 | 0.15838 | 759.958 | 16 | -0.6118 | 72.786 | -333.609 | 560.189 | 10.446 0.249 | -23.59 1.94 |
| 180.32901 | -0.11719 | 0.16487 | 794.589 | 30 | -0.3388 | 41.534 | -336.362 | 592.041 | 10.705 0.338 | -23.83 1.31 |
| 178.78691 | -1.38413 | 0.16501 | 795.626 | 16 | -0.6118 | 32.816 | -356.024 | 581.868 | 10.435 0.277 | -23.20 1.61 |
| 180.16246 | -0.18769 | 0.16582 | 800.753 | 24 | -0.4357 | 40.484 | -340.196 | 595.317 | 10.808 0.564 | -23.27 1.90 |
| 180.45227 | -0.19379 | 0.16650 | 804.397 | 16 | -0.6118 | 43.725 | -340.034 | 598.320 | 10.821 0.547 | -23.47 2.24 |
| 180.38699 | -0.33766 | 0.16750 | 809.411 | 23 | -0.4542 | 44.113 | -343.511 | 600.583 | 10.699 0.385 | -23.67 1.50 |
| 176.98413 | -1.81699 | 0.16835 | 813.988 | 21 | -0.4937 | 16.852 | -376.398 | 586.029 | 10.683 0.353 | -23.51 1.60 |
| 180.81659 | -0.69076 | 0.16897 | 817.506 | 17 | -0.5855 | 51.154 | -347.351 | 604.818 | 10.561 0.298 | -23.61 1.56 |
| 179.99211 | -0.55074 | 0.16972 | 821.137 | 17 | -0.5855 | 41.700 | -351.930 | 605.973 | 10.806 0.464 | -23.36 2.79 |
| 185.27594 | 0.33258 | 0.17114 | 826.924 | 16 | -0.6118 | 93.311 | -315.461 | 624.766 | 10.685 0.287 | -24.05 1.54 |
| 182.29358 | -2.26009 | 0.17434 | 845.192 | 24 | -0.4357 | 78.304 | -363.445 | 616.225 | 10.663 0.337 | -23.73 1.24 |
| 182.19055 | -1.88694 | 0.17485 | 848.786 | 16 | -0.6118 | 75.216 | -362.011 | 620.680 | 10.586 0.309 | -23.84 2.03 |
| 182.30838 | -2.30924 | 0.17651 | 856.321 | 20 | -0.5149 | 79.650 | -367.925 | 622.937 | 10.755 0.393 | -23.64 1.90 |
| 180.00969 | -0.62016 | 0.17970 | 874.033 | 19 | -0.5372 | 44.650 | -372.000 | 639.177 | 10.531 0.334 | -23.55 2.03 |
| 181.96855 | -1.18609 | 0.17980 | 874.058 | 24 | -0.4357 | 70.264 | -365.904 | 640.343 | 10.698 0.334 | -23.71 1.59 |
| 181.91740 | -1.19125 | 0.18057 | 880.405 | 17 | -0.5855 | 70.181 | -368.678 | 644.428 | 10.649 0.287 | -23.86 2.19 |
| 182.52495 | -0.44414 | 0.18158 | 884.374 | 20 | -0.5149 | 72.637 | -359.116 | 652.658 | 10.944 0.403 | -24.05 2.46 |
| 183.64435 | 0.46899 | 0.18296 | 892.272 | 19 | -0.5372 | 80.198 | -345.987 | 665.421 | 10.633 0.337 | -23.83 1.97 |
| 185.70647 | -1.67666 | 0.18738 | 915.635 | 16 | -0.6118 | 119.950 | -361.732 | 670.386 | 10.599 0.345 | -23.56 2.49 |
| 182.16962 | 1.84763 | 0.18847 | 922.305 | 20 | -0.5045 | 56.076 | -351.223 | 689.741 | 10.683 0.312 | -23.77 3.30 |
| 183.11053 | 1.84660 | 0.19486 | 956.029 | 17 | -0.5775 | 69.247 | -355.991 | 713.207 | 10.521 0.229 | -23.95 1.61 |
| 176.12039 | -2.65271 | 0.20450 | 1010.023 | 18 | -0.5606 | 15.137 | -467.134 | 696.211 | 10.646 0.265 | -23.87 1.45 |
| 185.44620 | -1.15147 | 0.25964 | 1324.008 | 19 | -0.5372 | 154.749 | -488.320 | 917.828 | 11.086 0.401 | -24.61 2.02 |

All distances derived using $H_0 = 70\,{\rm km\,s^{-1}\,Mpc^{-1}}$.

**Table 3**
Large Scale Structures G12 Field: Largest overdensities in 20 Mpc spheres

| R.A. | Dec | $z_{\rm spec}$ | $D_L$ | N | density | X | Y | Z | Log $M_\star$± err | $M_{W1}$± err |
|---|---|---|---|---|---|---|---|---|---|---|
| deg | deg | — | Mpc | — | Log N Mpc$^{-3}$ | Mpc | Mpc | Mpc | $M_\odot$ | mag |
| 180.33488 | 0.50015 | 0.02045 | 87.560 | 332 | -1.1010 | 4.784 | -41.612 | 74.887 | 10.153 0.697 | -19.02 0.78 |
| 184.32265 | -0.04592 | 0.07412 | 335.529 | 163 | -1.4099 | 37.781 | -143.622 | 274.816 | 10.369 0.419 | -21.73 4.67 |
| 175.02951 | 0.92720 | 0.07531 | 341.980 | 215 | -1.2897 | -9.169 | -164.941 | 271.759 | 10.420 0.412 | -21.85 0.77 |
| 180.89407 | 0.81483 | 0.07771 | 353.678 | 303 | -1.1406 | 20.220 | -156.354 | 287.827 | 10.477 0.427 | -21.85 0.78 |
| 175.71800 | -1.52843 | 0.07812 | 354.261 | 271 | -1.1891 | 0.795 | -179.305 | 275.357 | 10.439 0.493 | -21.77 3.67 |
| 178.30112 | 0.40035 | 0.07829 | 355.287 | 170 | -1.3916 | 8.427 | -165.356 | 284.870 | 10.259 0.446 | -21.81 2.55 |
| 184.23534 | 0.63296 | 0.07850 | 357.366 | 169 | -1.3942 | 37.734 | -149.646 | 293.220 | 10.387 0.453 | -21.76 0.77 |
| 180.11440 | -1.41423 | 0.08204 | 374.283 | 260 | -1.2071 | 23.710 | -176.934 | 296.281 | 10.504 0.458 | -21.96 3.11 |
| 177.43248 | -1.21826 | 0.08230 | 374.500 | 194 | -1.3343 | 8.988 | -183.154 | 293.437 | 10.275 0.463 | -21.86 2.50 |
| 181.01277 | 0.78844 | 0.08275 | 380.585 | 147 | -1.4548 | 22.369 | -167.257 | 308.344 | 10.385 0.456 | -21.94 0.76 |
| 180.77678 | -0.50476 | 0.09487 | 435.652 | 142 | -1.4698 | 28.240 | -196.801 | 344.672 | 10.387 0.467 | -22.12 2.40 |
| 175.57585 | -1.67503 | 0.10689 | 495.514 | 313 | -1.1265 | 0.665 | -245.569 | 374.296 | 10.568 0.424 | -22.46 4.06 |
| 179.35757 | -0.15798 | 0.10714 | 499.831 | 222 | -1.2757 | 20.954 | -226.235 | 390.123 | 10.520 0.569 | -22.55 1.98 |
| 183.65300 | -2.44568 | 0.11078 | 515.948 | 141 | -1.4576 | 60.830 | -230.485 | 398.659 | 10.533 0.414 | -22.66 3.40 |
| 176.24292 | -1.54479 | 0.11803 | 553.756 | 203 | -1.3146 | 5.237 | -268.521 | 416.158 | 10.550 0.426 | -22.68 2.54 |
| 184.39169 | -0.69596 | 0.11821 | 553.955 | 176 | -1.3766 | 63.147 | -231.665 | 433.312 | 10.556 0.400 | -22.82 2.75 |
| 176.60522 | -1.92923 | 0.12214 | 572.785 | 156 | -1.4290 | 9.874 | -277.899 | 428.046 | 10.523 0.424 | -22.88 2.49 |
| 183.27890 | 0.69813 | 0.12580 | 589.858 | 157 | -1.4262 | 51.789 | -240.348 | 462.677 | 10.509 0.399 | -22.73 2.84 |
| 178.53511 | -0.52065 | 0.12713 | 596.850 | 159 | -1.4207 | 19.554 | -271.164 | 454.412 | 10.511 0.432 | -22.78 2.13 |
| 176.60103 | -2.33004 | 0.13027 | 612.363 | 183 | -1.3596 | 12.281 | -297.738 | 452.473 | 10.564 0.399 | -22.97 2.11 |
| 179.28078 | -0.25154 | 0.13048 | 616.375 | 262 | -1.2038 | 25.098 | -274.216 | 470.589 | 10.541 0.435 | -22.82 1.71 |
| 175.08627 | 1.34118 | 0.13398 | 632.891 | 147 | -1.4548 | -17.557 | -286.185 | 478.834 | 10.573 0.413 | -22.93 1.60 |
| 177.96019 | -1.19788 | 0.13653 | 645.595 | 233 | -1.2547 | 19.238 | -298.296 | 483.031 | 10.595 0.469 | -22.93 1.93 |
| 184.96928 | -0.25953 | 0.15707 | 755.438 | 150 | -1.4460 | 86.521 | -298.435 | 574.208 | 10.653 0.340 | -23.37 1.88 |
| 178.19872 | 1.34711 | 0.15833 | 758.650 | 141 | -1.4729 | 10.489 | -321.023 | 570.785 | 10.600 0.338 | -23.33 2.21 |
| 178.35277 | -1.80101 | 0.16428 | 791.505 | 150 | -1.4460 | 30.550 | -360.255 | 575.711 | 10.633 0.341 | -23.35 1.73 |
| 180.22833 | -0.20266 | 0.16543 | 797.140 | 216 | -1.2876 | 41.086 | -338.554 | 592.902 | 10.643 0.465 | -23.36 1.43 |
| 180.35797 | -0.44359 | 0.16971 | 820.634 | 146 | -1.4577 | 44.958 | -348.747 | 607.088 | 10.610 0.387 | -23.33 1.61 |
| 182.21306 | -2.20682 | 0.17526 | 850.452 | 195 | -1.3321 | 77.520 | -365.410 | 619.762 | 10.663 0.386 | -23.43 1.44 |
| 179.87987 | -0.41877 | 0.17882 | 868.384 | 142 | -1.4698 | 41.679 | -368.691 | 636.389 | 10.666 0.339 | -23.64 1.53 |
| 182.13396 | -0.83328 | 0.18022 | 876.443 | 174 | -1.3815 | 70.092 | -362.401 | 644.378 | 10.682 0.386 | -23.58 1.30 |
| 183.60109 | 0.24574 | 0.18402 | 896.635 | 176 | -1.3766 | 81.447 | -349.853 | 666.666 | 10.630 0.344 | -23.59 1.04 |
| 179.85704 | 0.79264 | 0.20046 | 987.146 | 147 | -1.4548 | 37.826 | -398.679 | 718.201 | 10.758 0.374 | -23.77 1.18 |
| 179.85704 | 0.79264 | 0.20046 | 987.146 | 147 | -1.4548 | 37.826 | -398.679 | 718.201 | 10.758 0.374 | -23.77 1.18 |
| 182.39772 | -1.61033 | 0.20110 | 991.975 | 113 | -1.5690 | 86.660 | -409.540 | 711.941 | 10.722 0.310 | -23.81 1.19 |
| 180.04552 | -0.39097 | 0.20218 | 997.541 | 84 | -1.6978 | 48.849 | -413.926 | 717.501 | 10.686 0.243 | -23.91 1.59 |
| 176.04239 | -2.59983 | 0.20400 | 1008.114 | 77 | -1.6879 | 13.743 | -466.355 | 695.271 | 10.607 0.271 | -23.76 1.42 |
| 178.03131 | -2.03186 | 0.22282 | 1115.575 | 76 | -1.7413 | 38.296 | -488.322 | 769.650 | 10.660 0.277 | -23.96 1.14 |
| 176.97334 | 1.41832 | 0.23393 | 1176.821 | 111 | -1.5764 | -3.145 | -475.200 | 826.894 | 10.715 0.268 | -23.97 1.05 |
| 185.57928 | 1.28286 | 0.23620 | 1187.643 | 80 | -1.6926 | 123.615 | -414.725 | 857.734 | 10.666 0.269 | -24.02 1.21 |
| 180.81497 | 1.29220 | 0.23785 | 1199.193 | 81 | -1.7136 | 54.608 | -456.062 | 852.962 | 10.822 0.293 | -24.21 1.32 |
| 183.74768 | 1.52506 | 0.24372 | 1230.890 | 79 | -1.7146 | 97.912 | -439.348 | 881.397 | 10.887 0.343 | -24.22 1.35 |
| 180.07970 | 0.03818 | 0.24914 | 1261.893 | 78 | -1.7300 | 56.342 | -498.026 | 877.108 | 10.923 0.394 | -24.29 1.20 |
| 180.61147 | -1.66670 | 0.24921 | 1263.331 | 87 | -1.6826 | 79.130 | -516.631 | 865.775 | 10.860 0.323 | -24.29 1.01 |
| 183.81076 | 0.87965 | 0.24983 | 1265.483 | 76 | -1.7413 | 106.688 | -457.590 | 896.902 | 10.828 0.350 | -24.17 1.50 |

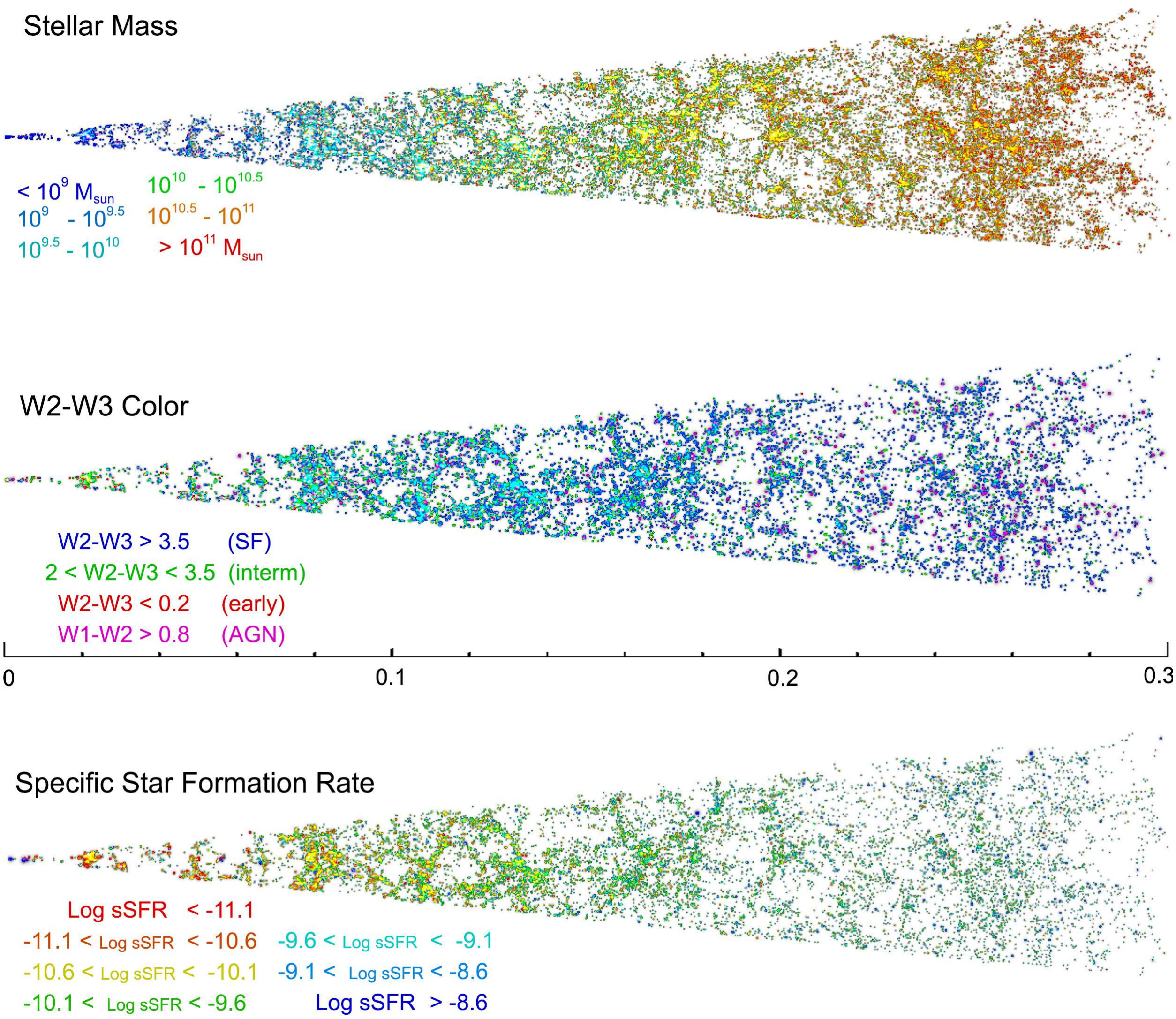

**Figure 21.** Three different views of the spatial cone distribution of *WISE* galaxies in G12. The upper panel is color-coded by the stellar mass (compare with Figure 11b), the middle panel by the *WISE* W2−W3 color, a proxy for morphological type: delineating early (spheroidal), intermediate (disks) and late-types (disks), and the bottom panel by the Specific Star Formation Rate (compare with Figure 13).

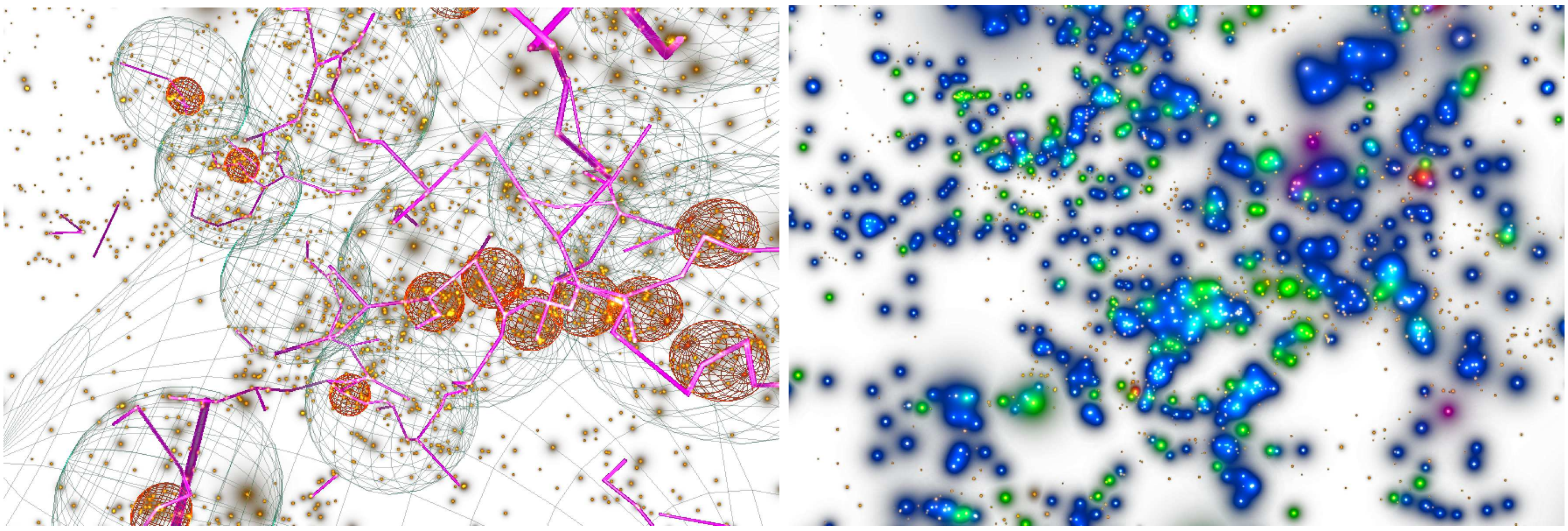

**Figure 22.** Close up view of a structure near ∼0.17 in redshift. The red spheres have a 5 Mpc diameter, and the larger green spheres are 20 Mpc in diameter; the magenta are filamentary connections (or bones), highlighting the different scale of structures from single galaxies, to groups and to superclusters. The second panel is color-coded by the *WISE* W2−W3 color (see e.g., Figure 11): spheroidals are depicted red, intermediate disks are green, late-type disks are blue, and AGN/LIRGS are magenta.

## 5. CONCLUSIONS & SUMMARY

Building on the early studies of *WISE* extragalactic populations (Jarrett et al. 2011; Yan et al. 2013; Assef et al. 2013) and the first GAMA-*WISE* study, Cluver et al. (2014), we have more broadly characterized the mid-infrared sources found in the 60 deg$^2$ GAMA G12 field, located within the Northern Galactic Cap, with the goal to map the 3D large scale structures, study the principal physical attributes of their host galaxies, and further explore the high redshift sources that are beyond the GAMA spectroscopic sensitivity. We choose mid-IR *WISE* as our base imaging survey because it fully covers the GAMA fields, provides a window to the nature of the host galaxies – notably stellar mass and star formation activity – that comprise the cosmic web, is sensitive to galaxies in the early Universe, and most importantly, because *WISE* will be one of the primary ancillary data sets used by the next generation radio, imaging and spectrocopic surveys owing to its 4-$\pi$ sky coverage and depth. It is our intent to characterize the behavior of source populations detected by *WISE* in anticipation of these future surveys.

To summarize, we have investigated the following:

- Cross-matched the ALLWISE catalog in G12, some 800,000 sources, with the GAMA redshift catalog, ∼60,000 sources, achieving over 95% matches for all GAMA sources, and with 981 sources from the LARGESS radio galaxy catalog drawn from FIRST (corresponding to a detection threshold of ∼0.75 mJy).

- The stellar and extragalactic nature of the *WISE* sources, delineating the sample by (<1%) resolved galaxies, sources with redshifts (<8%; GAMA galaxies), (∼26%) foreground stars and, the majority (74%), ∼591,400, likely extragalactic sources that extend to high redshifts.

- Global stellar masses. Stellar masses range from $10^7$ to $10^{12}$ $M_\odot$, with the most common value $\sim 10^{10.3}$, while for the resolved *WISE* galaxies, the average is much higher because resolved sources are a combination of low redshift (nearby) and large in angular size galaxies, translating to massive hosts.

- The SFR versus stellar mass distribution of galaxies. SFRs range from fully quenched to active >100 $M_\odot yr^{-1}$, but with most galaxies forming stars at >1 $M_\odot yr^{-1}$, consistent with the GAMA survey selection of SF blue galaxies and the *WISE* sensitivity to SF galaxies. Comparing the SFR and host mass – the specific SFR – lower mass galaxies are actively building their disks, intermediate-massed galaxies have ensemble or main-sequence building, but some are likely in their starburst phase, while massive galaxies have consumed their gas reservoirs and, for the most part, completed building their super-structure, evolving to a quiescent state. There is evidence for a 'merger' track, consisting of high mass galaxies or systems in a heightened state of SF.

- Radio galaxies from the LARGESS study exhibit infrared colors that have associations based on their spectroscopic type. AeBs strongly group in the Type-I QSO region of the *WISE* color diagram, LERGs and bulge-dominated galaxies are closely associated, while HERGs can be associated with SF and AGN groupings, and finally LARGESS SF-classified galaxies are consistent with *WISE*-identfied disk/spiral galaxies.

- Differential galaxy number counts in the W1 (3.4 $\mu$m) band, rise steadily to a peak of $10^{3.9}$ deg$^{-2}$ mag$^{-1}$ at W1 = 17.5 mag (31 $\mu$Jy). Compared to the equivalent counts from the *Spitzer* Deep-Wide Field (Bootes) Survey, the *WISE* counts are 2× lower, likely due to a combination of contaminant stars in the SDWFG, cosmic variance and growing incompleteness in the faint (and high redshift) *WISE* counts due to source blending and sensitivity variations (e.g., bright star halos). Comparing to deep K-band galaxy counts, converted to equivalent W1 using rest-frame color distribution models, the correspondence is reasonably good at the faint end, but the K-band shows a paucity at intermediate flux levels.

- At the faint end of the galaxy catalog, where redshifts are not available, we employ a luminosity function analysis to show that a substantial fraction, 27%, of sources are at high redshift, $z > 1$, although our models become highly uncertain at these depths due to our lack of understanding of how the LF evolves and changes for these early epochs. The *WISE* source counts are confirmed to be incomplete for W1 > 17th mag (49 $\mu$Jy).

- The galaxy selection function based on the GAMA redshifts, which is used to study the angular and radial clustering of the galaxy distribution to $z <$ 0.5.

- Two-point angular correlation functions, w($\theta$), for the sample delineated by measured brightness, stellar mass, color (morphological type) and redshift ranges. We find that brighter magnitudes cluster more strongly than fainter magnitudes, with a consistent decrease in the clustering at lower stellar masses. Bulge-dominated galaxies have the strongest clustering, intermediate disk (S0/Sa,Sb) also show clustering, while late-type spiral galaxies have the lowest amplitudes, consistent with the stellar mass results. At low redshifts, z$z < 0.3$, galaxies with AGN colors tend to have relatively low clustering amplitudes, and a scaling distribution that is flatter (in slope) than trend for all other samples; however, with low number statistics and the GAMA selection against QSOs, this clustering result is tentative.

- Two-point radial correlation functions, including $\xi(\Delta z)$, and the 2-dimensional parallel-to-transverse correlation, $\xi(\pi, \sigma)$ as a function of redshift shells to $z = 0.5$. The only solid correlation occurs for closely spaced galaxies, but there is an intriguing feature at ∼25 Mpc h$^{-1}$ at $z$ between 0.3 to 0.5, which is larger than the expected redshift distortion scale.

- 3D source over-densities using two different sampling scales: 5 Mpc and 20 Mpc spheres. We find a number of complexes and linked structures, including filamentary walls and super-structures. We investigate a connecting group-cluster at $z = 0.17$. There is reasonable correspondence between this simple LSS catalog and that of the GAMA Catalogue of Galaxy Groups. Finally, we map the structures using 3D visualization tools, exploring the LSS and local clustering may play a role with stellar mass and galaxy type.

## ACKNOWLEDGEMENTS

THJ thanks Rien van de Weygaert, Matthew Colless and David Hogg for many fruitful discussions, and Mark Subbarao for guidance/support with 3D rendering. We would like to thank the anonymous referee for the valuable insights that have greatly improved the work and manuscript presented here. MEC and THJ acknowledge support from the National Research Foundation (South Africa). M. Bilicki is supported by the Netherlands Organization for Scientific Research, NWO, through grant number 614.001.451, and through FP7 grant number 279396 from the European Research Council, as well as by the Polish National Science Center under contract #UMO-2012/07/D/ST9/02785. This publication makes use of data products from the Wide-field Infrared Survey Explorer, which is a joint project of the University of California, Los Angeles, and the Jet Propulsion Laboratory/California Institute of Technology, funded by the National Aeronautics and Space Administration.

GAMA is a joint European-Australasian project based around a spectroscopic campaign using the Anglo-Australian Telescope. The GAMA input catalog is based on data taken from the Sloan Digital Sky Survey and the UKIRT Infrared Deep Sky Survey. Complementary imaging of the GAMA regions is being obtained by a number of independent survey programs including GALEX MIS, VST KIDS, VISTA VIKING, *WISE*, Herschel- ATLAS, GMRT and ASKAP providing UV to radio coverage. GAMA is funded by the STFC (UK), the ARC (Australia), the AAO, and the participating institutions. The GAMA website is http://www.gama-survey.org/.